\documentclass[11pt,a4paper]{article}
\pdfoutput=1
\usepackage{amssymb,amsmath,amsfonts, mathtools, mathrsfs}
\usepackage[utf8]{inputenc} 
\usepackage[dvipsnames]{xcolor}

\newif\ifnatbibsort\natbibsorttrue

\relax

\ifnatbibsort\RequirePackage[numbers,sort&compress]{natbib}\else\RequirePackage[numbers,compress]{natbib}\fi
\RequirePackage[colorlinks=true
,urlcolor=blue
,anchorcolor=blue
,citecolor=blue
,filecolor=blue
,linkcolor=blue
,menucolor=blue
,pagecolor=blue
,linktocpage=true
,pdfproducer=medialab
,pdfa=true
]{hyperref}

\usepackage{graphicx}
\usepackage{caption}
\usepackage{tensor}
\usepackage{subfigure}
\usepackage{enumerate}
\usepackage{dsfont}
\usepackage{verbatim}
\usepackage{amsfonts,euscript,amssymb,stmaryrd,braket}
\usepackage{tikz}
\usetikzlibrary{arrows,decorations.markings,patterns}
\usepackage{slashed}

\def\clock{{\count0=\time
		\divide\count0 60
		\ifnum\count0<10 0\fi\the\count0
		\multiply\count0 -60 \advance\count0 \time
		:\ifnum\count0<10 0\fi \the\count0
}}
\newcommand{\timestamp}{{\small\vbox{\hbox{\tt\jobname.tex}
			\hbox{\the\day/\the\month/\the\year, \clock}}}}

\newcommand{\nn}{\nonumber}
\newcommand{\bea}{\begin{eqnarray}}
\newcommand{\eea}{\end{eqnarray}}

\newcommand{\be}{\begin{equation}}
\newcommand{\ee}{\end{equation}}

\makeatletter
\let\old@startsection=\@startsection
\let\oldl@section=\l@section
\renewcommand{\@startsection}[6]{\old@startsection{#1}{#2}{#3}{#4}{#5}{#6\mathversion{bold}}}
\renewcommand{\l@section}[2]{\oldl@section{\mathversion{bold}#1}{#2}}
\makeatother

\numberwithin{equation}{section}

\def \be {\begin{equation}}
\def \ee {\end{equation}}
\def \ba {\begin{array}}
\def \ea {\end{array}}
\def \bea{\begin{eqnarray}}
\def \eea{\end{eqnarray}}
\def \nn {\nonumber}

\def \Tr {{\textrm{Tr}}}

\def \and {{~\textrm{and}~}}

\usepackage{color}

\begin{document}
	\renewcommand{\thefootnote}{\arabic{footnote}}

	\overfullrule=0pt
	\parskip=2pt
	\parindent=12pt
	\headheight=0in \headsep=0in \topmargin=0in \oddsidemargin=0in

	\vspace{ -3cm} \thispagestyle{empty} \vspace{-1cm}
	\begin{flushright} 
		\footnotesize
		\textcolor{red}{\phantom{print-report}}
	\end{flushright}

%
	
\begin{center}
	\vspace{.0cm}
	
	
		\vspace{0cm} 
		{\Large\bf \mathversion{bold}
	Probing RG flows, symmetry resolution and quench dynamics
	}
	\\
	\vspace{.25cm}
	\noindent
	{\Large\bf \mathversion{bold}
	through the capacity of entanglement
	}

	\vspace{0.8cm} {
		Ra\'ul Arias$^{\,a,b}$,
		Giuseppe Di Giulio$^{\,c}$,
		Esko Keski-Vakkuri$^{\,d,e}$
		and Erik Tonni$^{\,f}$
	}
	\vskip  0.5cm
	
	\small
	{\em
		$^{a}\,$Instituto de F\'isica de La Plata, CONICET\\
Diagonal 113 e/63 y 64, CC67, 1900 La Plata, Argentina
		\vskip 0.05cm
		$^{b}\,$Departamento de F\'isica, Universidad Nacional de La Plata,\\ Calle 49 y 115 s/n, CC67, 1900 La Plata, Argentina
		\vskip 0.05cm
		$^{c}\,$Institute for Theoretical Physics and Astrophysics and W\"urzburg-Dresden Cluster of Excellence ct.qmat, Julius-Maximilians-Universit\"at W\"urzburg, Am Hubland, 97074 W\"{u}rzburg, Germany

		\vskip 0.05cm
		$^{d}\,$Department of Physics, University of Helsinki \\
		PO Box 64, FIN-00014 University of Helsinki, Finland
                     \vskip 0.005cm 
                      $^{e}\,$Helsinki Institute of Physics \\
                      PO Box 64, FIN-00014 University of Helsinki, Finland
		\vskip 0.05cm
		$^{f}\,$SISSA and INFN Sezione di Trieste, via Bonomea 265, 34136, Trieste, Italy 
	}
	\normalsize

\end{center}

\vspace{0.3cm}
\begin{abstract} 
We compare the capacity of entanglement with the entanglement entropy 
by considering various aspects of these quantities
for free bosonic and fermionic models in one spatial dimension, 
both in the continuum and on the lattice.
Substantial differences are observed in the subleading terms of these entanglement quantifiers 
when the subsystem is made by two disjoint intervals,
in the massive scalar field and in the fermionic chain. 
We define $c$-functions based on the capacity of entanglement
similar to the one based on the entanglement entropy,
showing through a numerical analysis that they display a monotonic behaviour 
under the renormalisation group flow generated by the mass.
The capacity of entanglement and its related quantities
are employed to explore the symmetry resolution.
The temporal evolutions of the capacity of entanglement and of the corresponding contour function after a global quench are also discussed. 
\end{abstract}

\newpage

\tableofcontents

\section{Introduction}
\label{sec:intro}

The reduced density matrix $\rho_A$ of a subsystem $A$ provides the entanglement entropy 
\be
\label{defentropy}
S_A=-\textrm{Tr}\big(
\rho_A\ln \rho_A
\big)\,,
\ee
that quantifies the bipartite entanglement between $A$ and its complement.
This entanglement measure can also be obtained from the R\'enyi entropies $S^{(n)}_A$, 
a family of infinitely many entanglement quantifiers, as follows
\be\\
S^{(n)}_A=\frac{1}{1-n}\ln\textrm{Tr}\rho_A^n\,,
\qquad
S_A=\lim_{n\to 1} S_A^{(n)}\,.
\ee
There are fascinating connections between thermodynamic entropy,  entanglement entropy, and gravity in asympotically anti-de Sitter spacetimes. 
Recently, analogous connections have been studied between the thermodynamic
heat capacity and concepts of quantum information theory and gravity. 

In the context of entanglement in many-body physics and quantum field theory, {\em capacity of entanglement} $C_A(\rho_A )$, 
as introduced in \cite{Yao:2010woi} and \cite{schliemann},  was first modeled after the definition of thermal heat capacity, and proposed to detect different phases in topological matter. As thermodynamic heat capacity is related to the variance of thermodynamical entropy, it was realized that
capacity of entanglement is equal to the variance of the entanglement Hamiltonian $K_A = -\ln \rho_A$, and can also be derived from the R\'enyi entropies  \cite{Perlmutter:2013gua, deBoer:2018mzv, Nakaguchi:2016zqi} as follows
\be
\label{defcapacity}
C_A\,=\,
\partial^2_n \big( \log \textrm{Tr}\rho_A^n \big)\big|_{n=1}
=\,
\partial_n^2 \big( \textrm{Tr}\rho_A^n \big)\big|_{n=1}
-
\big[ \partial_n \big( \textrm{Tr}\rho_A^n \big)\big]^2 \big|_{n=1}
=
\,\langle K_A^2 \rangle-\langle K_A \rangle^2 \, .
\ee

Meanwhile, in quantum information theory, variance has appeared e.g.  
in subleading corrections to Landauer inequality \cite{Reeb_2014}, in the context of majorizing state transitions \cite{boes2020}, and in state interconvertibility in
finite systems \cite{Chubb2018}.

Recently, in the context of quantum field theories, gravity, and random states, there has been growing interest in capacity of entanglement \cite{Nakaguchi:2016zqi, deBoer:2018mzv, Verlinde:2019ade, Arias:2020qpg, deBoer:2020snb, Zurek:2020ukz, Kawabata:2021hac, Okuyama:2021ylc, Kawabata:2021vyo, Iso:2021dlj, Nandy:2021hmk, Prihadi:2021mtu, Banks:2021jwj, Bhattacharjee:2021jff, Caputa:2021sib, Boudreault:2021pgj, Patramanis:2021lkx, Huang:2021qzt, Allameh:2021moy, Bianchi:2021aui, Mintchev:2022xqh, SomDun2022,Zurek:2022xzl,Gukov:2022oed,Wei:2022bed,Chiriaco:2022cjf,Shrimali:2022bvt,PRXQuantum.3.030201,Verlinde:2022hhs,Li:2022mvy}. In part, the interest arises from the holographic duality between conformal field theories (CFTs)
and quantum gravity in asymptotically anti-de Sitter spacetimes. In this setting, the area law of entanglement entropy in a CFT was found to have a geometrical interpretation as the area of a minimal surface in the 
bulk spacetime \cite{Ryu:2006bv}. Later,  also R\'enyi entropies were interpreted in this context, taking into account gravitational backreaction \cite{Dong:2016fnf}.
It was then anticipated that variance of entanglement entropy is associated with gravitational fluctuations, and an interpretation based on \cite{Dong:2016fnf} and (\ref{defcapacity}) was developed in 
\cite{Nakaguchi:2016zqi, deBoer:2018mzv}. There are also other proposals to relate capacity of entanglement (alternatively called modular fluctuations) to quantum fluctuations in e.g.  \cite{Verlinde:2019ade,Banks:2021jwj,Verlinde:2022hhs} motivating suggestions that fluctuations may accumulate to give rise to possibly observable effects e.g.  in laser interferometry \cite{Verlinde:2019ade,Zurek:2020ukz,Banks:2021jwj,Zurek:2022xzl,Li:2022mvy}.  Another context where capacity of entanglement has been explored is the Page curve
of Hawking radiation. In the spirit of \cite{Yao:2010woi}, capacity of entanglement is seen to have a qualitatively different behaviour in different phases: it marks the transition peaking
at the Page time where the black hole develops an island region \cite{Kawabata:2021hac}. Further work in this direction can be found in 
\cite{Kawabata:2021vyo, Okuyama:2021ylc}, qualitatively related observations have also been made in the context of time evolution of local operators \cite{Nandy:2021hmk} and in phase transitions of entanglement spectrum. 
Finally, we note that, due to work in many different areas and the relative novelty, nomenclature has not yet been established: the same quantity is referred to as capacity of entanglement, entanglement capacity, entropy variance, varentropy, variance of surprisal, and modular fluctuations. This variance in terminology hampers somewhat the task of identifying relevant literature.    

The goal of this work is to further compare capacity of entanglement and entanglement entropy,  both by direct computations and via extending other entropy-based concepts to analogous concepts with definitions based on capacity. Our arena will be
that of simple two-dimensional CFTs and related discrete models,  allowing explicit calculations. 
Our direct comparisons begin from the area law of capacity of entanglement, which was found in \cite{deBoer:2018mzv} to hold for a global ground state in conformal field theories. 
In 1+1 dimensional CFTs on the line, in their ground state 
and for an interval $A$ of length $\ell$,
the law takes the sharpest form, where, 
in the series expansion as the UV cutoff $\epsilon \to 0$,
the leading term behavior of capacity of entanglement equals that of entanglement entropy, 
\be\label{CAequalsSA}
C_A = S_A = \frac{c}{3} \log \!\left( \frac{\ell}{\epsilon}\right) + O(1)\,,
\ee
where $c$ is the central charge of the model. 
We begin by exploring some other cases, such as two intervals, to identify subleading finite modifications to the above equality.  

The leading term equality $C_A = S_A$ was also found to hold for a finite system, and even for the time evolution
after a global or local quench \cite{deBoer:2018mzv}.  In the case of the quench, the equality is in conflict with the heuristic explanation of entanglement spreading by maximally entangled pairs of quasiparticles created at the quench propagating in opposite directions \cite{Calabrese_2005}. Tracing out one member of a pair, the entanglement entropy carried by a quasiparticle into the interval is maximal, while
it carries zero capacity. In \cite{deBoer:2018mzv}, this contradiction was amended by the suggestion that instead of the pairs being maximally entangled, they are randomly entangled. 

In this work we will study discretized models by employing the results of \cite{AlbaCalabrese17}, where it has been suggested that the distribution
 of created quasiparticles can be reconstructed from the Generalized Gibbs Ensemble that results after equilibration. We calculate the time evolution of $C_A$ and $S_A$ and compare the results with those computed from the quasiparticle model of \cite{AlbaCalabrese17}, finding agreement. In the discrete models, conformal invariance is broken, and as a result the quasiparticles propagate
in opposite directions with a whole spectrum of quasimomentum-dependent velocities, in contrast to moving at the speed of light in a CFT. As a result, in discrete models different features of entanglement are carried by the quasiparticles at different speed: we see this as capacity and entropy in an interval both growing linearly (before saturation), but at different rates. Recovery
of the CFT result in the continuum limit is expected, but we leave it for future more detailed study\footnote{Computation of results for discrete models requires a numerical fit of a parameter $\tau_0$ in the rate, see
e.g.  Fig. 3 in \cite{Coser:2014gsa}, which needs to be treated very carefully to recover the continuum limit CFT prediction.}. 

For another view on how these two measures of entanglement develop in time, we define a contour of capacity, modeled after the definition of contour of entanglement entropy \cite{Botero04,Chen_2014,FrerotRoscilde15,Coser:2017dtb}. The time evolution of contours can be roughly understood as wavefronts of entropy and capacity propagating in the system.
The fronts of the contour of entropy and of the contour of capacity have distinct features that can be traced back to the different distributions of entropy and capacity of the quasiparticles. 

A related concept is the monotonic readjustment of entanglement in the ground state of a system under a Renormalization Group (RG) flow. 
This idea is manifested by the entropic $c$-function, which was 
introduced for relativistic unitary QFTs in \cite{Casini:2006es}, based on the entanglement entropy of a subsystem\footnote{A generalization for higher
dimensional relativistic unitary QFTs is given by the $\mathcal{F}$-function, based on a renormalized entanglement entropy \cite{liumezei_2013, casinihuerta_2012}. See \cite{Nishioka:2018khk} for
a review of entanglement $c$-functions.}. 
The entropic $c$-function shows that a Lorentz invariant theory and a quantity satisfying the strong subadditivity (SSA) gives a rigorously monotonic $c$-function. However, as far as we know, there
is no rigorous proof for the {\em necessity} of the SSA to be satisfied, in order to find monotonicity. Furthermore, as the example of the R\'enyi entropy based generalization \cite{Casini:2005rm,Casini:2005zv} shows,
some functions may show monotonicity even when a rigorous proof has not been found. We would like to call the R\'enyi entropy based function an example of
an {\em accidental} $c$-function: one that behaves monotonically in a class of theories, while there exists no rigorous proof for this monotonicity. In contrast, the entanglement entropy determines a {\em rigorous} $c$-function. It may be that at least for some theories, where the majorization order\footnote{\label{footnote:majorization} Consider two $n\times n$ density matrices $\rho, \sigma$ with eigenvalues collected to ordered vectors $\vec{\lambda},\vec{\mu}$ with components in descending order, e.g.  $\lambda_1\geq \lambda_2 \geq \cdots$. If the partial sums of components satisfy  $\sum^k_{i=1}\lambda_i \geq \sum^k_{i=1}\mu_i \ \forall k=1,\ldots ,n$, then $\rho$ {\em majorizes} $\sigma$ and we denote $\rho \succ \sigma$.} of reduced density matrices under RG flow \cite{latorre2003ground, Orus:2005jq,Riera:2006vj} is true, a $c$-function based on a Schur concave\footnote{A quantifier $Q(\rho)$ is {\em Schur concave} if $\rho \succ \sigma \Rightarrow Q(\rho) \leq Q(\sigma)$.} quantifier, such as the R\'enyi entropies, can be shown to be monotonic.  Or, there
may be other reasons for finding monotonicity, yet to be discovered and understood.

We construct entanglement candidate $c$-functions and explore their monotonicity. 
One construction is based on capacity of entanglement $C_A$. 
Another construction is
based on  the quantifier investigated in \cite{boes2020}, 
defined as\footnote{This measure and its generalization are defined for general finite dimensional systems and states. In this work our focus is
on bipartite systems and reduced states in the subsystem $A$, hence the subscripts $A$ in $M_A(\rho_A)$ {\em etc.}  Note also that Ref. \cite{boes2020} uses the convention where definitions involve the binary logarithm $\log_2(x)$.
We prefer to follow the physics convention and use the natural logarithm in definitions.  Note that in this work we follow the convention of many of our references and denote the natural logarithm by 
$\log (x)$ instead of $\ln (x)$.  Since $\log(x) = (\log 2)\log_2 (x)  $, denoting below quantities defined with $\log_2$ by tildes (e.g.  $\tilde{S}=-\Tr [\rho \log_2 \rho]$), we have
$S = (\log 2)\tilde{S},\ C= (\log 2)^2\tilde{C} $ and $M=(\log 2)^2\tilde{M}$ with 
$ \tilde{M}(\rho) = \tilde{C}(\rho) + \big(\tilde{S}(\rho) + \tfrac{1}{\ln 2}\big)^2$, 
as given in \cite{boes2020}.}
\be\label{Mdefinition}
 M_A(\rho_A ) =C_A(\rho_A) + \big[S_A(\rho_A) + 1 \big]^2  ,
\ee   
which was shown to be Schur concave. 
This property implies that in quantum processes involving two states $\rho$ and $\sigma$ with the majorization order $\rho \succ \sigma$,
any Schur concave quantifier applied to the two states leads to an inequality. For example, von Neumann entropy is Schur concave, with $S(\rho )\leq S(\sigma)$ and likewise for $M$, i.e. $M(\rho )\leq M(\sigma)$. 
In this work we only consider mixed states characterised by reduced density matrices.


The expression (\ref{Mdefinition}) can be  generalised by introducing the {\em moments of shifted modular Hamiltonian} as follows \cite{Arias22Monotones}
\be
\label{Mn Tr Fn}
M_A^{(n)}(\rho_A) = \Tr \big[\rho_A \,(-\log \rho_A+ b_n)^n\big] -b^n_n\,,
\ee
for $n\geq 1$, with $M_A^{(2)}(\rho_A)\big|_{b_n = 1}=M_A (\rho_A)-1$. 
The properties of the sequence (\ref{Mn Tr Fn})
and other related sequences in the context of quantum information theory have been investigated in \cite{Arias22Monotones}.
The $M^{(n)}_A$ in (\ref{Mn Tr Fn}) 
can also be computed from the R\'enyi entropies, through the generating function formula
\bea
\label{Mn from Sn}
M^{(n)}_A(\rho_A) \,=\, 
 e^{b_n}(-1)^n  \frac{d^n}{d\alpha^n} \Big[\exp\Big\{\!-\alpha b + (1-\alpha)S_A^{(\alpha)} (\rho_A)\Big\}\Big]\Big|_{\alpha =1, b=b_n}
\!\! - \,
b_n^n\, .
\eea

Let us now the compare the properties of the quantities mentioned above. 
The Rényi entropies $S_A^{(\alpha)}$ are Schur concave for $\alpha>0$, but concave only for
$0<\alpha \leq 1$  \cite{GeomQuantumStates_book}. On the other hand, by the generating function formula (\ref{Mn from Sn}) they can be converted to $M^{(n)}_A$ which are concave for all $n\geq 1, b_n\geq n-1$ and thus
can be used to define entanglement monotones \cite{Arias22Monotones}. 
However, we do not expect them to satisfy SSA for $n\geq 2$. 
In contrast, the capacity $C_A$ does not satisfy any of the concavity properties or SSA. It is therefore somewhat surprising that both $C_A$ and $M_A$ turn
out to give accidental $c$-functions: they behave monotonically at least in massive free theories, and at the fixed point of the flow reduce to a constant determined by the central
charge of the theory, just like what was previously found for the  $c$-functions based on  Rényi entropies \cite{Casini:2005rm,Casini:2005zv}.

Finally, we consider the way in which the entanglement splits into different charge sectors
of a theory with a global symmetry, quantitatively determined by the symmetry-resolved
entanglement measures. This phenomenon has attracted significant attention, sparked
by some recent theoretical \cite{Lukin-19,Vitale:2021lds,Neven:2021igr} and experimental \cite{Goldstein:2017bua,Xavier:2018kqb} results, and has been studied in several contexts as lattice systems \cite{LaFlorencie2014,Bonsignori:2019naz,Fraenkel:2019ykl,Murciano:2019wdl,Parez:2020vsp,Azses:2021wav,Murciano:2022lsw,Piroli:2022ewy}, quantum field theories \cite{Goldstein:2017bua,Xavier:2018kqb,Murciano:2020vgh,Horvath:2020vzs,Calabrese:2021wvi} and holography \cite{Zhao:2020qmn}.

We define symmetry-resolved versions of the capacity of entanglement and
of the $n^{th}$ moments of shifted modular Hamiltonian and discuss some of their
properties. 
In particular, we find that in systems endowed with a global $U(1)$ symmetry, the moments of shifted modular Hamiltonian $M^{(n)}$
can be written as a sum over the charge sectors of a certain combination
of the symmetry-resolved $M^{(k)}(q)$, with $k\leqslant n$. We compare the symmetry-resolved
entanglement entropy and capacity of entanglement of an interval in a Luttinger liquid CFT
(massless compact boson), observing that they have the same dominant logarithmic behaviour,
while are different at order $\log(\log \ell)$.

This paper is organized as follows.  In Sec.\,\ref{CFT}, we study modifications to the equality (\ref{CAequalsSA}) in more general settings.  
In Sec.\,\ref{massivebosons}, we use  $C_A$ and $M_A$ to define entanglement $c$-functions. 
We then move to consider time evolution after a global quench (Sec.\,\ref{sec:quench} and Sec.\,\ref{sec:contour}). 
In Sec.\,\ref{sec:quench}, we compute and compare the time evolution of $S_A$ and $C_A$ after a global quench in CFTs and free chains. We then
show how the results for the free chains can be obtained and explained through the quasiparticles picture. 
In Sec.\,\ref{sec:contour},  we define a contour function
for the capacity of entanglement, and then compare its time evolution to the one of the contour function of entanglement entropy, after a global quench. 
In Sec.\,\ref{sec:SymResCapacity} we define the symmetry-resolved capacity of entanglement 
and the symmetry-resolved moments of shifted modular Hamiltonian, 
discuss their properties and provide explicit results in a Luttinger liquid CFT. 
We end with a summary and outlook in Sec.\,\ref{sec:conclusions}.
Additional details about the computations and further discussions are reported in Appendices\;\ref{app-lattice}, \ref{app:3qubits} and \ref{app:CTMdetails}.

\section{Capacity of entanglement in 2D QFTs and fermionic chains}
\label{CFT}

In this section we evaluate $S_A$ and $C_A$ for some bipartitions of one dimensional 
and  translation invariant  systems.

\subsection{Some known CFT results}

In this section we study $S$, $C$ defined in (\ref{defcapacity}) and $M$ given in (\ref{Mdefinition}) in simple bosonic and fermionic conformal field theories, and related discrete models (which can be mapped to fermionic chains).   Considering a 2D CFT
for certain states and when the subsystem $A$ is a single interval of length $\ell$,  it has been found that \cite{Callan:1994py,Holzhey:1994we,Calabrese:2004eu,Cardy:2016fqc} 
\be
\label{Trrhon_CFT}
\textrm{Tr}\rho_A^n
\,=\,
c_n \;
e^{-\frac{c}{12}\big(n-\frac{1}{n}\big)W_A}\,,
\ee
where $c$ is the central charge of the CFT, 
$W_A$ is a function of $\ell$ 
that depends also on the state and on the geometry of the entire system 
and which diverges as the UV cutoff $\epsilon \to 0$.
The constant $c_n$ is model dependent and $c_1 = 1$ 
(because of the normalisation of $\rho_A$).
For instance, when the system is on the infinite line and in its ground state,
when the system is on the circle of length $L$ and in its ground state
or when the system is on the infinite line and at finite temperature $1/\beta$,
for $W_A$ we have respectively
\be
\label{W_A CFT-known}
W_A=2\log\!\bigg(\frac{\ell}{\epsilon}\bigg)\,,
\;\;\qquad\;\;
W_A=2\log\!\bigg(\frac{L}{\pi\epsilon} \, \sin\frac{\pi \ell }{L}\bigg)\,,
\;\;\qquad\;\;
W_A=2\log\!\bigg(\frac{\beta}{\pi\epsilon} \, \sinh\frac{\pi \ell}{\beta}\bigg)\,.
\ee

By employing (\ref{Trrhon_CFT})  into the definitions (\ref{defentropy}) and (\ref{defcapacity}),
it is straightforward to find that  \cite{deBoer:2018mzv}
\be
\label{capacityequalentropy}
C_A = S_A = \frac{c}{6}\,W_A + O(1)\,,
\ee 
and that $C_A$ and $S_A$ 
differ at the subleading order $O(1)$ determined by the non-universal constant 
$c_n$\footnote{In higher dimensional CFTs and more general quantum field theories, 
the relation is more ambiguous; indeed the UV cutoff in the two quantities appears in a power law
and the quantities become more dependent on the regularization scheme (see \cite{deBoer:2018mzv} for more discussion).}.
We explore various cases where $S_A - C_A$ is UV finite and non-trivial.

In the following we report the expression of $ M^{(n)} (\rho;b_n )$ defined in (\ref{Mn Tr Fn}) for a CFT on the line in its ground state and  
an interval $A$ of length $\ell$. 
By using (\ref{Trrhon_CFT}) and (\ref{Mn from Sn}), for the leading term we find
\be
\label{Mn LeadingCFT}
 M^{(n)}_A (b_n)
=
\left(
\frac{\log (\ell / \epsilon)}{3} 
\right)^n
+
O\Big(\big(\log (\ell / \epsilon)\big)^{n-1}\Big)\,,
\ee
where the subleading terms in $\ell / \epsilon$ depend both on the non-universal constants and on the parameter $b_n$. 
For instance, in the special case of $n=2$ we get
\be
 M^{(2)}_A (b_2)=\left(
\frac{\log (\ell / \epsilon)}{3} 
\right)^2
+
\frac{1}{3}\left(
1
+
2 b_2
-2 c'_1
\right)
\log (\ell / \epsilon)
+O(1)\,,
\ee
where the subleading terms that we have neglected are finite as $\epsilon$ vanishes. 
Throughout this manuscript, with a slight abuse of notation,  
we denote by $\ell$ both the number of consecutive sites in a block $A$ and the length of the corresponding interval $A$ in the continuum. 
This convention is adopted also for the number of sites of a finite chain 
and for the finite size of the corresponding system in the continuum limit, both denoted by $L$. 

\subsection{Ground state, two disjoint intervals}
\label{subsec:twointerval}

Another important class of examples where $C_A$ and $S_A$ are significantly different 
corresponds to subsystems $A$ made by the union of disjoint intervals. 
In the following we consider the simplest case where 
$A= A_1 \cup A_2$ is the union of two disjoint intervals $A_j =(u_j , v_j)$ on the line
and the entire CFT  is in its ground state. 
The moments of the reduced density matrix
can be written as a four-point function of branch point twist fields
\cite{Caraglio:2008pk,Furukawa:2008uk,Calabrese:2009ez,Calabrese:2010he,Coser:2013qda}
\begin{equation}
\label{kn FT relation 2int}
\textrm{Tr}\rho_A^n
\,=\,
c_n^2
\bigg(
\frac{\epsilon^2\, |u_1-u_2||v_1-v_2|}{|u_1-v_1||u_2-v_2||u_1-v_2||u_2-v_1|}
\bigg)^{\frac{c}{6}(n-\frac{1}{n})}
\mathcal{F}_n(x)\,,
\end{equation}
where $\mathcal{F}_n(x)$ is a model dependent function of the cross ratio of the four endpoints
\begin{equation}
\label{harmonic-ratio}
x=\frac{(u_1-v_1)(u_2-v_2)}{(u_1-u_2)(v_1-v_2)}\,,
\end{equation}
and $c_n$ is the constant occurring in (\ref{Trrhon_CFT}).
Explicit expressions for $\mathcal{F}_n(x)$ for generic integer $n$ are known only for few models 
\cite{Casini:2005rm,Calabrese:2009ez,Calabrese:2010he,Coser:2013qda,Grava:2021yjp}.

From (\ref{defentropy}), (\ref{defcapacity})  and (\ref{kn FT relation 2int}), 
for the entanglement entropy one finds 
\begin{equation}
\label{entropy 2 int}
S_A
\,=\,
\frac{c}{3}\log
\bigg(\frac{|u_1-v_1||u_2-v_2||u_1-v_2||u_2-v_1|}{|u_1-u_2||v_1-v_2| \epsilon^2}
\bigg)-\partial_n\big[\log\mathcal{F}_n(x)\big]\big|_{n=1}-2c'_1\,,
\end{equation}
while the capacity of entanglement reads
\begin{equation}
\label{capacity 2 int}
C_A
\,=\,
\frac{c}{3}\log
\bigg(\frac{|u_1-v_1||u_2-v_2||u_1-v_2||u_2-v_1|}{|u_1-u_2||v_1-v_2| \epsilon^2}
\bigg)
+
\partial_n^2\big[\log\mathcal{F}_n(x)\big]\big|_{n=1}
+
2\big[\partial^2_n(\log c_n)\big] \big|_{n=1}\,.
\end{equation}
We may cancel the divergences and construct an UV finite combination by the difference
\be
\label{SminusC2disjoint}
S_A - C_A 
\,=\,
- \,\partial_n^2 \big[\log\! \big( \mathcal{F}_n(x)\big)\big] \big|_{n=1}
-
\partial_n \big[\log\! \big( \mathcal{F}_n(x)\big)\big] \big|_{n=1}
-
2\,\big[\partial^2_n(\log c_n)\big] \big|_{n=1}
-
2\,c'_1\,,
\ee
which is a non-trivial function of $x$, but difficult to find analytically because 
the analytic continuation in $n$ is usually not accessible.
Numerical analyses based on extrapolations can be performed \cite{Agon:2013iva,DeNobili:2015dla}.
%

For the sake of simplicity, let us consider two equal intervals of length $\ell$ and indicate with $d$ the separation between them. 
In this case (\ref{capacity 2 int}) and (\ref{entropy 2 int}) simplify respectively to
\bea
\label{capacity 2 int eq int}
C_A
&=&
\frac{2 c}{3}\log\frac{\ell}{\epsilon}+\frac{c}{3}\log(1-x)
+\partial_n^2 \big[\log\mathcal{F}_n(x) \big]\big|_{n=1}+2\big[\partial^2_n(\log c_n)\big] \big|_{n=1}\,,
\\
\rule{0pt}{.8cm}
\label{entropy 2 int eq int}
S_A
&=&\frac{2 c}{3}\log\frac{\ell}{\epsilon}+\frac{c}{3}\log(1-x)-\partial_n\big[\log\mathcal{F}_n(x)\big]\big|_{n=1}-2c'_1\,,
\eea
where the cross ratio reads
\begin{equation}
\label{cross ration d ell}
x=\frac{1}{(1+d/\ell)^2}\,.
\end{equation}
From (\ref{capacity 2 int eq int}) and (\ref{entropy 2 int eq int}) it is evident that for two equal intervals (\ref{capacityequalentropy}) holds up to $O(1)$ corrections depending on $x$.

For instance, the massless Dirac fermion is a CFT with $c=1$ and  $\mathcal{F}_n(x)=1$ identically \cite{Casini:2005rm};
hence (\ref{capacity 2 int eq int}) and (\ref{entropy 2 int eq int}) drastically simplify respectively to
\bea
\label{capacity 2 int eq int FF}
C_A
&=&
\frac{2}{3}\log\frac{\ell}{\epsilon}+\frac{1}{3}\log(1-x)+2\big[\partial^2_n(\log c_n)\big] \big|_{n=1}\,,
\\
\rule{0pt}{.8cm}
\label{entropy 2 int eq int FF}
S_A
&=&
\frac{2}{3}\log\frac{\ell}{\epsilon}+\frac{1}{3}\log(1-x)-2\,c'_1\,.
\eea
These expressions tell us that $S_A - C_A$ is independent of $x$,
which is ultimately a consequence of the triviality of $\mathcal{F}_n(x)$ for this model.

\begin{figure}[t!]
\vspace{.2cm}
\hspace{-1.1cm}
\centering
\includegraphics[width=1.05\textwidth]{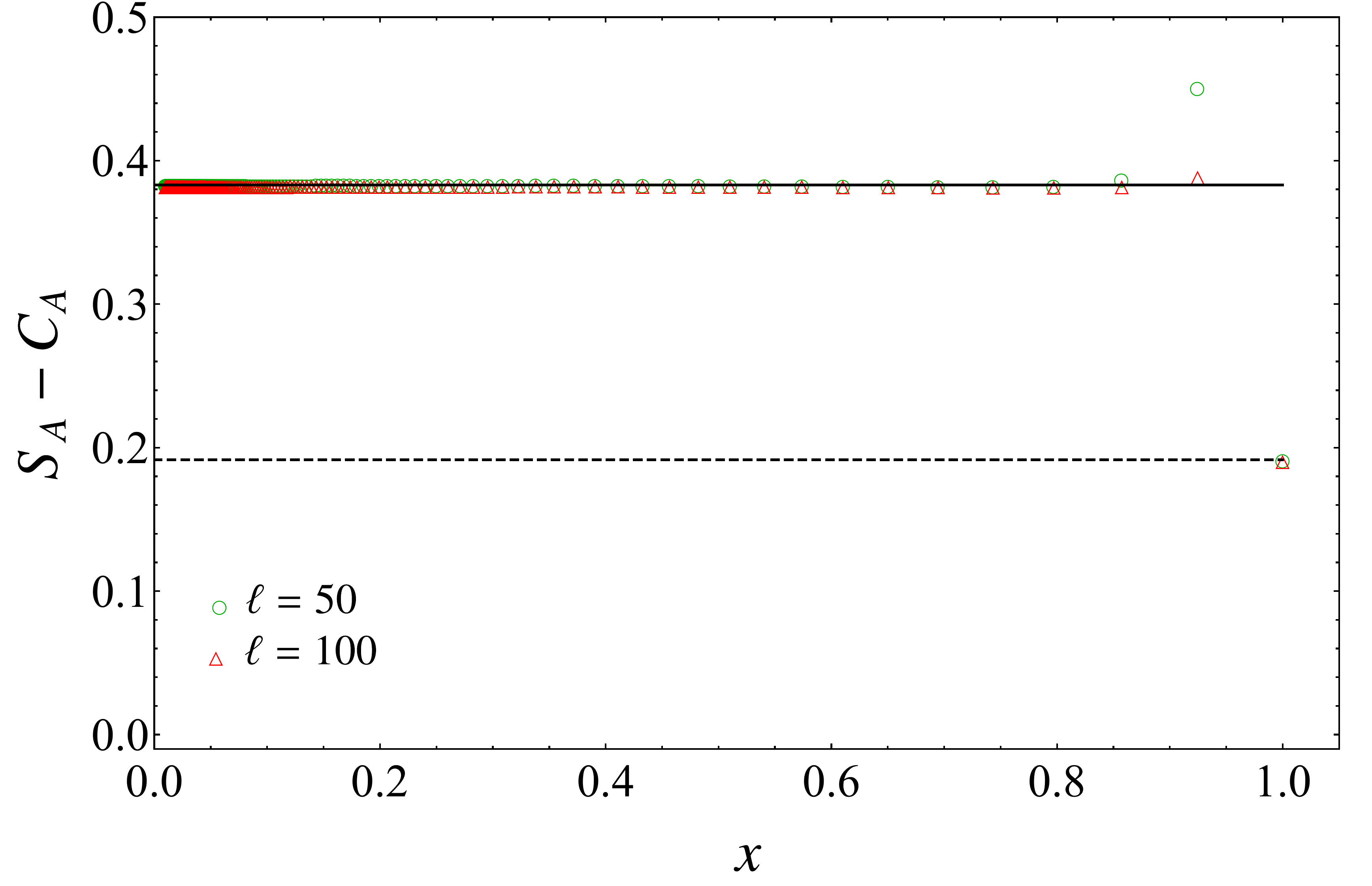}
\caption{
$S_{A} - C_{A}$ in terms of the cross ratio $x$ for two equal disjoint intervals of length $\ell$ in a free fermionic chain, where $\mathcal{F}_n(x)=1$ identically. 
The horizontal solid line is obtained from (\ref{SminusC2disjoint}) and by using the values reported in the text for the non-universal constant terms. 
The height of the dashed line is half the height of the solid line.}
\label{fig:FF_2int_CEvsEE}
\end{figure}

A CFT where the function $\mathcal{F}_n(x)$ is non trivial is the free compactified massless scalar, whose action reads
\begin{equation}\label{compactifiedboson}
I = \frac{g}{4\pi}\int \partial_\mu \phi \,\partial^\mu \phi\,d^2 x \,,
\end{equation}
with a field compactification radius $R$ such that $\phi\sim \phi + 2\pi j  R$, $j\in\mathbb{Z}$.
The function $\mathcal{F}_n(x)$ for this model has been found in  \cite{Calabrese:2009ez} and
its analytic continuation in $n$ is not know for any value of the compactification radius.
In the decompactification regime $ g R^2 \ll 1$, this function becomes
\begin{equation}
\label{log Fn decomp regime}
\log\mathcal{F}_n(x)
=
\frac{1-n}{2}\log\! \big(g R^2\big) - \frac{D_n(x)+D_n(1-x)}{2}\,,
\,\,\qquad\,\,
D_n(x)= \sum_{k=1}^{n-1} \log F_{k/n}(x)\,,
\end{equation} 
where $F_{y}(x)\equiv\,_2F_1(y,1-y,1;x)$.
The analytic continuation of (\ref{log Fn decomp regime}) has been performed by employing that  
\begin{equation}
\label{log Fn decomp regime integral}
D_n(x)=\frac{n}{2\textrm{i}}\oint_\mathcal{C'} \cot(\pi z n)\log\! \big[ F_z (x) \big]\, dz\,,
\end{equation}
being $\mathcal{C'}$ defined as 
the rectangle in the complex plane whose vertices are $\big\{ 1-\textrm{i}L\, ,1+\textrm{i}L\, ,\textrm{i}L\, ,-\textrm{i}L \big\}$.
Taking the derivative of (\ref{log Fn decomp regime integral}) 
and observing that the horizontal contributions in the contour integral vanish as $L\to\infty$,
one obtains \cite{Calabrese:2009ez}
\begin{equation}
\label{d1prime}
D_1'(x)=
\frac{\pi}{\textrm{i}}
\int_{-\textrm{i} \infty}^{\textrm{i} \infty} \frac{z}{[\sin(\pi z )]^2} \, \log F_z (x)\, dz\,.
\end{equation}
A similar computation leads to
\begin{equation}
\label{d1two}
D_1''(x)=
\frac{2\pi}{\textrm{i}}
\int_{-\textrm{i} \infty}^{\textrm{i} \infty} \frac{z}{[ \sin(\pi z )]^2}\big[1-\pi z \cot(\pi z ) \big]\log F_z (x)\, dz\,.
\end{equation}
By employing (\ref{d1prime}) and (\ref{d1two}) respectively 
into (\ref{entropy 2 int eq int}) and (\ref{capacity 2 int eq int})  with $c=1$,
one finds 
\bea
\label{entropy 2int compact}
S_A
&=&\frac{2 }{3}\log\!\left( \frac{\ell}{\epsilon} \right)
+\frac{1}{3}\log(1-x) + \frac{D'_1(x)+D'_1(1-x)}{2}+\frac{1}{2}\log (g R^2) -2c'_1\,,
\\
\rule{0pt}{.7cm}
\label{capacity 2int compact}
C_A
&=&\frac{2 }{3} \log\!\left( \frac{\ell}{\epsilon} \right)
+\frac{1}{3}\log(1-x)-\frac{D''_1(x)+D''_1(1-x)}{2}+2\big[\partial^2_n(\log c_n)\big] \big|_{n=1}\,.
\eea

A numerical check of the capacity of entanglement  (\ref{capacity 2 int eq int FF}) can be performed 
by employing the infinite chain of free fermions, which provides the lattice discretisation of the massless Dirac field on the line. The Hamiltonian describing this chain reads
\be
\label{Ham FF}
H=-\,\sum_{n=-\infty}^{+\infty}
\!  \bigg[\,
\hat{c}^\dag_n \,\hat{c}_{n+1}+\hat{c}^{\dag}_{n+1}\,\hat{c}_n
\bigg]\,,
\ee
where $\hat{c}_n^\dag$ and $\hat{c}_n$ satisfy the canonical anticommutation relations
$\{\hat{c}_n^\dag,\hat{c}_m^\dag\}=\{\hat{c}_n,\hat{c}_m\}=0$ and $\{\hat{c}_n,\hat{c}^\dag_m\}=\delta_{m,n}$. 
In Fig.\,\ref{fig:FF_2int_CEvsEE} we show $S_A - C_A$ in terms of the cross ratio $x$ for two disjoint equal blocks made by $\ell$ consecutive sites in the fermionic chain
described by (\ref{Ham FF}).
The numerical procedure to obtain the data points (reported in terms of the cross ratio $x$)
 is described in Appendix \ref{subapp:fermionlattice}.
From (\ref{capacity 2 int eq int FF}) and (\ref{entropy 2 int eq int FF}), a constant value is expected for this quantity when $\ell$ is large enough. 
This is confirmed by the numerical lattice results in Fig.\,\ref{fig:FF_2int_CEvsEE}, 
where the data points lie on the horizontal black solid line determined by $-2(c_1'+[\partial^2_n(\log c_n)] \big|_{n=1})\simeq 0.3830 $, 
whose value has been obtained in \cite{Arias22Monotones} using the results of \cite{JinKorepin04}.
The height of the last data point (which corresponds to $x=1$) is half the height of the other data points. 
This is due to the fact that, in the limit where the two intervals become adjacent, 
the non universal constant must be taken into account only once; 
hence it is given by $-(c_1'+[\partial^2_n(\log c_n)] \big|_{n=1})\simeq 0.1915 $ (horizontal black dashed line).

\subsection{Free massive fields}
\label{subsec:cap-massive}


\subsubsection{Scalar field}
\label{subsec:massivebosonFTcapacity}

Consider a 1+1-dimensional real scalar field theory with mass $m$ in its ground state and on the infinite line,
which is bipartite into an interval $A$ and its complement. 
By employing the replica approach to the entanglement entropies,  we have that
\begin{equation}
\label{eq:guess1}
\textrm{Tr}\rho_A^n
=
\dfrac{Z_n}{Z_1^n}\,,
\end{equation}
where $Z_n$ is the partition function in the imaginary time 
on the $n$-sheeted Riemann surface obtained by gluing cyclically $n$ copies of the spacetime along the cut $A$.
The partition function $Z_n$ can be computed as follows \cite{Casini:2005zv}
\begin{equation}
\label{eq:sum}
\log Z_n=\sum_{k=0}^{n-1}\log  \zeta_{k/n}\,,
\end{equation}
where $\zeta_a$ is the partition function of a real scalar field in a plane with boundary conditions
$\Phi(x)_{+}=e^{2\pi i a}\, \Phi(x)_{-}$ along the upper part and the lower part of the cut. 
By introducing 
\begin{equation}
\label{eq:mainCH}
w_a
=
\ell \, \partial_\ell\log \zeta_a\,,
\end{equation}
we have that
\begin{equation}
\label{eq:cnalpha}
\mathcal{C}_n\equiv
\sum_{k=0}^{n-1} w_{\frac{k}{n}}
\,=\,
\ell \, \partial_\ell \log Z_n
\qquad
\Longrightarrow
\qquad
\log Z_n=\int_{\log \epsilon}^{\log \ell} \mathcal{C}_n \, d (\log \ell')\,,
\end{equation}
where $\epsilon$ is the UV cutoff.
The function $w_a$ defined in (\ref{eq:mainCH}) can be written as 
\begin{equation}
w_a(\eta) \,=\,
-\! \int_{\eta}^{\infty} y \,u_a^2(y) dy\,,
\end{equation}
where $\eta \equiv m \ell$ and $u_a$ is the solution of the Painl\'eve V differential equation
\begin{equation}
\label{eq:PainleveBoson}
u_a''+\frac{u_a'}{\eta}
\,=\,
\frac{u_a}{1+u_a^2}\,\big(u_a' \big)^2+u_a(1+u_a^2)+\frac{4 (a-1/2)^2}{\eta^2}\; u_a(1+u_a^2)\,,
\end{equation}
whose solution is not known analytically. 
However, in the $\eta \to 0$ regime, it is known that
\begin{equation}
\label{eq:mainCH bosons}
w_a=\ell \, \partial_\ell\log \zeta_a
\,=\,
-\,a(1-a)-\dfrac{1}{2\log (\eta)}+O\big(\log^{-2}(\eta)\big)\,.
\end{equation}
By using (\ref{eq:mainCH}), (\ref{eq:sum}) and the fact that $Z_1$ is independent of $\ell$, 
we get 
(see also Eq.\,(86) in \cite{Casini:2005zv}, with $\mathcal{C}_n=(1-n)c_{n,\textrm{\tiny there}}$)
\begin{equation}
\label{eq:tointegrate}
\mathcal{C}_n
\,=\,
\frac{\partial \log \textrm{Tr}\rho_A^n}{\partial \log (\ell/\epsilon)}
\,=\,
\frac{\partial \log Z_n}{\partial \log (\ell/\epsilon)} - n\,\frac{\partial \log Z_1}{\partial \log(\ell/\epsilon)}
\,=\,
\frac{1-n^2}{6n}+ \frac{1-n}{2\log \eta } + \dots\,,
\end{equation}
which can be integrated when $\ell \gg \epsilon$, finding 
 \be
 \label{massiveTrrhon}
 \log \! \big[ \textrm{Tr}\rho_A^n \big]
 \,=\,
 \frac{1-n^2}{6n} \, \log (\ell / \epsilon)
 +
 \frac{1-n}{2}\,
 \Big[
 \log\!\big(\!-\log(m\ell)\big) - \log\!\big(\!-\log(m\epsilon)\big) 
 \Big] 
 + 
 \dots\,,
\ee
where the dots denote subleading terms originating from the terms neglected in (\ref{eq:mainCH bosons}).
From (\ref{massiveTrrhon}), for the leading terms of the entanglement entropy (\ref{defentropy}) one obtains \cite{Casini:2005zv}
\be
\label{SAmasslessscalarFT}
S_A=
\frac{1}{3}\, \log (\ell / \epsilon)
+
\frac{1}{2}  \,
\Big[
 \log\!\big(\!-\log(m\ell)\big) - \log\!\big(\!-\log(m\epsilon)\big) 
 \Big] 
 +\dots\,,
\ee
while for the capacity of entanglement (\ref{defcapacity}) we have that
\be
\label{CEmasslessscalarFT}
C_A=
\frac{1}{3} \log (\ell / \epsilon) + \dots\,.
\ee
Thus, while the leading terms of $S_A$ and $C_A$ are the same,
we observe a substantial difference in the subleading terms.
Indeed, the double logarithmic correction (due to the zero mode)
occurring in the entanglement entropy  \cite{Casini:2005zv}
is not present in the expansion (\ref{CEmasslessscalarFT})
of the capacity of entanglement.

\subsubsection{Dirac fermion}
\label{subsec:massiveDiracFTcapacity}

An analysis similar to the one discussed in Sec.\,\ref{subsec:massivebosonFTcapacity} 
can be carried out for the free massive Dirac fermion, following closely \cite{Casini:2005rm}. 
In \cite{Casini:2005rm} it has been found that (we remind that $\mathcal{C}_n=(1-n)c_{n, \textrm{\tiny there}}$)
\be
\label{cn_FreeDirac}
\mathcal{C}_n
=
\frac{\partial \log \textrm{Tr}\rho_A^n}{\partial \log (\ell/\epsilon)}=\frac{1-n^2}{6n}
\big[
1-\eta^2 (\log \eta)^2
\big]
+
O(\eta^2 \log \eta)\,.
\ee
Integrating this expression first and then using (\ref{defentropy}) and (\ref{defcapacity}), we obtain (recall $\eta=m\ell$)
\be
\label{CapacityEntropyMassiveFermion}
S_A
=
C_A
\,=\,
\frac{1}{3} \log(\ell / \epsilon)
-
\frac{1}{6}
\big[ m\ell \log (m\ell) \big]^2
+O\big((m\ell)^2 \log (m\ell)\big)\,,
\ee
where $S_A$ and $C_A$ differ at subleading orders.
Differently from the results (\ref{SAmasslessscalarFT}) and (\ref{CEmasslessscalarFT}) for the scalar field, 
for the Dirac field the first correction due to the non vanishing mass in  $S_A$ and $C_A$ is the same.

\subsection{Oscillating terms}
\label{subsec:oscillations}

In the previous examples we have seen that
the difference between $S_A$ and $C_A$ comes from the subleading terms,
when $A$ is a single interval.
In the following we show this fact in specific  models where $S_A - C_A$ can be evaluated analytically.

Consider the free fermionic chain whose Hamiltonian is
\be
\label{xxtext}
H=-\,\sum_{n=-\infty}^{+\infty}
\!  \bigg[\,
\hat{c}^\dag_n \,\hat{c}_{n+1}+\hat{c}^{\dag}_{n+1}\,\hat{c}_n
-
2 h \bigg(\hat{c}^\dag_n \,\hat{c}_{n}-\frac{1}{2}\bigg)\bigg]\,,
\ee
which reduces to (\ref{Ham FF}) when $h=0$.
The ground state of this model is a Fermi sea with a Fermi momentum $k_{\textrm{\tiny F}}=\arccos|h|$. 
Considering the subsystem $A$ made by a block of $\ell$ consecutive sites, it has been found that
\be
\label{log-Tr-oscillator}
\log \textrm{Tr} \rho_A^n 
\,=\,
-\,\frac{1}{6} \left(n - \frac{1}{n} \right) \log \ell
+ 
\log (c_n)
+  
\frac{b_n\, \cos(2 k_{\textrm{\tiny F}} \ell)}{|\, 2\ell  \sin (k_{\textrm{\tiny F}} ) \,|^{2/n}}
+
\frac{d_n}{|\, 2\ell  \sin (k_{\textrm{\tiny F}} ) \,|^{2}}
+
\dots\,,
\ee
\\
where $c_n$ is the constant obtained in \cite{JinKorepin04}
through the Fisher-Hartwig conjecture
and in the subleading terms
which have been computed in
\cite{Calabrese:2009us,CalabreseEssler_10_XXchain}
through the generalised Fisher-Hartwig conjecture,
the coefficients read
\be
b_n \equiv 2  \Bigg( \frac{\Gamma\big(\tfrac{1}{2}(1+ 1/n)\big)}{\Gamma\big(\tfrac{1}{2}(1- 1/n)\big)} \Bigg)^2\,,
\;\;\qquad\;\;
d_n \equiv
\frac{1-n^2}{285 n^3}\, \Big[15(3n^2-7)+(49-n^2) \, (\sin k_\textrm{\tiny F})^2\Big]\,.
\ee
The dots in (\ref{log-Tr-oscillator}) and in subsequent equations correspond to 
higher order subleading terms that have been neglected.

As highlighted in \cite{Calabrese:2009us,CalabreseEssler_10_XXchain},
the entanglement entropy does not contain oscillating subleading terms; indeed
\be
\label{entropy no oscillation}
S_A 
\,=\,
\frac{1}{3} \, \log \ell
+ c_1'+
\frac{8}{95}\; \frac{4 (\sin k_\textrm{\tiny F})^2 - 5}{|\, 2\ell  \sin (k_{\textrm{\tiny F}} ) \,|^{2}}+
\dots\,,
\ee
while the R\'enyi entropies contain subleading oscillatory terms at order $\ell^{-2/n}$ (see (\ref{log-Tr-oscillator})).
As for this qualitative feature, 
the capacity of entanglement (\ref{defcapacity}) is more similar to the R\'enyi entropies.
Indeed, by using that $b_n \to 0$ and $\partial_n b_n\to 0$ as $n \to 1$,  we get
\be
\label{Capacity_oscillations}
C_A 
\,=\,
\frac{1}{3} \, \log \ell
+ 
[\partial^2_n(\log c_n)] \big|_{n=1}
+
\frac{\cos(2 k_{\textrm{\tiny F}} \ell)}{|\, 2\ell  \sin (k_{\textrm{\tiny F}} ) \,|^{2}}
-
\frac{240-122(\sin k_\textrm{\tiny F})^2}{285\, \ell^2 |\, \sin (k_{\textrm{\tiny F}} ) \,|^{2}}+
\dots\,,
\ee
which contains a subleading oscillatory term.

In the context of quantum field theories, 
a similar behaviour has been found for the family of non-relativistic Lifshitz spinless fermion fields $\psi(t,x)$ 
satisfying the equal time canonical anticommutation relations 
and whose time evolution is 
\begin{equation}
\left[ \,\textrm{i}\,\partial_t -\frac{1}{(2m)^{2z-1}} \,(-\,\textrm{i}\, \partial_x)^{2z} \, \right]\psi (t,x) = 0 \,,
\;\;\;\qquad \;\;\;
z \in {\mathbb N}\,,
\end{equation} 
which becomes the familiar Schr\"odinger equation for $z=1$.
The state of the entire system is characterised by zero temperature and non-vanishing chemical potential $\mu$.
Focussing on the Schr\"odinger field theory for simplicity, i.e. $z=1$,
the R\'enyi entropies and the entanglement entropy of an interval 
either on the line or at the beginning of the semi-infinite line 
have been studied in \cite{Mintchev:2022xqh, Mintchev:2022yuo}.

In order to compare with the lattice model results discussed above, 
let us consider the interval $A = (-R,R) $ on the line \cite{Mintchev:2022xqh}.
In this case the moments $\textrm{Tr}\rho_A^n$ are UV finite 
and they are single-variable functions of the dimensionless variable $k_\textrm{\tiny F} R$, 
being $k_\textrm{\tiny F} \equiv \sqrt{2m\mu}$ defined as the Fermi momentum.
The whole regime $k_\textrm{\tiny F} R \in (0, +\infty)$ has been explored,
finding analytic expressions for some terms of the expansions 
both at large $k_\textrm{\tiny F} R$ and at small $k_\textrm{\tiny F} R$.
It has been shown also that $S_A$ is a strictly increasing function,
while the R\'enyi entropies display oscillations. 
Similarly to the fermionic chain considered above,
also in this fermionic Schr\"odinger field theory
$C_A$ is qualitatively more similar to the R\'enyi entropies.
Indeed, some oscillations occur in $C_A$ in the regime of small values of $k_\textrm{\tiny F} R$,
as shown in Fig.\,15 of \cite{Mintchev:2022xqh},
where both $S_A$ and $C_A$ are reported.
A quantitative analysis of these oscillations can be carried out, but it is beyond the scope of this manuscript.

\section{Entanglement $c$-functions along the RG flow}
\label{massivebosons}

An attractive idea is that entanglement in the ground state of a system readjusts itself under RG flow. This idea is manifested by the entropic $c$-function, which was 
introduced for $1+1$-dimensional relativistic unitary QFTs in \cite{Casini:2006es}, based on the entanglement entropy of a subsystem. A generalization for higher
dimensional relativistic unitary QFTs is given by the $\mathcal{F}$-function, based on a renormalized entanglement entropy \cite{liumezei_2013, casinihuerta_2012} (see also the review \cite{Nishioka:2018khk}).
Our focus will be in 1+1 dimensional relativistic QFTs. One
takes the subsystem to be an interval of length $\ell$, then the $c$-function is given by
\be
\label{CfunctS}
\mathcal{C}_S=\ell\frac{d S_A}{d \ell}=\frac{d S_A}{d \log(\ell/\epsilon)}\,,
\ee
which becomes $c/3$ at the fixed points. One can show \cite{Casini:2006es} that $\mathcal{C}_S$
is monotonically decreasing, $\partial_\ell \mathcal{C}_S \leqslant 0$, and the monotonicity with respect to $\ell$ corresponds to monotonicity of $\mathcal{C}_S$ under readjusting of the couplings $g_i$ of the theory. The proof of monotonicity 
is based on the Lorentz invariance of the theory and strong subadditivity (SSA) of entanglement entropy.
\\
Similar functions were studied starting from the R\'enyi entropies $S^{(n)}_A$, which do not satisfy SSA \cite{Casini:2005zv,Casini:2005rm}. 
There is no rigorous argument to expect the R\'enyi entropy based functions to be monotonic in a generic QFT, yet they were observed to behave monotonically in some theories. 
A stronger proposal for the readjustment of entanglement is the concept of fine-grained entanglement loss along
renormalization group flows \cite{latorre2003ground, Orus:2005jq,Riera:2006vj}: according to this idea, the reduced density matrix $\rho_A$ of the ground state follows a majorization ordering along the RG flow.


In this section we explore whether an entanglement candidate $c$-function based on capacity of entanglement $C_A$ or the second moment of shifted modular Hamiltonian $M_A$ can exhibit monotonicity in some theories. As we showed, $M_A$ satisfies a stronger property than Schur concavity of R\'enyi entropies: it is concave and an entanglement monotone. However, as explained in more details in Appendix \ref{app:3qubits}, it violates SSA. In contrast, $C_A$ does not satisfy any of the concavity properties or SSA. It is therefore somewhat surprising that both $C_A$ and $M_A$ turn
out to give accidental $c$-functions: they behave monotonically at least in massive free theories, and at the fixed point of the flow reduce to a constant determined by the central
charge of the theory.

As our theories we consider the massive free scalar field and massive Dirac field theories. For numerical calculations we discretize the theories, the first one to the harmonic
chain, and the latter to a free fermionic chain.  We begin by deriving some analytic results, first for the harmonic chain.

%
%
%

%
%
%
%
%

\subsection{Capacity of entanglement for the harmonic chain: CTM approach}
\label{sec:CTM}

A very powerful tool for computing the entanglement in gapped lattice models is the corner transfer matrix (CTM) \cite{PeschelKaulke99,it-87,nishino,nishino2}.
Exact results for the massive harmonic chain can be obtained from \cite{Peschel91Gaussian,Peschel}, while results for the XXZ chain and the Ising model are contained in \cite{Calabrese:2004eu,cal2010,Alba2018A}.

Consider the infinite harmonic chain 
with nearest neighbour spring-like interaction described by the Hamiltonian
\be
\label{HC ham}
\widehat{H}_{\textrm{\tiny HC}} = 
\sum_{i=-\infty}^{+\infty} 
\left(
\frac{1}{2\mu}\,\hat{p}_i^2+\frac{\mu\omega^2}{2}\,\hat{q}_i^2 +\frac{\lambda}{2}(\hat{q}_{i+1} -\hat{q}_i)^2
\right) ,
\ee
where the position and the momentum operators  $\hat{q}_i$ and $\hat{p}_i$
are Hermitean operators satisfying the canonical commutation relations
$[\hat{q}_i , \hat{q}_j]=[\hat{p}_i , \hat{p}_j] = 0$ 
and $[\hat{q}_i , \hat{q}_j]= \textrm{i} \delta_{i,j}$
(we set $\hbar =1$ throughout this manuscript). 
The canonical transformation given by 
$\hat{q}_i \to  \hat{q}_i / \sqrt[4]{\mu\lambda}$ and $\hat{p}_i \to \sqrt[4]{\mu\lambda}\, \hat{p}_i$ 
allows us to write this Hamiltonian as
\be
\label{HC ham v2}
\widehat{H}_{\textrm{\tiny HC}}
= \frac{\sqrt{\lambda/\mu}}{2}\,
\sum_{i=-\infty}^{+\infty} 
\left(
\hat{p}_i^2+ \frac{\omega^2}{\lambda/\mu} \, \hat{q}_i^2 +  (\hat{q}_{i+1} -\hat{q}_i)^2
\right),
\ee
which naturally leads us to introduce
\be
\label{omega_tilde}
\tilde{\omega}^2 = \frac{\omega^2}{\lambda/\mu} \,.
\ee

We assume that the system is in its ground state and take the subsystem $A$ first to be half of the chain (we will later consider a finite interval as a subsystem). 
The CTM approach relies on the possibility of relating  the corner transfer matrix of the two-dimensional integrable Gaussian model to the reduced density matrix of the subsystem $A$ and allows us to write the entanglement Hamiltonian associated to the latter as \cite{Peschel91Gaussian,Peschel}
\be
\label{EH_CTM}
H_\textrm{CTM}=\sum_{j=0}^\infty \varepsilon_j n_j
=
\sum_{j=0}^\infty \varepsilon (2j+1) n_j\,,
\ee
where $n_j$ are bosonic number operators and
\be
\label{eps-omega-peschel}
\varepsilon=\varepsilon(\tilde{\omega}) \equiv \frac{\pi\,K\!\big(\sqrt{1-\kappa(\tilde{\omega})^2}\,\big)}{K(\kappa(\tilde{\omega}))}\,,
\qquad
\qquad
\kappa(\tilde{\omega})
\equiv
\frac{2+\tilde{\omega}^2 - \tilde{\omega} \,\sqrt{\tilde{\omega}^2 + 4}}{2}\,,
\ee
being $K$ the complete elliptic integral of the first kind and $\tilde{\omega}$ defined in (\ref{omega_tilde}).
The knowledge of the entanglement Hamiltonian in (\ref{EH_CTM}) (and therefore of the reduced density matrix) allows us to write \cite{ep-rev}
\be
\label{ZnCTM_HC}
 \log \textrm{Tr}\rho_A^n
=\sum_{j=0}^\infty
\left[
n\log\left(1-e^{-(2j+1)\varepsilon}\right)-\log\left(1-e^{-(2j+1)n\varepsilon}\right)
\right] .
\ee
From (\ref{defentropy}), the CTM result for the entanglement entropy reads \cite{ep-rev}
\be
\label{EE_CTM_HC}
S_A=
\sum_{j=0}^\infty
\left[\frac{\varepsilon (2j+1)}{e^{(2j+1)\varepsilon}-1} -\log\left(1-e^{-(2j+1)\varepsilon}\right)\right]\,.
\ee
As for the capacity of entanglement, according to (\ref{defcapacity}),
taking two derivatives of (\ref{ZnCTM_HC}) with respect to $n$ and evaluating the result in $n=1$, we get
\be
\label{CoE_CTM_HC}
C_A=
\sum_{j=0}^\infty
\left(\frac{\varepsilon (2j+1)}{e^{(2j+1)\varepsilon}-1}\right)^2 e^{(2j+1)\varepsilon}=\sum_{j=0}^\infty
\left(\frac{\varepsilon (2j+1)}{2 \sinh\left(\frac{\varepsilon}{2}(2j+1)\right)}\right)^2 \,.
\ee
We stress that these expressions for $S_A$ and $C_A$ are different, while, as discussed in Appendix \ref{sec:criticalCTM}, they become equal in the critical regime.
The entanglement entropy (\ref{EE_CTM_HC}) can be written in a closed form, as done in \cite{ep-rev}. It reads
\be
\label{EE_CTM_HC_elliptic-ep}
S_A=-\frac{1}{24}\bigg[\log\!\bigg(\frac{16 (\kappa')^4}{\kappa^2}\bigg)- \big(1+\kappa^2\big)\frac{4 K(\kappa)K(\kappa')}{\pi}
\bigg] \, , 
\ee
where $\kappa$ is defined in (\ref{eps-omega-peschel}) and $\kappa' \equiv \sqrt{1-\kappa^2}$.
The derivation of (\ref{EE_CTM_HC_elliptic-ep}) has been reviewed in Appendix \ref{app:derivationCTM}.
In Appendix \ref{app:derivationCTM} we also exploit some of the properties of the Jacobi theta functions $\theta_r(q)\equiv\theta_r(0,q)$ 
with $r \in \{2,3,4\}$ in order to obtain a closed form for the capacity of entanglement and another expression for the entanglement entropy. 
We find
\be
\label{EE_CTM_HC_elliptic}
S_A=-\frac{1}{6}\bigg[\log 2+ \log\!\bigg( \frac{ \theta_4^2(e^{-\varepsilon})}{\theta_2(e^{-\varepsilon})\theta_3(e^{-\varepsilon})}\bigg)-\frac{\varepsilon}{4}\big( \theta_2^4(e^{-\varepsilon})+\theta_3^4(e^{-\varepsilon})\big)\bigg]\,,
\ee
and
\be
\label{CE_CTM_HC_elliptic}
C_A=\frac{\varepsilon^2}{6} \, e^{-\varepsilon}
\big[\,
\theta_3^3(e^{-\varepsilon}) \, \theta'_3(e^{-\varepsilon})
+\theta_2^3(e^{-\varepsilon}) \, \theta'_2(e^{-\varepsilon})
\,\big]\,,
\ee
where we have defined $\theta_r'(q)=\partial_q\theta_r(q)$, with  $r \in \{2,3,4\}$\footnote{The derivative of elliptic theta functions with respect to the variable $u$ can be written as the following series expansions
\be
\theta'_3(q)=2\sum_{k=1}^\infty k^2 q^{k^2-1}\,,
\,\,\qquad\,\,
\theta'_2(q)=\frac{\theta_2(q)}{4q}+2\sum_{k=1}^\infty k(k+1) q^{k(k+1)-\frac{3}{4}}\,,
\ee
which have been used to check the validity of (\ref{CE_CTM_HC_elliptic}).}.
The CTM techniques are also employed in Appendix \ref{XXZCTM} to compute the entanglement entropy and the capacity of entanglement in XXZ spin chains.

Next, we take the subsystem $A$ to be an interval of length $\ell$. In this case, since the global ground state is a Gaussian state, 
we will compute $S_A$ and $C_A$ by the method of finding the symplectic eigenvalues of the reduced covariance matrix. This method is
reviewed in Appendix \ref{subapp:entanglementHC}, and in the end we compute $S_A$ and $C_A$ numerically. We may also compare the finite interval
case with the half-infinite subsystem, by taking the limit $\ell \rightarrow \infty$ where we expect to recover twice the constants predicted by the CTM calculations (\ref{EE_CTM_HC_elliptic}) and (\ref{CE_CTM_HC_elliptic}).

\begin{figure}[t!]
\vspace{.2cm}
\hspace{-1.4cm}
\includegraphics[width=.57\textwidth]{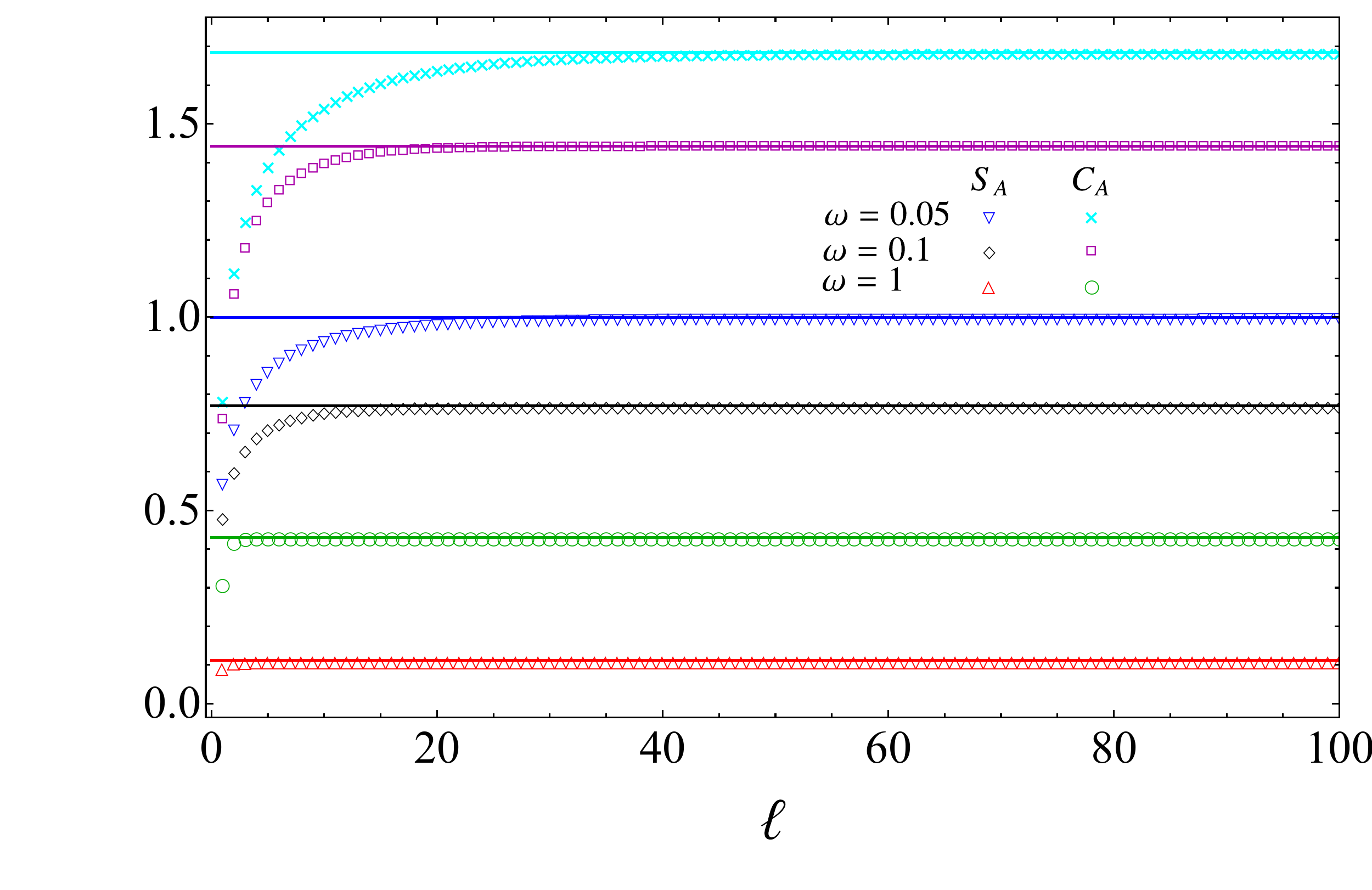}
\hspace{-0.5cm}
\includegraphics[width=.57\textwidth]{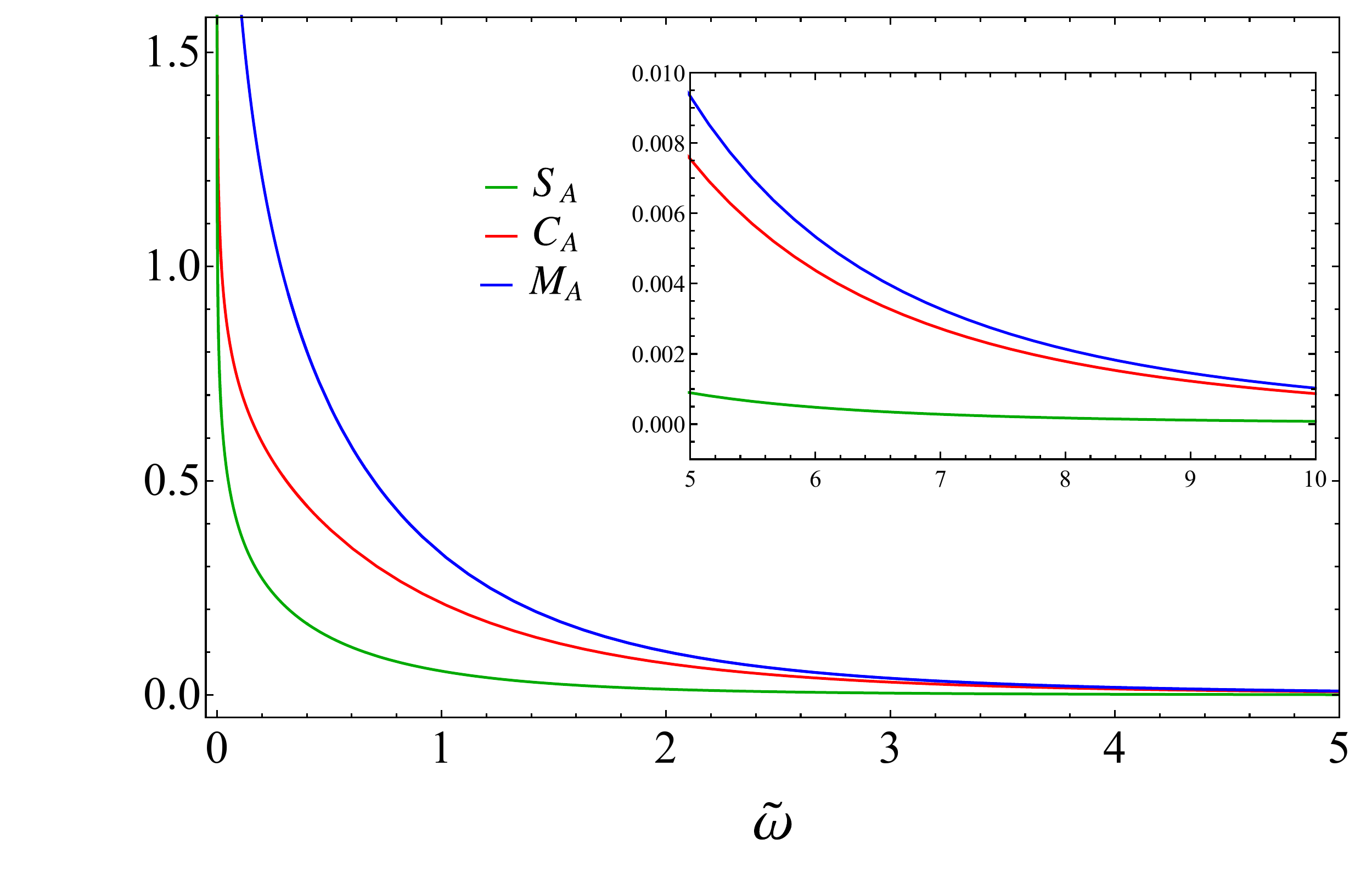}
\vspace{-.5cm}
\caption{
In the left panel we report $S_A$ and $ C_A$ of an interval in an infinite harmonic chain as a function of the length of the interval $\ell$.
The solid lines correspond to twice the constants predicted by the CTM calculations (\ref{EE_CTM_HC_elliptic}) and (\ref{CE_CTM_HC_elliptic}). The curves in the right panel have been obtained from (\ref{EE_CTM_HC_elliptic}), (\ref{CE_CTM_HC_elliptic}) and (\ref{Mdefinition}).
}
\vspace{.5cm}
\label{fig:HC_1intmassive_CEvsEE}
\end{figure}

In the left panel of Fig.\,\ref{fig:HC_1intmassive_CEvsEE} we show the results for $S_A$ and $ C_A$, for an interval in an infinite harmonic chain as a function of the length of the interval $\ell$ for three values of $\tilde{\omega}=\omega$.
All the numerical data points for the harmonic chains displayed in this manuscript have been obtained by setting $\mu=1$ and $\lambda=1$.
The data are obtained as explained in Appendix \ref{subapp:entanglementHC}.
The numerical curves saturate to a constant: the value of $\ell$ at which the saturation is reached is smaller for bigger values of $\omega$.
The saturation constants are very well predicted by the CTM calculations (\ref{EE_CTM_HC}) and (\ref{CoE_CTM_HC}) (up to a factor two due to the number of endpoints) and in the panel correspond to the horizontal solid lines.
In the right panel of Fig.\,\ref{fig:HC_1intmassive_CEvsEE} we plot $S_A$, $C_A$ and $M_A$ as functions of $\tilde{\omega}$ using (\ref{EE_CTM_HC_elliptic}), (\ref{CE_CTM_HC_elliptic}) and these two results in (\ref{Mdefinition}) respectively. All these functions are decreasing in $\tilde{\omega}$ and can be regarded as $c$-functions along the RG flow.
The decreasing behaviour of $S_A$ and $M_A$ as functions of $\tilde{\omega}$ can be justified exploiting their Schur concavity and the computation reported in Appendix \ref{app:majorizationHC}, while we do not have an {\it a priori} argument for the behaviour of $C_A$.

\subsection{$c$-functions from $C_A$ and $M_A$}

Inspired by the definition (\ref{CfunctS}), 
from $C_A$ and $M_A$ defined in (\ref{defcapacity}) and (\ref{Mdefinition}), we introduce respectively 
\be
\label{CfunctC}
\mathcal{C}_C=\ell \, \frac{d C_A}{d \ell}=\frac{d C_A}{d \log(\ell/\epsilon)}\,,
\ee
and
\be
\label{CfunctM}
\mathcal{C}_M=\ell \, \frac{d M_A}{d \ell}=\frac{d M_A}{d \log(\ell/\epsilon)}\,.
\ee
While $\mathcal{C}_C$ at the fixed point gives $c/3$,
for $\mathcal{C}_M$ we obtain $\tfrac{2c^2}{9} \log(\ell/\epsilon)$.
In order to obtain a finite result at the fixed point, let us consider
\be
\label{CfunctMtilde}
\widetilde{\mathcal{C}}_M
\,=\,
\frac{d M_A}{d [\log(\ell/\epsilon)]^2}
\,=\,
\frac{1}{2\log(\ell/\epsilon)} \; \ell \, \frac{\partial M_A}{\partial \ell}
\,=\,
\frac{\mathcal{C}_M}{2\log(\ell/\epsilon)}\,,
\ee
which gives $c^2/9$ at the fixed points.
From (\ref{Mn LeadingCFT}) with $n=2$ into (\ref{CfunctMtilde}) it is straightforward to observe that, 
at the fixed point, 
corrections of order $1/\log(\ell/\epsilon)$ 
arise in $\widetilde{\mathcal{C}}_{M}$.

\subsubsection{$c$-functions in 1+1 dimensional free QFT}
\label{subsec:cfunction}

In the following subsection we compute (\ref{CfunctC}), (\ref{CfunctM}) and (\ref{CfunctMtilde}) 
in the context of 1+1 dimensional free massive QFT.

\begin{figure}[t!]
\vspace{.2cm}
\hspace{2.7cm}
 \subfigure
 {
\includegraphics[width=.57\textwidth]{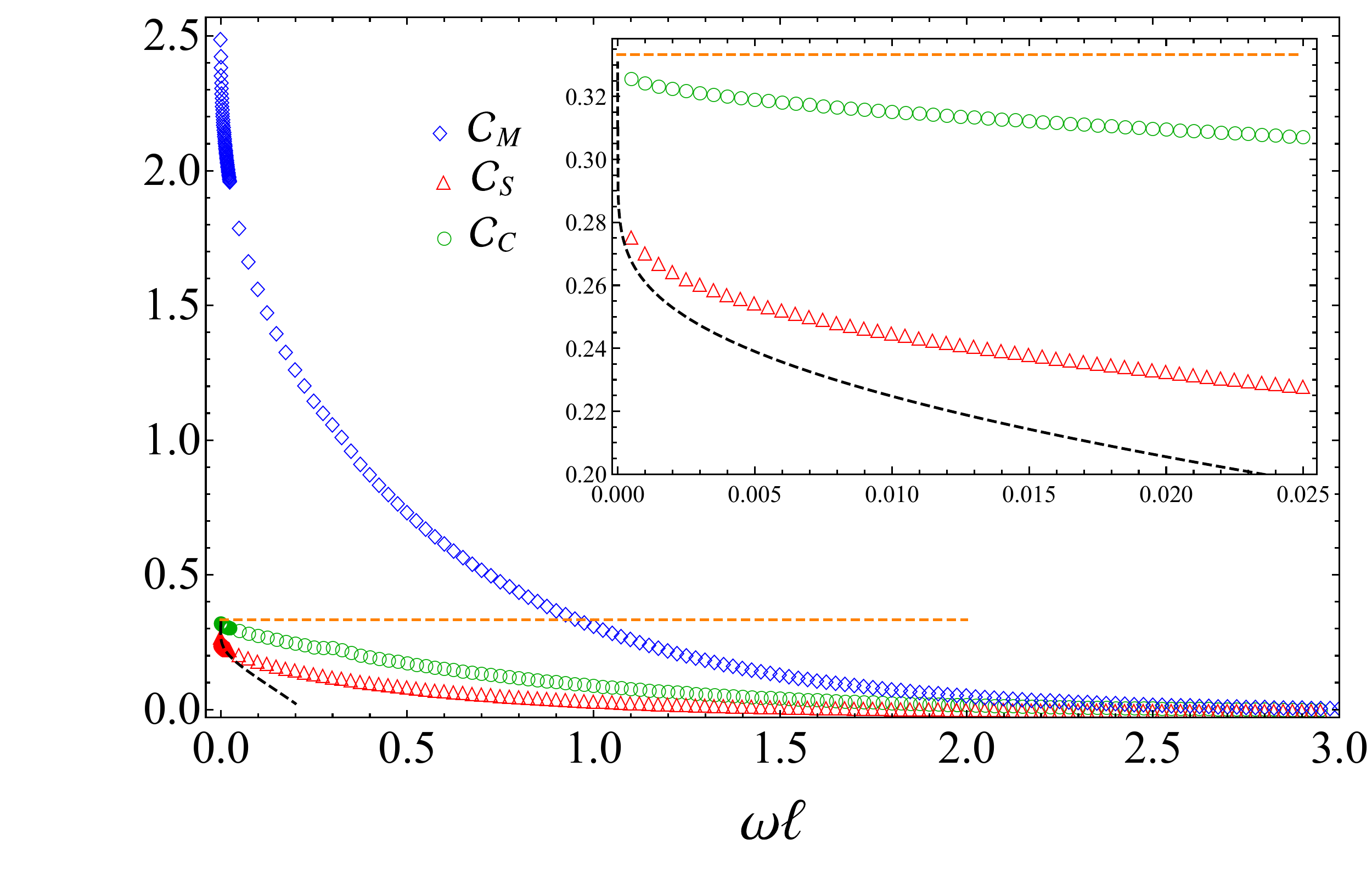}
 }
 \\
\subfigure
{
 \hspace{-1.4cm}
 \includegraphics[width=.57\textwidth]{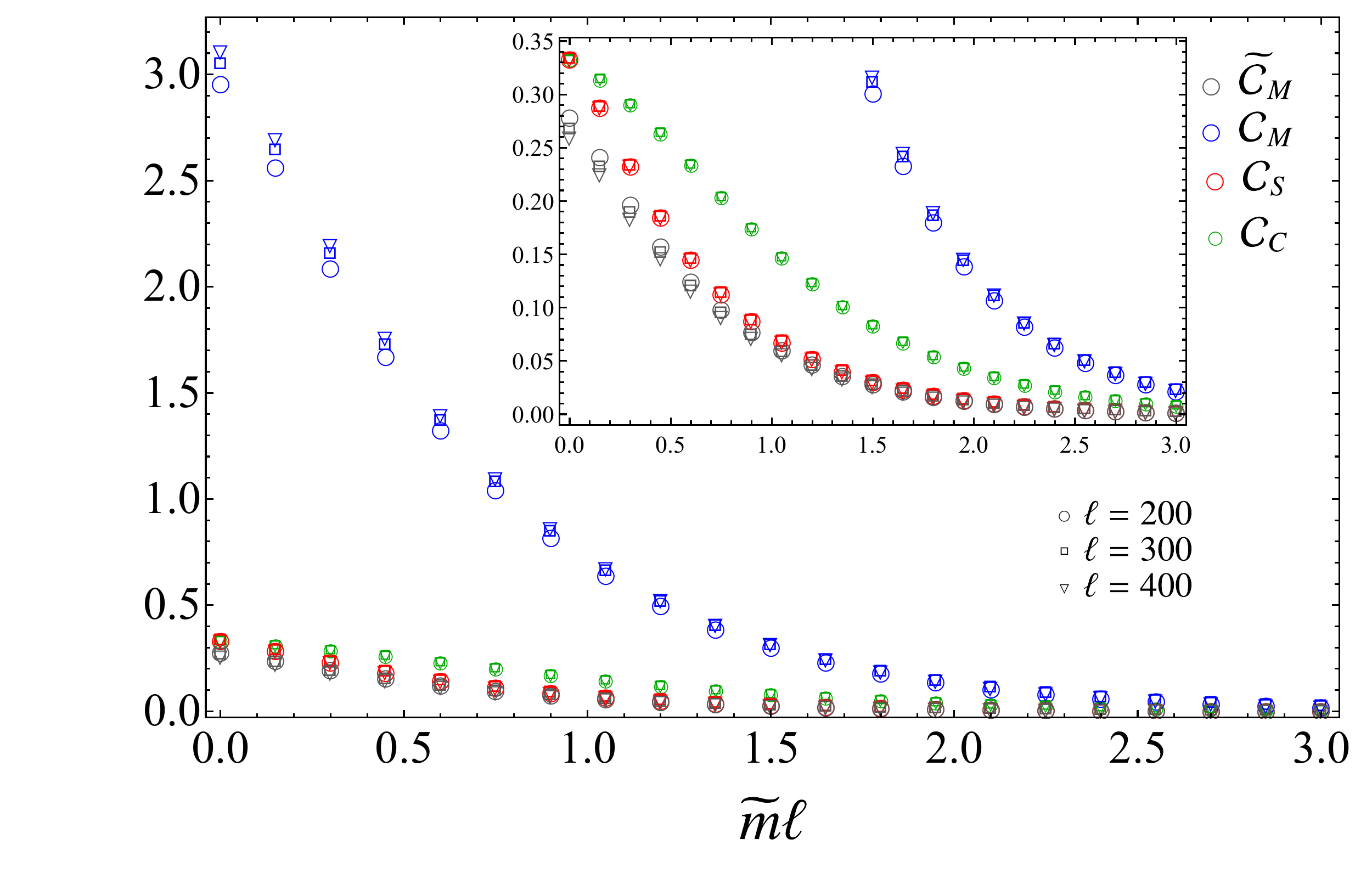}
}
\subfigure{
\hspace{-.7cm}
\includegraphics[width=.57\textwidth]{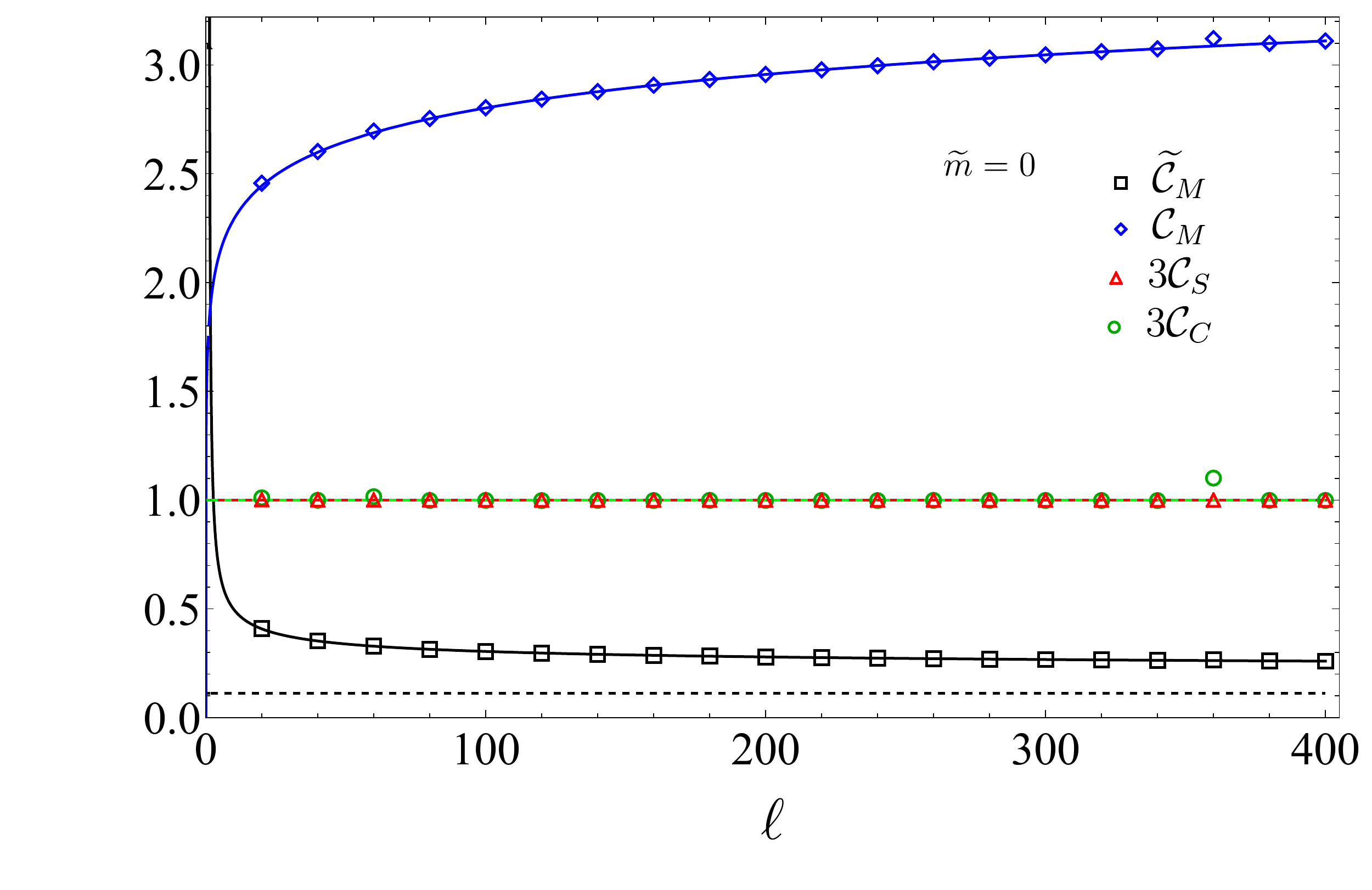}
}
\vspace{-.5cm}
\caption{
The $c$-functions (\ref{CfunctS}), (\ref{CfunctC}), (\ref{CfunctM}) and (\ref{CfunctMtilde}). In the top panel we compute them for the harmonic chain as function of $\omega\ell$ with $\ell=100$.
The black dashed curve is obtained from (\ref{c-funct}), while the orange dashed line corresponds to the constant $\frac{1}{3}$.
In the bottom panels we consider the discretisation of the massive Dirac field theory. In the bottom left panel the data points are plotted as function of $\widetilde{m}\ell$, while in the bottom right panel are reported as function of $\ell$ with $\widetilde{m}=0$. The green and red horizontal lines correspond to the constant 1, while the blue and the black curves are given by the the first and the second expression in (\ref{c_M massless}) respectively with $\epsilon=1$. The horizontal black dashed line is the constant $\frac{1}{9}$. 
In all the panels for the  discrete derivatives of the numerical data (\ref{num_der}) has been employed.
}
\vspace{-.2cm}
\label{fig:Cfunctions}
\end{figure}

{\it Massive scalar field theory.}
We consider a free scalar field theory with mass $m$ and its discretization  
through the harmonic chain described by the Hamiltonian (\ref{HC ham}). 
We set $\mu=1$ and $ \lambda=1$ in (\ref{HC ham}); hence $\omega$ is identified in the continuum limit with the mass $m$ of the scalar field. 
In this setup, one can compute numerically the quantities (\ref{CfunctS}), (\ref{CfunctM}) and (\ref{CfunctMtilde})\footnote{
The numerical results displayed in Fig.\,\ref{fig:Cfunctions}, both for the bosonic and the fermionic models, 
have been obtained by computing $S_A$, $C_A$ and $M_A$ numerically first
and then taking the discrete derivative as \cite{Casini:2005rm}
\be
\label{num_der}
f'(j+1/2)=\frac{1}{4}\big[f(j+2)+f(j+1)-f(j)-f(j-1)\big]\,.
\ee}.
The numerical results displayed in the top panel of Fig.\,\ref{fig:Cfunctions} (for the details see Appendix \ref{subapp:entanglementHC})
show that all these three quantities are decreasing functions of $\omega\ell$.

At QFT level, the results of \cite{Casini:2005zv}, reviewed in Sec.\,\ref{subsec:cap-massive}, provide the behaviour of these quantities for small values of $\eta=m\ell$. Indeed, $\mathcal{C}_S$ and $\mathcal{C}_C$ can be rewritten as
\be
\label{c-funct-def}
\mathcal{C}_S=-\big(\partial_n \mathcal{C}_n\big)\big|_{n=1}\,,
\,\,\qquad\,\,
\mathcal{C}_C=\big(\partial^2_n \mathcal{C}_n\big)\big|_{n=1}\,,
\ee
where $\mathcal{C}_n$ is given in (\ref{eq:tointegrate}).
Applying these definitions to (\ref{eq:tointegrate}) we obtain
\be
\label{c-funct}
\mathcal{C}_S=\frac{1}{3}+\frac{1}{2\log \eta}+O\big(\log^{-2}(\eta)\big)\,,
\,\,\qquad\,\,
\mathcal{C}_C=\frac{1}{3}+O\big(\log^{-2}(\eta)\big)\,.
\ee
From (\ref{c-funct}) we can observe that, while we can say that in the small mass limit $\mathcal{C}_S$ is a decreasing function of $\eta$, the same cannot be concluded for $\mathcal{C}_C$ since we know only the constant term. However the monotonic decreasing behaviour of $\mathcal{C}_C$ can be observed from the numerical calculation reported in the top panel of Fig.\,\ref{fig:Cfunctions}.

While $\mathcal{C}_S$ and $\mathcal{C}_C$ converges to $\frac{1}{3}$ when $m\ell\to 0$, $\mathcal{C}_M$ seems to diverge. 
This can be explained by computing $M_A$ from (\ref{Mdefinition}) using (\ref{SAmasslessscalarFT}) and (\ref{CEmasslessscalarFT}): 
denoting by $-c_1'$ the non universal constant in the expression of the entanglement entropy, one finds
\bea
\nonumber
M_A
&=&
\frac{1}{9}\log^2 \left(\frac{\ell}{\epsilon}\right)
+\bigg[1+\frac{2}{\log 2}-2 c_1'+\log\!\big(-\log(m\ell)\big)\bigg]\frac{\log(\ell/\epsilon)}{3}
\\
&&
+\,
\bigg(\frac{1}{\log 2}-c_1'\bigg)\log\!\big(-\log(m\ell)\big)
+
\frac{1}{4}\log^2\big(-\log(m\ell)\big)+O(1)\,.
\eea
Applying the definition in (\ref{CfunctM}) we obtain 
\bea
\label{cfunct M scalar}
\mathcal{C}_M
&=&
\bigg(\frac{2}{9}+\frac{1}{3\log(m\ell)}\bigg)\log \left(\frac{\ell}{\epsilon}\right)+
\frac{1}{2\log(m\ell)}\bigg[\frac{2}{\log 2}-2 c_1'+\log\!\big(-\log(m\ell)\big)\bigg]
\\
&&
+\,
\frac{1}{3}\log\!\big(-\log(m\ell)\big)+O(1)\,,
\nonumber
\eea
that is divergent when $m\ell \to 0$ because of the last term.
To avoid the divergence, we can apply the definition (\ref{CfunctMtilde}) to (\ref{cfunct M scalar}), finding
\be
\widetilde{\mathcal{C}}_M=
\frac{1}{9}+\frac{1}{6\log(m\ell)}+
\frac{\frac{2}{\log 2}-2 c_1'+\log\!\big(-\log(m\ell)\big)}{4\log(\ell/\epsilon)\log(m\ell)}+
\frac{\log\!\big(-\log(m\ell)\big)}{6\log(\ell/\epsilon)}
+O\big(1/\log(\ell/\epsilon)\big)\,.
\ee
If we take the limit $\ell/\epsilon\to\infty$ before $m\ell\to 0$ we obtain
\be
\widetilde{\mathcal{C}}_M=\frac{1}{9}+\frac{1}{6\log(m\ell)}\,,
\ee
that at the conformal fixed point for $m\ell\to 0 $ gives $\widetilde{\mathcal{C}}_M(m\ell\to 0)=\frac{1}{9}$.

{\it Massive Dirac field theory.}
Another example of the calculation of the $c$-functions (\ref{CfunctS}), (\ref{CfunctM}) and (\ref{CfunctMtilde}) concerns the $1+1$ dimensional massive Dirac field theory.
 We discretise the massive Dirac field theory with mass $m$ on the lattice
through a free fermionic chain described by the following Hamiltonian 
\cite{Casini:2005rm} 
\begin{equation}\label{eq:hamiltonianF}
H=-\frac{\mathrm{i}}{2}\sum_{j=0}^{N-1}
\Big(\hat{c}^{\dagger}_{j+1}\hat{c}_j-\hat{c}^{\dagger}_j \hat{c}_{j+1}\Big)
+
\widetilde{m}\sum_{j=0}^{N-1}(-1)^j \hat{c}^{\dagger}_j \hat{c}_j\,,
\end{equation}
where  $\hat{c}_j$ satisfy the anti-commutation relations $\{ \hat{c}_j,\hat{c}^{\dagger}_k \}=\delta_{jk}$, the number of sites of the chain is given by $N$
and $\widetilde{m}$ is the discrete counterpart of the mass $m$.
 In Appendix \ref{app:massiveDirac} we report the correlators of this model when $N\to\infty$ and we derive an analytic expression in terms of hypergeometric functions.

As we can see in the bottom left panel of Fig.\,\ref{fig:Cfunctions}, also in this case the three functions are decreasing in $\widetilde{m}\ell$. While $\mathcal{C}_S$ and $\mathcal{C}_C$ converge to $\frac{1}{3}$ when $\widetilde{m}\to 0$ (as expected from CFT), $\mathcal{C}_M$ goes to a different value in the conformal limit. 
Given that for the lattice fermionic model we are considering it is possible to set $\widetilde{m}=0$ sharply, in the bottom right panel we have studied the $c$-functions as functions of $\ell$ when $\widetilde{m}=0$.
The functions $\mathcal{C}_S$ and $\mathcal{C}_C$ are constantly equal to $\frac{1}{3}$, while the behaviour of $\mathcal{C}_M(\widetilde{m}=0)$ is non trivial: it can be argued as follows.
 For a $1+1$ dimensional Dirac theory with $m=0$, which is a CFT with central charge equal to one, we expect (using also the non universal constant terms reported in \cite{Arias22Monotones})
\be
M_A=\frac{1}{9} \big[ \log(\ell / \epsilon) \big]^2+1.77917\, \log(\ell / \epsilon)+O(1)\,.
\ee
Then, using (\ref{CfunctM}) and (\ref{CfunctMtilde})
\be
\label{c_M massless}
\mathcal{C}_M(m=0)=\frac{2}{9} \log(\ell / \epsilon) +1.77917\,,
\,\,\qquad\,\,
\widetilde{\mathcal{C}}_M(m=0)=\frac{1}{9}+\frac{0.889587}{\log(\ell/\epsilon)}\,,
\ee
where we stress that, in the limit $\ell\to\infty$, $\widetilde{\mathcal{C}}_M(m=0)$ is constant and therefore scale invariant at the fixed point.
Identifying the mass of the field $m$ and the lattice parameter $\widetilde{m}$ in the continuum limit, we have that the behaviours in (\ref{c_M massless}) are
nicely reproduced by the data (blue and black points) in the the bottom right panel of Fig.\,\ref{fig:Cfunctions} (in Fig.\,\ref{fig:Cfunctions} we have set $\epsilon=1$).

\section{Capacity of entanglement after a global quantum quench}\label{quench}
\label{sec:quench}

The global quantum quench is a protocol that has been largely studied during the past years to explore 
the dynamics of quantum systems out of equilibrium
(see the reviews \cite{Calabrese:2016xau,Essler:2016ufo} for an exhaustive list of references).
Given a system prepared in the ground state $| \psi_0 \rangle$ of the hamiltonian $\widehat{H}_0$, 
at $t=0$ a sudden global change is performed such that
the unitary evolution of $| \psi_0 \rangle$ is induced by the hamiltonian $\widehat{H}$, namely
\be
\label{time_evolved_state}
| \psi(t) \rangle =  e^{-\textrm{i} \widehat{H} t} \, | \psi_0 \rangle\,,
\qquad
t>0\,.
\ee
Since $\widehat{H}_0$ and $\widehat{H}$ do not commute in general, the time evolution in (\ref{time_evolved_state}) is highly non trivial.

In this section we study the time evolution of the capacity of entanglement of an interval in an infinite line after a global quantum quench.
We consider a fermionic model and a bosonic model,
in the simple cases where all the Hamiltonians involved are free.

The temporal evolutions of various entanglement quantifiers 
after these quenches have been considered in the literature: we mention
the entanglement Hamiltonians \cite{Cardy:2016fqc,Fagotti_2013,Wen:2018svb,DiGiulio:2019lpb},
the entanglement spectra \cite{Torlai_2014,Cardy:2016fqc,DiGiulio:2019lpb,Surace:2019mft},
the contours of entanglement \cite{Chen_2014,DiGiulio:2019lpb,Kudler-Flam:2019oru} 
and the entanglement negativity \cite{Coser:2014gsa}.
Also the temporal evolution after a quantum quench of the circuit complexity between two reduced density matrices 
have been recently considered \cite{DiGiulio:2020hlz,DiGiulio:2021oal}.

\subsection{CFT approach}
\label{subsec:CFTquench}

For critical evolutions, CFT methods have been developed to study the 
temporal evolution of the R\'enyi entropies after a global quench \cite{Calabrese:2016xau}.

When $A$ is a semi-infinite line, a linear growth has been found,
both by using the twist fields correlators \cite{Calabrese_2005} 
and the entanglement Hamiltonian \cite{Cardy:2016fqc}.
In this case (\ref{Trrhon_CFT}) holds with 
\be
\label{W global quench}
W_A=\log\!\bigg[\frac{\tau_0}{\pi \epsilon}\cosh(2\pi t/\tau_0)\bigg]\,,
\ee 
where $\tau_0$ is a parameter that encodes some properties of the initial state. 
Taking $t/\tau_0\gg 1$ in (\ref{W global quench}), one obtains
\be
\label{W global quench large t}
W_A=\log\!\bigg(\frac{\tau_0}{2\pi \epsilon}\bigg)+\frac{2\pi t}{\tau_0}\,.
\ee

When $A$ is an interval of length $\ell$ in an infinite line,
in the regime where $\frac{\ell}{\tau_0}\gg 1 $ and $\frac{t}{\tau_0}\gg 1$, 
the twist field approach has lead to the following temporal evolution \cite{Calabrese_2005} 
\be
\label{Trrhon quench interval}
\textrm{Tr}\rho_A^n
=\,
\tilde{c}_n \left( \frac{2 \pi}{\tau_0} \right)^{\frac{c}{12}(n-\frac{1}{n})}\left(\frac{e^{2\pi \ell/\tau_0}+e^{4 \pi  t/\tau_0}}{e^{2\pi \ell/\tau_0}e^{4\pi  t/\tau_0}} \right)^{\frac{c}{12}(n-\frac{1}{n})}\,,
\ee
where $\tilde{c}_n$ is a non-universal constant such that $\tilde{c}_1=1$.
By employing this result into (\ref{defentropy}) and (\ref{defcapacity}), we get respectively 
 \be
  \label{quench-SC-cft}
 S_A
\,=\,
\textrm{const} +
\left\{
\begin{array}{ll}
\displaystyle
\frac{2\pi c}{3 \tau_0} \;t
\hspace{.7cm}
&
t< \ell / 2
\\
\rule{0pt}{.8cm}
\displaystyle
\frac{\pi c \,\ell}{3 \tau_0}
&
t > \ell / 2
\end{array}
\right.
\;\;\qquad\;\;
 C_A
\,=\,
\textrm{const} +
\left\{
\begin{array}{ll}
\displaystyle
\frac{2\pi c}{3 \tau_0} \;t
\hspace{.7cm}
&
t< \ell / 2
\\
\rule{0pt}{.8cm}
\displaystyle
\frac{\pi c \,\ell}{3 \tau_0}
&
t > \ell / 2
\end{array}
\right.
 \ee
 where the constant term is
  $- \tilde{c}_1'-\frac{c}{6}\log(2\pi /\tau_0)$ for $S_A$
 and  $[\partial^2_n(\log \tilde {c}_n)]\big|_{n=1} -\frac{c}{6}\log(2\pi /\tau_0)$ for $C_A$.
The saturation regime in (\ref{quench-SC-cft})  is induced by the finiteness of the subsystem. 
In order to get rid of the constant term, it is convenient to consider
\be
\Delta S_A(t)=  S_A(t)-  S_A(0)\,,
 \qquad \qquad
\Delta C_A(t)=  C_A(t)-  C_A(0)\,.
\ee
Thus, if the evolution Hamiltonian is critical, from (\ref{quench-SC-cft}) one expects $\Delta S_A(t)=\Delta  C_A(t)$.
Notice that the linear growth for both the quantities in (\ref{quench-SC-cft}) is consistent 
with the one obtained by combining (\ref{capacityequalentropy}) and (\ref{W global quench large t}), 
up to a factor $2$ due to the occurrence of two entangling points.

The CFT result in (\ref{quench-SC-cft}) tells us that
$S_A$ and $C_A$ grow linearly in time with the same slope until $t\simeq \ell/2$.
Since the numerical analysis performed in Sec.\,\ref{sec:quenchHC} provide different slopes 
for the linear growth of these two quantities, 
let us introduce a possible dependence on $n$ in $\tau_0$, as already done in \cite{Coser:2014gsa} (in this paper, see the right panel of Fig.\,3 and its inset).
Evaluating (\ref{defentropy}) and (\ref{defcapacity}) through (\ref{Trrhon quench interval}) with $\tau_0 = \tau_0(n)$,
we find that, while $S_A$ remains equal to (\ref{quench-SC-cft}), 
for $C_A$ we have
\be
C_A
=
-\frac{c}{6}
\log\!\bigg(e^{-\frac{2\pi\ell}{\tau_0(1)}}+e^{-\frac{4\pi t}{\tau_0(1)}}\bigg)
-
\frac{c\pi \,\tau'_0(1)}{3\,\tau^2_0(1)}
\bigg[
-
(\ell+2t)
+
(\ell-2 t)
\tanh\!\bigg(\frac{\pi(\ell-2 t)}{\tau_0(1)}\bigg)\bigg]
+
\textrm{const}\,, 
\ee
where the constant reads $\big[\partial^2_n(\log \tilde {c}_n) \big] \big|_{n=1} -\frac{c}{6}\log(2\pi /\tau_0)+\frac{c\,\tau'_0(1)}{3\tau_0(1)}$.
In the regimes of small and large $t$, we have respectively
\be
\label{CA-tau0-prime}
 C_A
\,=\,
\textrm{const} +
\left\{
\begin{array}{ll}
\displaystyle
\frac{2\pi c}{3\, \tau_0(1)}
\bigg(1+\frac{2 \tau'_0(1)}{\tau_0(1)} \bigg) \, t
\hspace{.7cm}
&
t< \ell / 2
\\
\rule{0pt}{.8cm}
\displaystyle
\frac{\pi c }{3 \,\tau_0(1)}
\bigg(1+\frac{2 \tau'_0(1)}{\tau_0(1)} \bigg)\,\ell
&
t > \ell / 2\,,
\end{array}
\right.
\ee
which has a different slope for the initial linear growth with respect to $S_A$.

\subsection{Quantum quenchs in free chains}
\label{sec:quenchHC}

\subsubsection{Quasi-particle picture}

In order to explain the qualitative temporal behaviour (\ref{quench-SC-cft}),
where a linear growth is followed by a saturation,
a quasi-particle picture has been introduced \cite{Calabrese_2005, Calabrese:2016xau}.
When the initial state has very high energy with respect to the ground state of the 
Hamiltonian governing the temporal evolution,
it can be seen as a source of quasi-particle excitations. 
In one spatial dimension, it is assumed that at $t= 0$ each spatial point of the system 
emits in the same way a pair of entangled quasi-particles with opposite momenta $k$ and $- k$ 
according to certain probability distribution that depends on both the initial state and the evolution hamiltonian.  
Only the particles emitted at the same point are entangled.

Considering the spatial bipartition $A\cup B$,
since only the particles emitted at the same point are entangled,
at time $t$ a point in $A$ is entangled with another one in $B$ 
if they are reached simultaneously by two quasi-particles emitted from the same point at $t=0$. 
When $A$ is an interval of length $\ell$ in the line,
since $S_A$ is proportional to the number of quasi-particles entangling the two subsystems, 
for the entanglement entropy at time $t$ one finds  
\cite{Calabrese_2005, Calabrese:2016xau,AlbaCalabrese17}
\be
\label{entropy_QPpict}
\Delta S_A(t)=\,2 \,t \int_{2|v_k|t<\ell}
  \!\! \tilde{s}(k)   \,v_k   \, dk
  +
  \ell \int_{2|v_k|t>\ell}  \!\! \tilde{s}(k)\,dk\,,
\ee
where $v_k$ is the velocity of the quasi-particles with momentum $k$ 
and $\tilde{s}(k)$ denotes the product of the momentum distribution function 
and the contribution of the pair of quasi-particles with momenta $k$ and $-k$ to the entanglement entropy.
The explicit expressions of $v_k$ and $\tilde{s}(k)$ are model dependent
and the initial value of the entanglement entropy 
cannot be obtained from this qualitative description.

A straightforward extension of the above description to the capacity of entanglement gives
\be
\label{capacity_QPpict}
\Delta C_A(t)=\,2 \,t \int_{2|v_k|t<\ell}    \!\!\tilde{c}(k)   \,v_k   \,dk
  +
  \ell \int_{2|v_k|t>\ell}  \!\! \tilde{c}(k)\,dk\,,
\ee
where $v_k$ is the same velocity occurring in (\ref{entropy_QPpict})
and $\tilde{c}(k)$ is the model dependent function 
given by  the product  between the momentum distribution function 
and the contribution of the pair of quasi-particles with momenta $k$ and $-k$ to the capacity of entanglement.

By adapting the analysis reported in \cite{ac-18-qp-quench},
where $\tilde{s}(k)$ has been computed from the asymptotic state for $t\to\infty$ of the model,
in the following we evaluate $\tilde{c}(k)$.
In free and integrable models infinitely many conserved quantities occur;
hence the stationary state reached at $t \to \infty$ is characterised by a  Generalised  Gibbs  Ensemble (GGE)
\cite{Rigol_07,Sotiriadis:2014uza,Ilievski_2015,Ilievski:2016fdy}$\;$(see the review \cite{Vidmar_2016} for an extensive list of references).
In particular, for free models the GGE reads
\cite{Calabrese:2007rg,ac-18-qp-quench}
\be
\label{partitionGGE}
\rho_{\textrm{\tiny GGE}}
=
\frac{e^{-\sum_{k}\lambda_k \hat{n}^{(\textrm{\tiny B,F})}_k}}{Z}\,,
\;\;\;\qquad\;\;\;
Z=\prod_k \big({1\mp e^{-\lambda_k}}\big)^{\mp 1},
\ee
where $\hat{n}^{(\textrm{\tiny B})}_k$ and $\hat{n}^{(\textrm{\tiny F})}_k$ are bosonic and fermionic number operators respectively. 
The upper and the lower signs in the partition function $Z$, 
which is written in terms of the Lagrange multipliers $\lambda_k$
and guarantees the normalisation to one of the density matrix,
correspond to the bosonic and the fermionic case respectively. 
The expectation value on the GGE state of $\hat{n}^{(\textrm{\tiny B,F})}_k$ is
\be
n_k\equiv\langle\hat{n}^{(\textrm{\tiny B,F})}_k\rangle=-\frac{\partial \log Z}{\partial \lambda_k}
=\frac{1}{e^{\lambda_k}\mp 1}\,.
\ee

In order to compute the entropy and the capacity in the GGE, 
one introduces $Z_n$ by rescaling $\lambda_k\to n\lambda_k$ in (\ref{partitionGGE}), namely
\be
Z_n\equiv \mathrm{Tr}e^{-n\sum_{k}\lambda_k \hat{n}^{(\textrm{\tiny B,F})}_k} =
\prod_k \big(1\mp e^{-n\lambda_k}\big)^{\mp 1}\,.
\ee
Then, since $\textrm{Tr} \rho_A^n = Z_n / Z^n$, one finds
\bea
\label{entropyGGE}
S_\textrm{\tiny GGE}
&=&
-\,\partial_n (\log Z_n)\big|_{n=1} + \log Z 
\,=\,
 \sum_k \big[ (n_k\pm 1 )\,\log (1\pm n_k ) -n_k \log n_k \big]\,,
 \\
\label{capacityGGE}
C_\textrm{\tiny GGE}
&=&
\partial^2_n (\log Z_n)\big|_{n=1}
 = 
 \sum_k (1\pm n_k )\,n_k   \bigg[ \!\log\! \left(\frac{1\pm n_k}{n_k} \right)\!\bigg]^2\,.
\eea

In the thermodynamic limit $L \to \infty$ the sum over the momenta becomes an integral over a model dependent domain $\mathcal{K}$
that depends on the quench we are considering.
In this limit (\ref{entropyGGE}) and (\ref{capacityGGE})  become respectively
\bea
\label{entropyGGETD}
S_\textrm{\tiny GGE}
&=&
 L \int_{\mathcal{K}}  
\big[ (n_k\pm 1 )\,\log (1\pm n_k ) -n_k \log n_k \big]\, \frac{dk}{|\mathcal{K}|} \,,
\\
\rule{0pt}{.8cm}
\label{capacityGGETD}
C_\textrm{\tiny GGE}
&=&
 L \int_{\mathcal{K}} 
(1\pm n_k )\,n_k   \bigg[ \!\log\! \left(\frac{1 \pm n_k}{n_k} \right)\!\bigg]^2 \frac{dk}{|\mathcal{K}|} \,,
\eea
where $|\mathcal{K}|$ denotes the size of the model dependent domain $\mathcal{K}$.

When the subsystem $A$ is an interval with length $\ell<L$,
the density of thermodynamic entropy in the GGE 
coincides with the one of the entanglement entropy along the evolution after the quench \cite{ac-18-qp-quench}. 
For free models, this holds also for Rényi entropies \cite{Alba:2017kdq};
hence the validity of this property is expected also for the capacity of entanglement.
The densities of entropy and capacity of entanglement occurring in 
(\ref{entropy_QPpict}) and (\ref{capacity_QPpict}) are
\be
\label{h-c-function_HC}
\tilde{s}(k)
=
\frac{1}{|\mathcal{K}|} \, \Big[ \left(n_k\pm 1 \right)\log \left(1\pm n_k \right)-n_k\log n_k \,\Big]\,,
\;\;\qquad\;\;
\tilde{c}(k)
=
\frac{(1\pm n_k )\,n_k}{|\mathcal{K}|}   \bigg[ \!\log\! \left(\frac{1\pm n_k}{n_k} \right)\!\bigg]^2\,.
\ee
These densities provide the saturation constants of $S_A$ and $C_A$ as follows
\be
\label{larget_SA_CA}
\lim_{t\to\infty} \frac{\Delta S_A}{\ell} \,=\int_{\mathcal{K}}  \tilde{s}(k)\, dk\,,
\;\;\;\;\qquad\;\;\;\;
\lim_{t\to\infty} \frac{\Delta C_A}{\ell} \, =\int_{\mathcal{K}}  \tilde{c}(k)\, dk\,.
\ee

In the free fermionic and bosonic systems that we are considering,
the reduced density matrices for a block made by $\ell$ consecutive sites 
are Gaussian states which can be written as follows 
\cite{Peschel,Peschel03,ep-rev}
\be
\label{rhoA_Gaussian}
\rho_A
=
\frac{e^{-\sum_{k=1}^\ell\varepsilon_k \hat{b}_k^\dagger \hat{b}_k}}{\mathrm{Tr}\big(e^{-\sum_{k=1}^\ell\varepsilon_k \hat{b}_k^\dagger \hat{b}_k}\big)}
=
\frac{e^{-\sum_{k=1}^\ell\varepsilon_k \hat{b}_k^\dagger \hat{b}_k}}{\prod_{k=1}^\ell \left(1 \mp e^{-\varepsilon_k}\right)^{\mp 1}}\,,
\ee
where $\hat{b}_k^\dagger, \hat{b}_k$ are bosonic (fermionic) creation and annihilation operators 
and the upper (lower) signs in the last expression correspond to the bosonic (fermionic) case respectively. 
The occupation number is determined by single-particle entanglement energies $\varepsilon_k$
\be
\label{nk_Gaussian}
\mathrm{Tr}\big( \rho_A \hat{b}_k^\dagger \hat{b}_k\big)
=
\frac{1}{e^{\varepsilon_k}\mp 1}
\equiv
\tilde{n}_k\,.
\ee

In the simplest configuration, the chain is made by two sites and $\ell=1$ in (\ref{rhoA_Gaussian}) and (\ref{nk_Gaussian}).
In this case, by employing (\ref{nk_Gaussian}), we have that 
the spectrum of $\rho_A$ in (\ref{rhoA_Gaussian}) for fermions and bosons read respectively
\be
\label{ES_1fermion}
\left\lbrace \frac{e^{-\varepsilon}}{1+e^{-\varepsilon}}\,,\frac{1}{1+e^{-\varepsilon}} \right\rbrace
=
\big\lbrace \tilde{n}\,, 1-\tilde{n} \big\rbrace\,,
\ee
and
\be
\label{ES_1boson}
\left\lbrace e^{-s\varepsilon}
\left(1-e^{-\varepsilon} \right) ; s\in \mathbb{N}_0 \right\rbrace
=\,
\left\lbrace
\frac{1}{1+\tilde{n}}\left(\frac{\tilde{n}}{1+\tilde{n}}\right)^s; \, s\in \mathbb{N}_0
 \right\rbrace\,,
\ee
where we have set $\varepsilon_1\equiv\varepsilon$ and $\tilde{n}_1\equiv\tilde{n}$.
These spectra provide the corresponding entanglement entropy and capacity of entanglement.
From (\ref{ES_1fermion}) and (\ref{ES_1boson}) we get respectively 
\be
\label{SA-CA-1-fermion}
S^{(\textrm{1f})}_A
=
-\tilde{n}\log\tilde{n}-(1-\tilde{n})\log(1-\tilde{n})\,,
\,\,\qquad\,\,
C^{(\textrm{1f})}_A
=
(1- \tilde{n} )\,\tilde{n}  \bigg[ \!\log\! \left(\frac{1- \tilde{n}}{\tilde{n}} \right)\!\bigg]^2\,,
\ee
and 
\be
\label{SA-CA-1-boson}
S^{(\textrm{1b})}_A
=
-\tilde{n}\log\tilde{n}+(1+\tilde{n})\log(1+\tilde{n})\,,
\,\,\qquad\,\,
C^{(\textrm{1b})}_A
=
(1+ \tilde{n} )\,\tilde{n}  \bigg[ \!\log\! \left(\frac{1+ \tilde{n}}{\tilde{n}} \right)\!\bigg]^2\,.
\ee
Notice that  these expressions for $S_A$ and $C_A$
coincide with the corresponding densities in (\ref{h-c-function_HC}), once we identify $n_k=\tilde{n}$. 
This is consistent with the quasi-particles picture after a quench of free theories,
where the quasi-particles can be seen as pairs of counter-propagating fermions or bosons with the same momenta.

\subsubsection{Harmonic chain}
\label{sec-quench-hc-CA}

Considering the harmonic chain (\ref{HC ham}),
in the following we explore the temporal evolution of the capacity of entanglement after 
the global quench 
where  the initial state is the ground state of (\ref{HC ham}) for the value  $\omega_0$
and the evolution Hamiltonian is (\ref{HC ham}) with $\omega \neq \omega_0$.

\begin{figure}[t!]
\vspace{.2cm}
\hspace{-1.4cm}
\includegraphics[width=.57\textwidth]{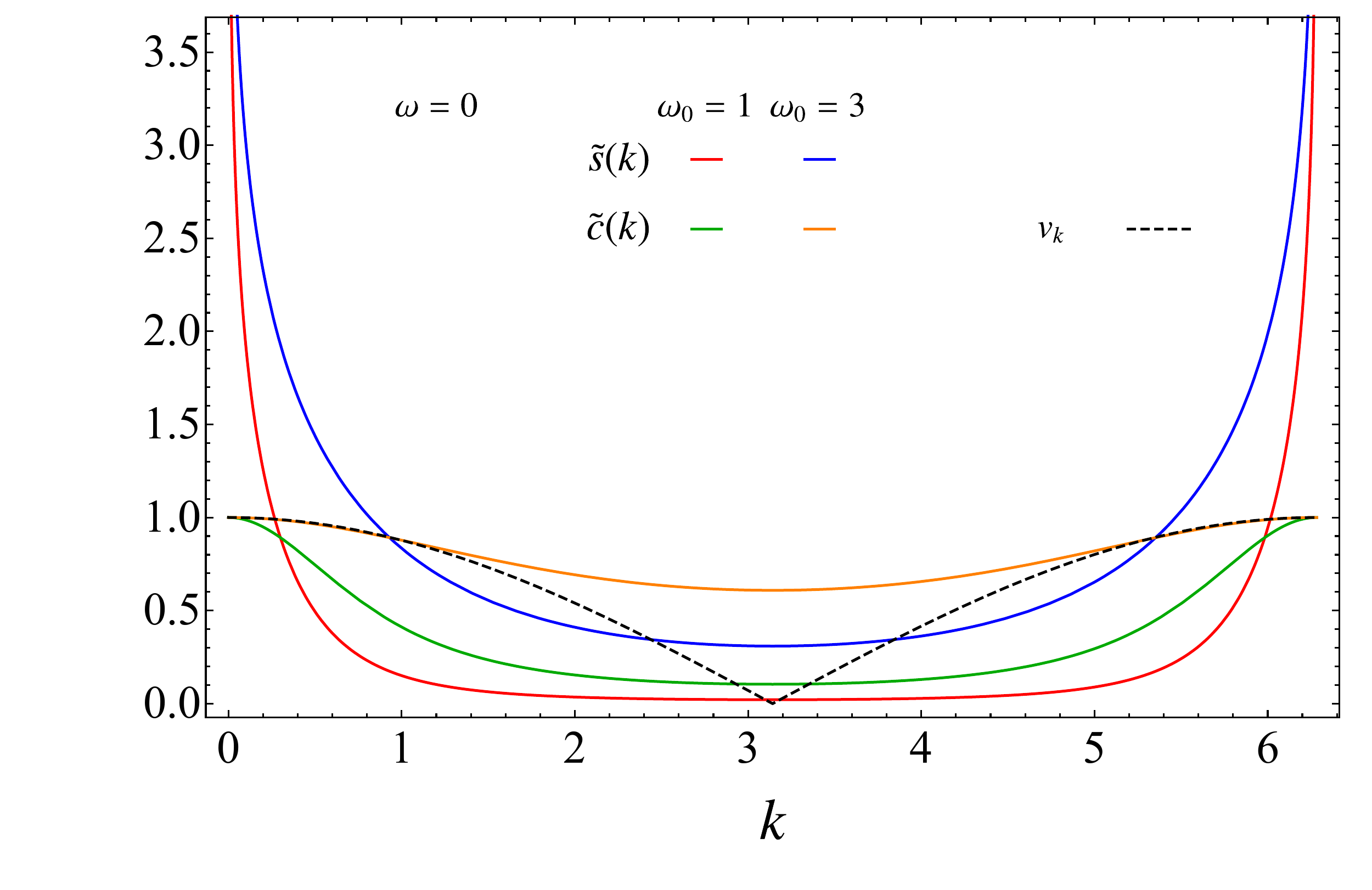}
\hspace{-.5cm}
\includegraphics[width=.57\textwidth]{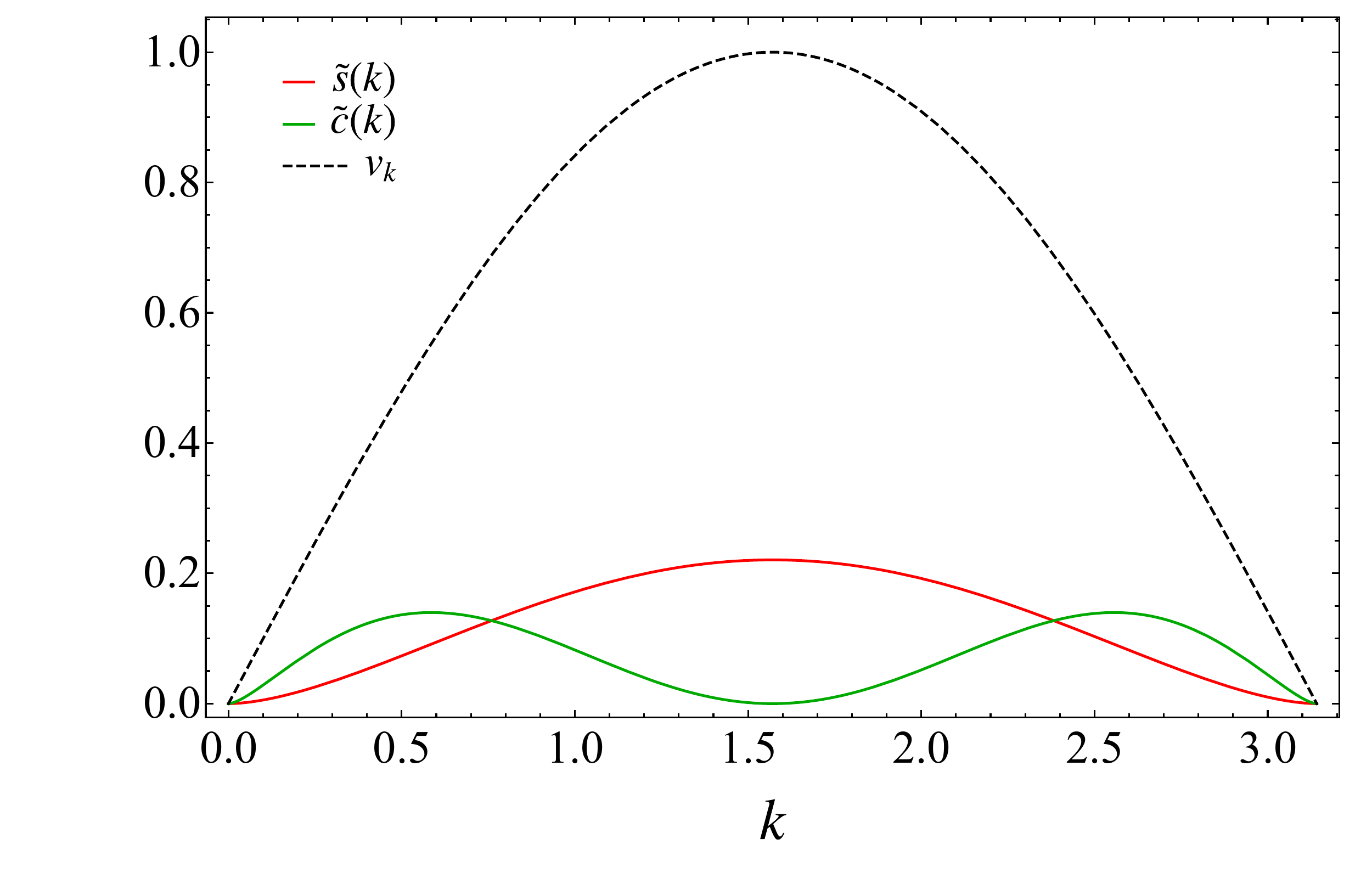}
\vspace{-.6cm}
\caption{
The densities of the quasi-particles (\ref{h-c-function_HC}) and the corresponding velocity 
for the quantum quenches 
discussed in Sec.\,\ref{sec:quench} and Sec.\,\ref{sec:contour}
for the bosonic (left panel) and the fermionic (right panel) chain,
in terms of the momentum $k$.
}
\vspace{.5cm}
\label{fig:densities_quench}
\end{figure}

The temporal evolutions of the entanglement entropy and of the capacity of entanglement 
after this global quench are given by (\ref{entropy_QPpict}) and  (\ref{capacity_QPpict}) respectively,
with $k\in[0,2\pi]$ and the densities in (\ref{h-c-function_HC}).
The occupation numbers $n_k$ to employ in these expressions read
\cite{Calabrese:2007rg}
\be
\label{nk}
n_k
=
\frac{1}{4}\left(\frac{\omega_k}{\omega_{0,k}}+\frac{\omega_{0,k}}{\omega_{k}} \right)-\frac{1}{2}\,,
\ee
where 
\be
\label{dispersionrelation}
\omega_k=\sqrt{\omega^2+ \frac{4\lambda}{\mu}\sin^2(k/2)}\,,
\ee
and $\omega_{0,k}$ is obtained by replacing $\omega$ with $\omega_0$ in this dispersion relation. 
The velocity of the quasi-particles after the quench is given by 
\be 
\label{vk}
v_k \equiv \frac{\partial \omega_k}{\partial k} = 
\, \frac{(\lambda/\mu)\sin(k)}{\sqrt{\omega^2+ 4(\lambda/\mu)\sin^2(k/2)}}\,.
\ee
Both the saturation constants in (\ref{larget_SA_CA}) depend on $\omega_0$ and $\omega$.
We cannot find analytic expressions for these constants, 
but their values can be obtained numerically case by case.

In Fig.\,\ref{fig:densities_quench} we show the functions (\ref{h-c-function_HC})
entering  in (\ref{entropy_QPpict}) and (\ref{capacity_QPpict}),
both for the bosons (left panel) and the fermions considered in the next subsection (right panel).
For the quench in the harmonic chain, 
the density functions $\tilde{s}$ and $\tilde{c}$ are defined in the domain $k\in [0,2\pi]$ and are obtained by employing (\ref{nk}) and (\ref{vk}).
In the left panel of Fig.\,\ref{fig:densities_quench}, 
we plot $\tilde{s}$ and $\tilde{c}$ for $\omega=0$ and two different values of $\omega_0$. 
A crucial difference occurs between $\tilde{s}$ and $\tilde{c}$\,:
for either  $k=0$ or $k = 2\pi$ 
the density $\tilde{s}$ diverges logarithmically, while $\tilde{c}$ is always finite. 
We remark that both $\tilde{s}$ and $\tilde{c}$ have maxima at $k=0$ and $k=2\pi$, 
where the velocity of the particles takes its maximum value.
In the right panel of Fig.\,\ref{fig:densities_quench}
we show the densities $\tilde{s}$ and $\tilde{c}$ 
(see (\ref{occupation Fermi GGE}) and (\ref{velocity Fermi GGE})) 
for the free fermionic quench.
In this case, while $\tilde{s}$ has a maximum in the center of the domain $k\in [0,\pi]$ and vanishes at its extrema,
the function $\tilde{c}$ vanishes for $k\in \{0,\pi/2,\pi\}$ and it has local maxima when $k=\bar{k},\pi-\bar{k}$, with $\bar{k}\simeq 0.585$.
For this fermionic quench, notice that 
$\tilde{s}$ and the velocity reach their  maximum at $k = \pi/2$, 
where $\tilde{c}$ is minimum.

\begin{figure}[t!]
\vspace{.2cm}
\hspace{-1.4cm}
\includegraphics[width=.57\textwidth]{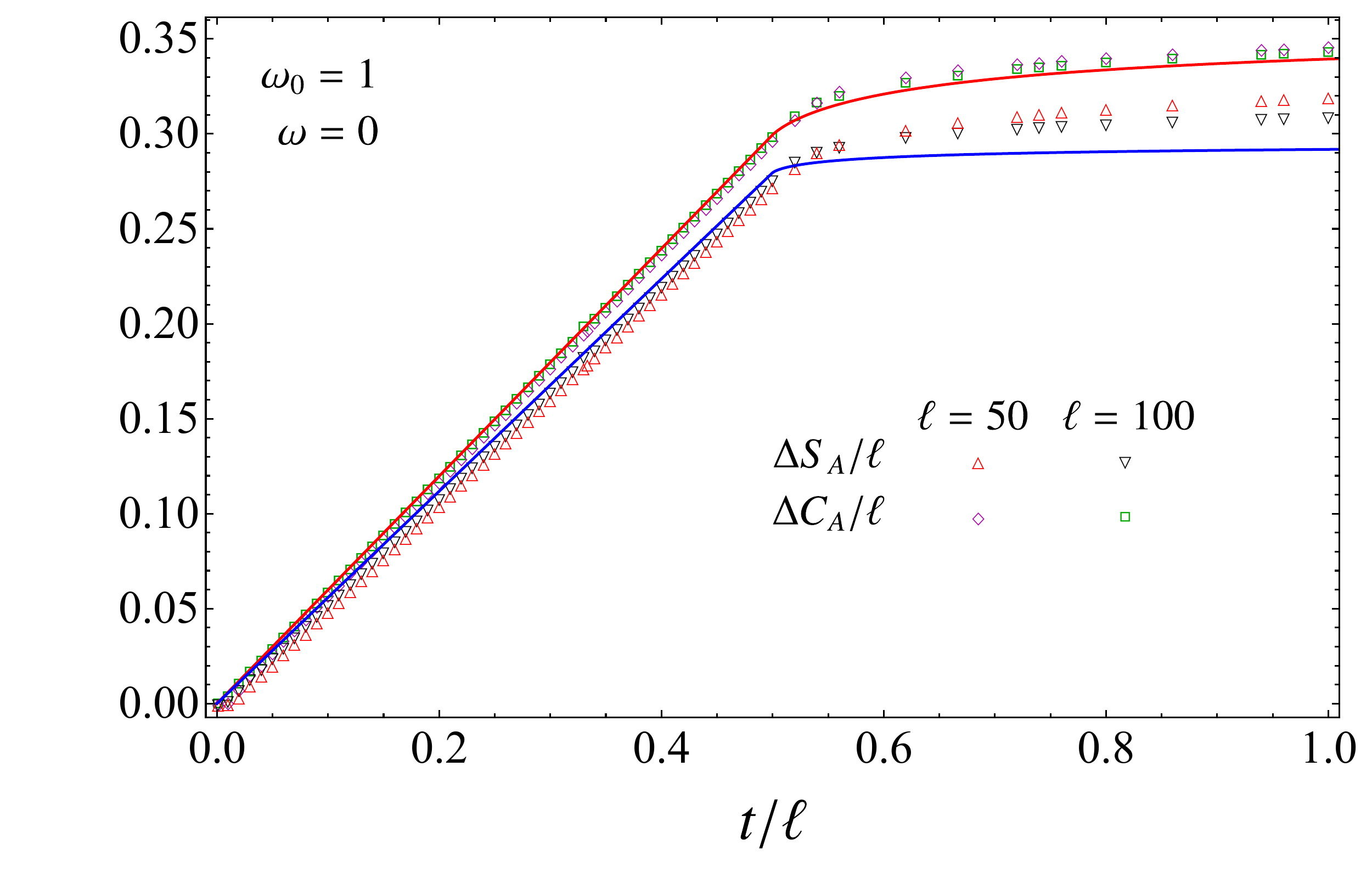}
\hspace{-.5cm}
\includegraphics[width=.57\textwidth]{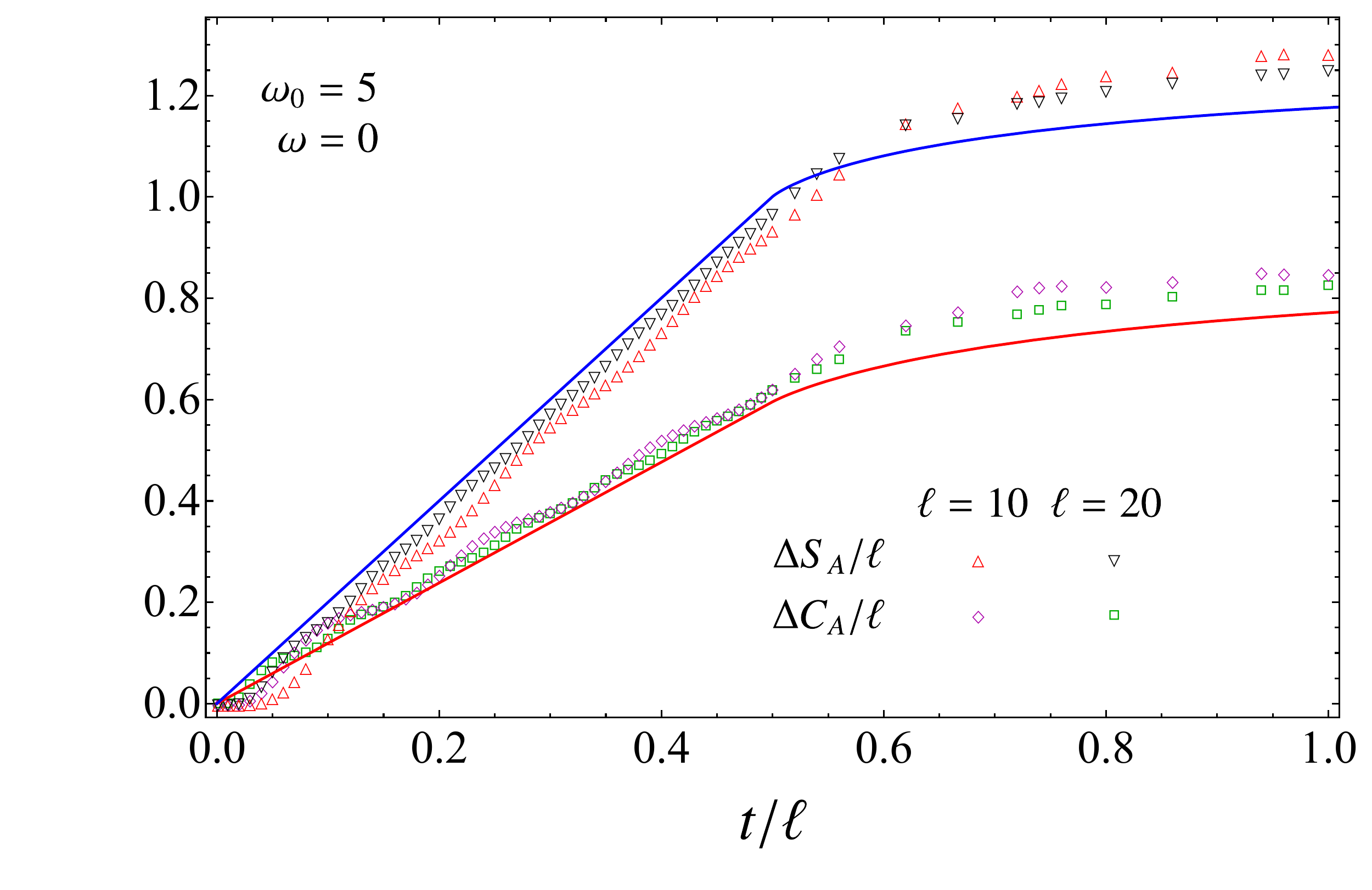}
\vspace{-.2cm}
\caption{
Temporal evolutions of $\Delta S_A$ and $\Delta C_A$ 
of a block made by $\ell$ consecutive sites in the infinite harmonic chain 
after a global quantum quench of the mass parameter from $\omega_0$ to $\omega$.
In the left panel $\omega_0=1$, while $\omega_0=5$ in the right panel. 
}
\vspace{.5cm}
\label{fig:HC_CEvsEEvsQPformulaforEE}
\end{figure}

In Fig.\,\ref{fig:HC_CEvsEEvsQPformulaforEE} we show the temporal evolutions of
$\Delta S_A$ and $\Delta C_A$ of an interval in an infinite harmonic chain after 
two global quantum quenches of the mass parameter from $\omega_0$ to $\omega=0$,
with  $\omega_0=1$ (left panel) and $\omega_0=5$ (right panel).
The numerical data shown in this figure have been obtained as explained in Appendix \ref{app-lattice}.
Since the evolution hamiltonian is massless, the CFT predictions can be employed. 
We have reported the data corresponding to two different lengths for the interval 
and the solid lines are obtained from the expressions derived from the quasi-particle picture, namely (\ref{entropy_QPpict}), (\ref{capacity_QPpict}), 
(\ref{h-c-function_HC}), (\ref{nk}) and (\ref{vk}).
In both the panels of this figure the data reported have either $\omega_0\ell=50$ or $\omega_0\ell=100$.
Both $\Delta S_A$ and $\Delta C_A$ exhibit a linear growth in the regime $t/\ell<1/2$ and a saturation for $t/\ell>1/2$, as expected from CFT 
(see (\ref{quench-SC-cft})).
However, the slopes of the linear growth for $\Delta S_A$ and $\Delta C_A$
are different and, consequently, also the saturation value.
%
Thus, (\ref{quench-SC-cft}) is not confirmed by the numerical data at quantitative level. 
Following \cite{Coser:2014gsa},
this discrepancy can be explained by assuming that $\tau_0$ depends on $n$,
which leads to (\ref{CA-tau0-prime}).
Now, the different slopes for the linear growths for $\Delta S_A$ and $\Delta C_A$
can be reproduced by fitting $\tau_0(1)$ and $\tau'_0(1)$.
Furthermore, notice that the slopes of the linear growths 
depend on the value of $\omega_0$. 
Let us remark that we can have either 
$\Delta S_A<\Delta C_A$ (see e.g. the left panel) 
or $\Delta S_A> \Delta C_A$ (see e.g. the left panel),
depending on the value of $\omega_0$.

\begin{figure}[t!]
\vspace{.2cm}
\hspace{-1.1cm}
\includegraphics[width=1.05\textwidth]{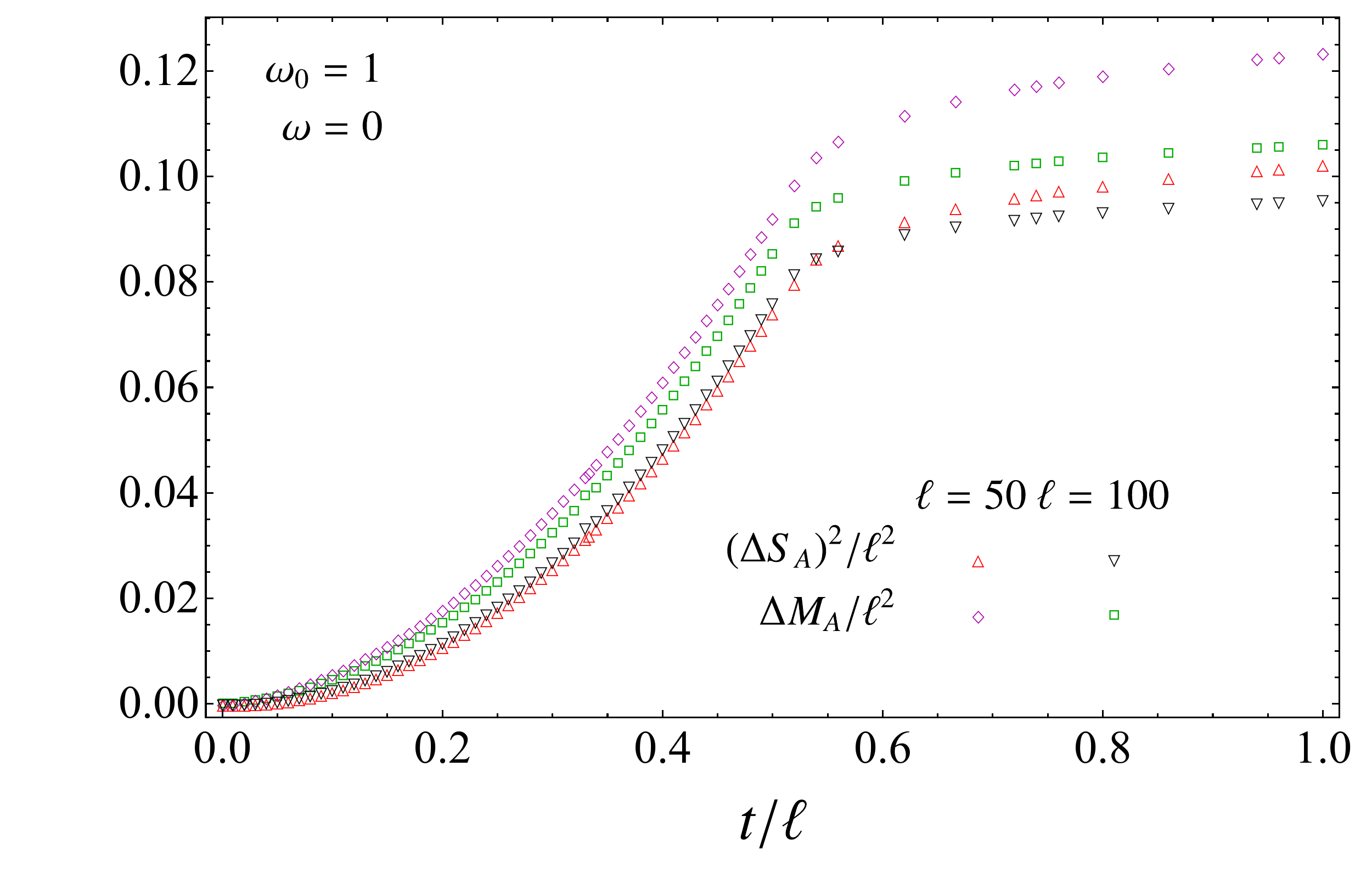}
\vspace{-.2cm}
\caption{
Temporal evolutions of $\Delta M_A$ and of $(\Delta S_A)^2$ 
of a block of $\ell$ consecutive sites in an infinite harmonic chain 
after a global quantum quench of the mass parameter.
}
\vspace{.5cm}
\label{fig:HC_Mfuncafterquench}
\end{figure}

From the definition of $M_A$ in (\ref{Mdefinition}), we have that
$\Delta M_A\equiv M_A(t)-M_A(0)$ reads
\\
\be
\label{Delta-MA-SA-CA}
\Delta M_A
=
\big(\Delta S_A\big)^2
+
\Delta C_A
+
2\,\Delta S_A\bigg(S_A(0)+1\bigg)\,.
\ee
This expression and
the data collapses of $\Delta S_A/\ell$ and $\Delta C_A/\ell $ 
(see Fig.\,\ref{fig:HC_CEvsEEvsQPformulaforEE}) for large $\ell$
suggest to consider $\Delta M_A /\ell^2$
and compare it with $(\Delta S_A /\ell)^2$.
Moreover, (\ref{Delta-MA-SA-CA}) tells us also that $\Delta M_A / \ell^2 \to (\Delta S_A / \ell)^2$ for large $\ell$.
This is supported by the data in Fig.\,\ref{fig:HC_Mfuncafterquench}
showing the temporal evolution of $\Delta M_A$.
The values of $\ell$ in this figure are not large enough to display this collapse.
However, the data for $\Delta M_A /\ell^2$ approach the ones for $(\Delta S_A /\ell)^2$ for increasing $\ell$, as expected.

\subsubsection{Free fermionic chain}
\label{subsec:quenchFF}

In the following we consider 
the temporal evolution of the capacity of entanglement in a free fermionic chain
after the global quench introduced in \cite{ep-07-local-quench, ep-rev}.

\begin{figure}[t!]
\vspace{.2cm}
\hspace{-1.1cm}
\includegraphics[width=1.05\textwidth]{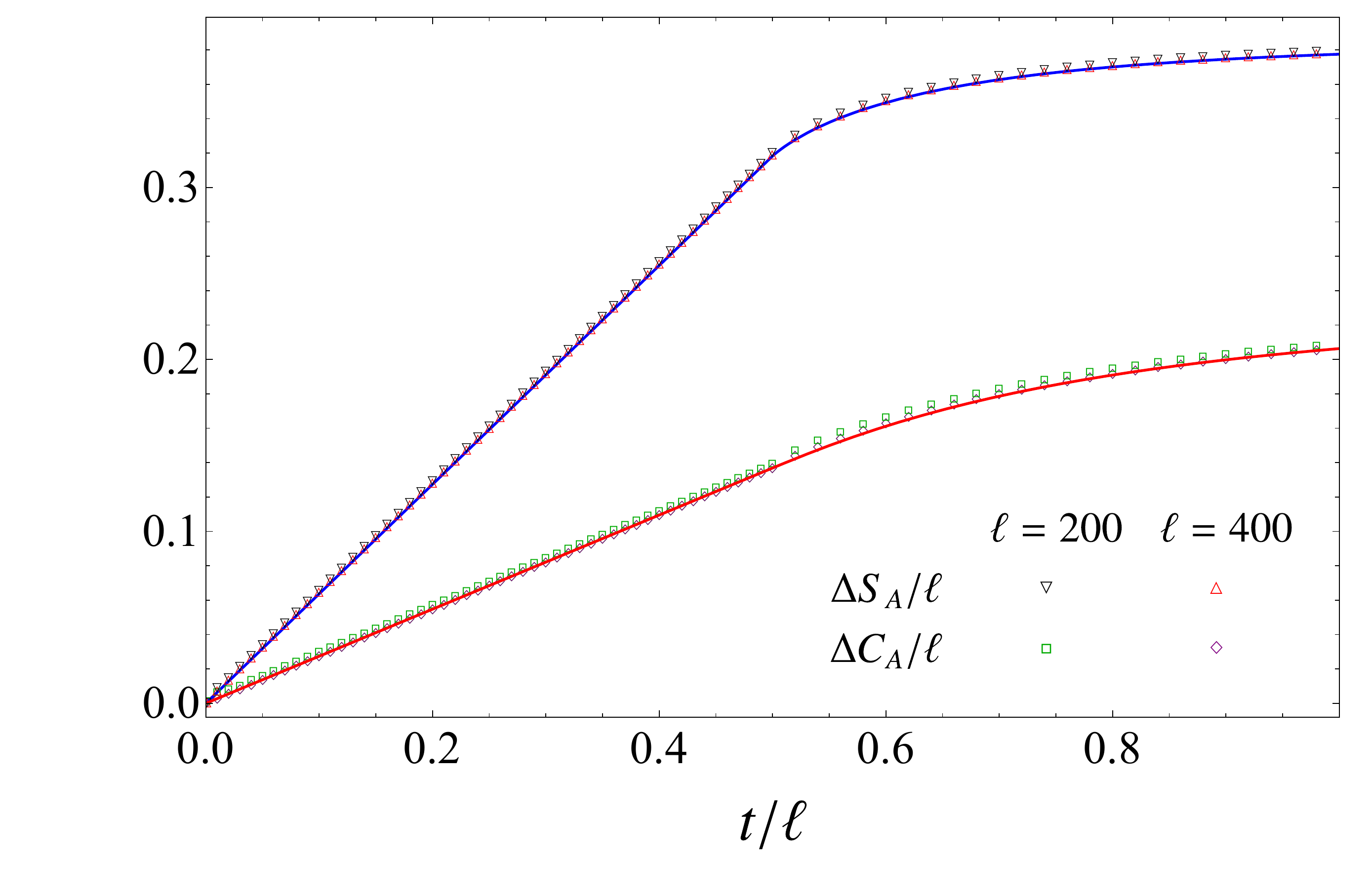}
\vspace{-.5cm}
\caption{
Temporal evolution of $\Delta S_A$ and $\Delta C_A$ 
of a block of $\ell$ consecutive sites in the infinite free fermionic chain 
after a global quench from the fully dimerised chain 
to the homogeneous gapless chain.
Two different values of $\ell$ are considered. 
 The blue and the red solid lines are obtained from (\ref{entropy_QPpict}) and (\ref{capacity_QPpict}) respectively, 
 by using (\ref{h-c-function_HC}), (\ref{occupation Fermi GGE}) 
 and (\ref{velocity Fermi GGE}).
}
\vspace{.5cm}
\label{fig:Cvstfermions}
\end{figure}

Consider the following inhomogeneous free fermionic Hamiltonian 
\cite{ep-07-local-quench}
\be
\label{H_ff inhomo}
\widehat{H}_0= -\,\frac{1}{2}\sum_{n=-\infty}^{+\infty}
\! t_n \Big(\hat{c}^\dag_n \,\hat{c}_{n+1}+\hat{c}^{\dag}_{n+1}\,\hat{c}_n\Big)\,,
\ee
in terms of the fermionic creation and annihilation operators $\hat{c}_n^\dagger$ and $\hat{c}_n$
(which satisfy the standard anticommutation relations $\{\hat{c}_n^\dag,\hat{c}_m^\dag\}=\{\hat{c}_n,\hat{c}_m\}=0$ and $\{\hat{c}_n,\hat{c}^\dag_m\}=\delta_{m,n}$),
with  $t_{2n}=1$ and $t_{2n+1}=0$ (i.e. a fully dimerized chain).
The system is half filled and prepared in the ground state 
$| \psi_0 \rangle$ of $\widehat{H}_0$ and,
at $t=0$, the inhomogeneity is removed by setting all $t_n=1$;
hence the unitary time evolution of $| \psi_0 \rangle$ 
is governed by the translation invariant hopping Hamiltonian 
(tight binding model at half filling)
\be
\label{H_ff homo}
\widehat{H}= -\,\frac{1}{2}\sum_{n=-\infty}^{+\infty}
\! \! \Big(\hat{c}^\dag_n \,\hat{c}_{n+1}+\hat{c}^{\dag}_{n+1}\,\hat{c}_n\Big)\,,
\ee
which is gapless; hence 
in the continuum limit the CFT predictions can be used.
Notice that the Hamiltonian (\ref{H_ff homo}) is equal to the one in (\ref{Ham FF}) up to a prefactor $1/2$. We introduce this rescaling, which does not alter the physical properties of the model, for fixing conventionally the maximal velocity of the excitations equal to one (see \ref{velocity Fermi GGE}).

\begin{figure}[t!]
\vspace{.2cm}
\hspace{-1.1cm}
\includegraphics[width=1.05\textwidth]{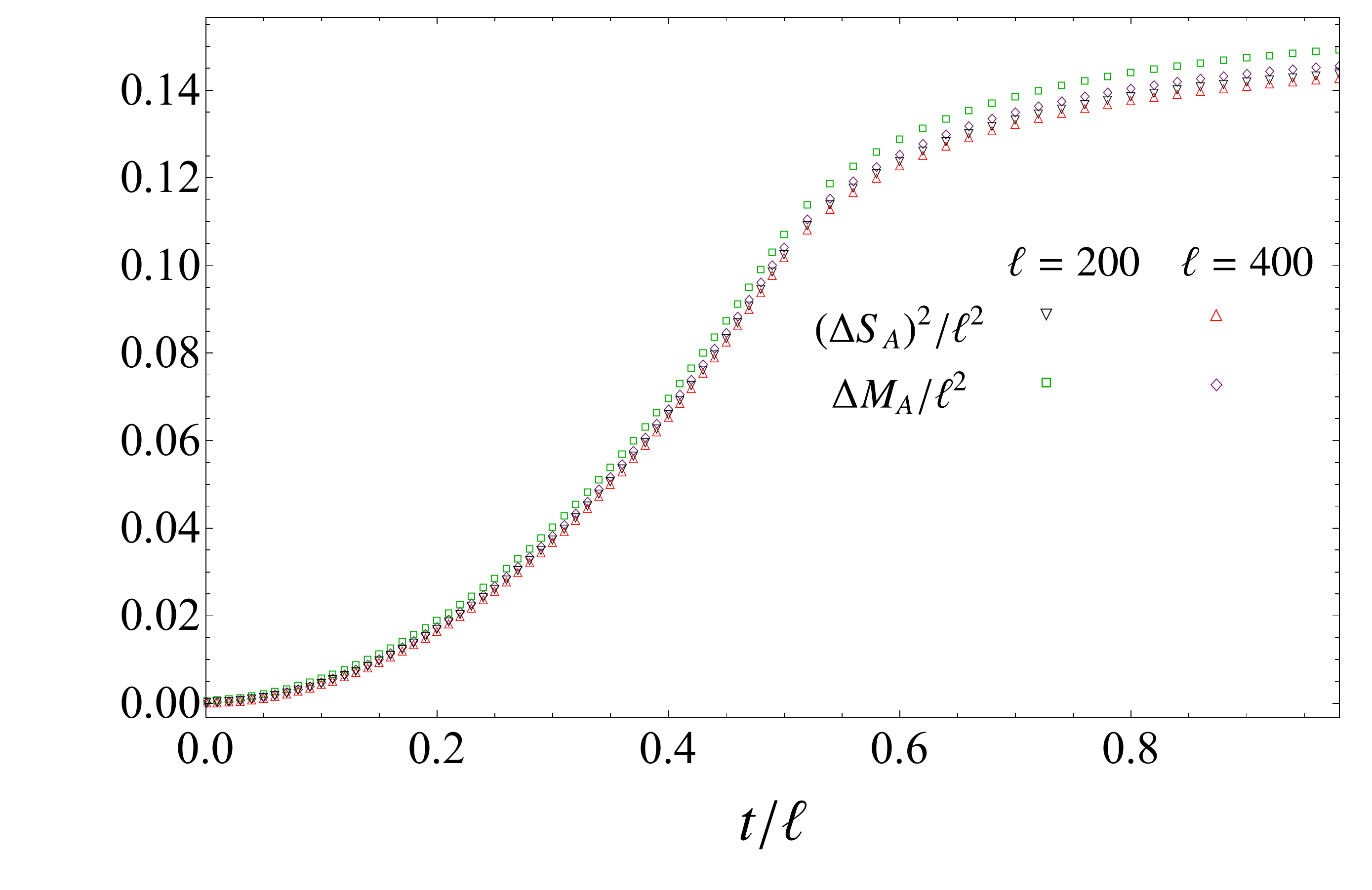}
\vspace{-.2cm}
\caption{
Temporal evolution of $\Delta M_A$ and of $(\Delta S_A)^2$ 
of a block of $\ell$ consecutive sites in the infinite fermionic chain,
after a quantum quench from the fully dimerised chain 
to the homogeneous gapless chain.
}
\vspace{.5cm}
\label{fig:FF_Mfuncafterquench}
\end{figure}

For the temporal evolutions of the entanglement entropy 
and of the capacity of entanglement,
the formulas (\ref{entropy_QPpict}) and  (\ref{capacity_QPpict})
from the quasi-particle picture
with $k\in[0,\pi]$ and (\ref{h-c-function_HC})
can be employed. 
In this fermionic quench, the occupation numbers $n_k$ read \cite{ep-07-local-quench}
\be 
\label{occupation Fermi GGE}
n_k=\frac{1+\cos k}{2}\,,
\;\;\;\;\qquad\;\;\;\;
k\in[0,\pi]\,,
\ee
whose range is between 0 and 1 (as expected from the fermionic statistics)
and the velocity of a quasi-particle with momentum $k$ after the quench is
\be 
\label{velocity Fermi GGE}
v_k=\sin k\,.
\ee
By using (\ref{occupation Fermi GGE}) and (\ref{velocity Fermi GGE}) 
in (\ref{larget_SA_CA}), for the saturation constants of $\Delta S_A$ 
and $\Delta C_A$ at large time one finds  respectively 
\be
\lim_{t\to\infty} \frac{\Delta S_A}{\ell}
\,=\,
\log4 -1\simeq 0.3863\,,
\;\;\;\qquad\;\;\;
\lim_{t\to\infty} \frac{\Delta C_A}{\ell}
\,=\,
\frac{\pi^2}{8}-1\simeq 0.2337\,.
\ee

In Fig.\,\ref{fig:Cvstfermions} we show the temporal evolutions of
$\Delta C_A$ and $\Delta S_A$ for the global quench of the free fermionic system 
described above. 
The numerical data in Fig.\,\ref{fig:Cvstfermions} 
are obtained as described in Appendix \ref{app-lattice} 
and are reported for two different values of the subsystem size $\ell$.
The linear growth predicted from CFT when $t/\ell<1/2$ is observed, 
but  $\Delta C_A$ and $\Delta S_A$ increases with
different slopes. 
Hence, like for the quench in the harmonic chain discussed in Sec.\,\ref{sec-quench-hc-CA}
the prediction (\ref{quench-SC-cft}) with $\tau_0$ independent of $n$ 
does not hold, but this behaviour can be explained by introducing a $n$ dependent parameter $\tau_0$.
The curves coming from the quasi-particle picture,
obtained by plugging (\ref{h-c-function_HC}), (\ref{occupation Fermi GGE}) 
and (\ref{velocity Fermi GGE}) into 
(\ref{entropy_QPpict}) and (\ref{capacity_QPpict}), 
correspond to the solid lines in Fig.\,\ref{fig:Cvstfermions}
and exhibit a very good agreement with the numerical 
data, also after the linear growth regime. 
The different slopes in the linear growths of $\Delta S_A$ and $\Delta C_A$ in Fig.\,\ref{fig:Cvstfermions}
can be understood within the quasi-particle picture. 
The coefficient of the linear growth of $\Delta S_A$ and $\Delta C_A$
is determined by the first term in (\ref{entropy_QPpict}) and (\ref{capacity_QPpict}) respectively. 
From the right panel of Fig.\,\ref{fig:densities_quench}, we observed that in this quench the entropy density $\tilde{s}$ has maximum when $v_k$ is maximum, 
while $\tilde{c}$ is minimum for this value of $k$; 
hence we expect that $\Delta S_A$ grows faster $\Delta C_A$, as confirmed by the data reported in Fig.\,\ref{fig:Cvstfermions}.

The temporal evolution of $\Delta M_A/\ell^2$ compared to the one of $(\Delta S_A/\ell)^2$ is shown in Fig.\,\ref{fig:FF_Mfuncafterquench}
and the expected convergence of $\Delta M_A/\ell^2$ to $(\Delta S_A/\ell)^2$ for large $\ell$ (discussed below (\ref{Delta-MA-SA-CA}))
is more evident with respect to the bosonic case (see Fig.\,\ref{fig:HC_Mfuncafterquench}).

\section{Contour functions for the capacity of entanglement}
\label{sec:contour}

The contour for the entanglement entropies is a function of the position defined in $A$ 
that describes the spatial structure of the bipartite entanglement inside the subsystem $A$ 
when the system is in a pure state. 
When the system is out of equilibrium, also a non trivial dependence on time typically occurs. 
In this section we discuss the contour function associated to the capacity of entanglement.

In a lattice model, 
the contour function for the entanglement entropy 
and the contour function for the capacity of entanglement
are $s_A:A\to\mathbb{R}$ and $c_A:A\to\mathbb{R}$ respectively such that
\be
\label{entropy contour_lattice}
S_A=\sum_{i\in A}s_A(i)\,,
\;\;\;\qquad\;\;
C_A=\sum_{i\in A} c_A(i)\,,
\ee
and satisfying the positivity constraint given by $s_A(i)\geqslant 0$ and $c_A(i)\geqslant 0$.
For $s_A(i)$, further requirements have been discussed in \cite{Chen_2014}.

It is straightforward to extend these notions to quantum field theories in the continuum 
by introducing positive real functions $s_A(x)$ and $c_A(x)$ defined for $x\in A$ such that 
\be
\label{ent-cap-contour_continuum}
S_A=\int_{ A}s_A(x)\,dx\,,
\;\;\;\qquad\;\;\;
C_A=\int_{ A}c_A(x)\,dx\,.
\ee
These contour functions can be easily obtained from the contour function $s_{A}^{(n)}(x)$ 
of the R\'enyi entropies, defined by $S_A^{(n)}=\int_{ A}s_A^{(n)}(x)\,dx$, as follows
\be
s_A(x) = - \big[ \partial_n s_{A}^{(n)}(x) \big] \!\big|_{n=1}\,,
\;\;\;\;\qquad\;\;\;\;
c_A(x) = \big[ \partial^2_n s_{A}^{(n)}(x) \big] \! \big|_{n=1}\,.
\ee

For CFT in one spatial dimension and for the bipartitions considered in \cite{Cardy:2016fqc},
which always involve an interval $A=(u,v)$ of length $\ell$,
the following function has been suggested for the contour function of the R\'enyi entropies \cite{Coser:2017dtb}
\be
\label{contour_renyi_CFT}
s_A^{(n)} 
= 
-\frac{c}{12}\bigg(n-\frac{1}{n}\bigg)f'(x)+\frac{\log c_n}{\ell}\,,
\ee
where $f(z)$ is the conformal mapping characterising the underlying physical case, which is related to $W_A$ in (\ref{Trrhon_CFT}) as follows
\be
\label{integrated annulus}
W_A=\int_{A_\epsilon} f'(x) \,dx\,,
\ee
being $A_\epsilon \equiv (u+\epsilon , v- \epsilon) \subset A$.
From (\ref{contour_renyi_CFT}), one obtains
\be
\label{contour_ent_cap_CFT}
s_A(x)
\,=\,
\frac{c}{6} \, f'(x)-\frac{c'_1}{\ell}\,,
\;\;\;\qquad\;\;\;
c_A(x)
\,=\,
\frac{c}{6} \, f'(x)+\frac{\big[\partial^2_n(\log c_n)\big]\! \big|_{n=1}}{\ell}\,.
\ee

When the conformal field theory is in its ground state and $A$ is an interval on the line, 
we have that $f(x)=\log\!\big(x/(\ell-x)\big)$; hence the contour functions in (\ref{contour_ent_cap_CFT}) become respectively
\be
\label{cont-s-c inverval}
s_A(x)=\frac{c}{6}\;\frac{\ell}{(\ell-x)x}-\frac{c'_1}{\ell}\,,
\;\;\;\qquad\;\;
c_A(x)=\frac{c}{6}\;\frac{\ell}{(\ell-x)x}+\frac{\big[\partial^2_n(\log c_n)\big]\!\big|_{n=1}}{\ell}\,.
\ee

\begin{figure}[t!]
\vspace{.2cm}
\subfigure
{
\hspace{-1.6cm}\includegraphics[width=.57\textwidth]{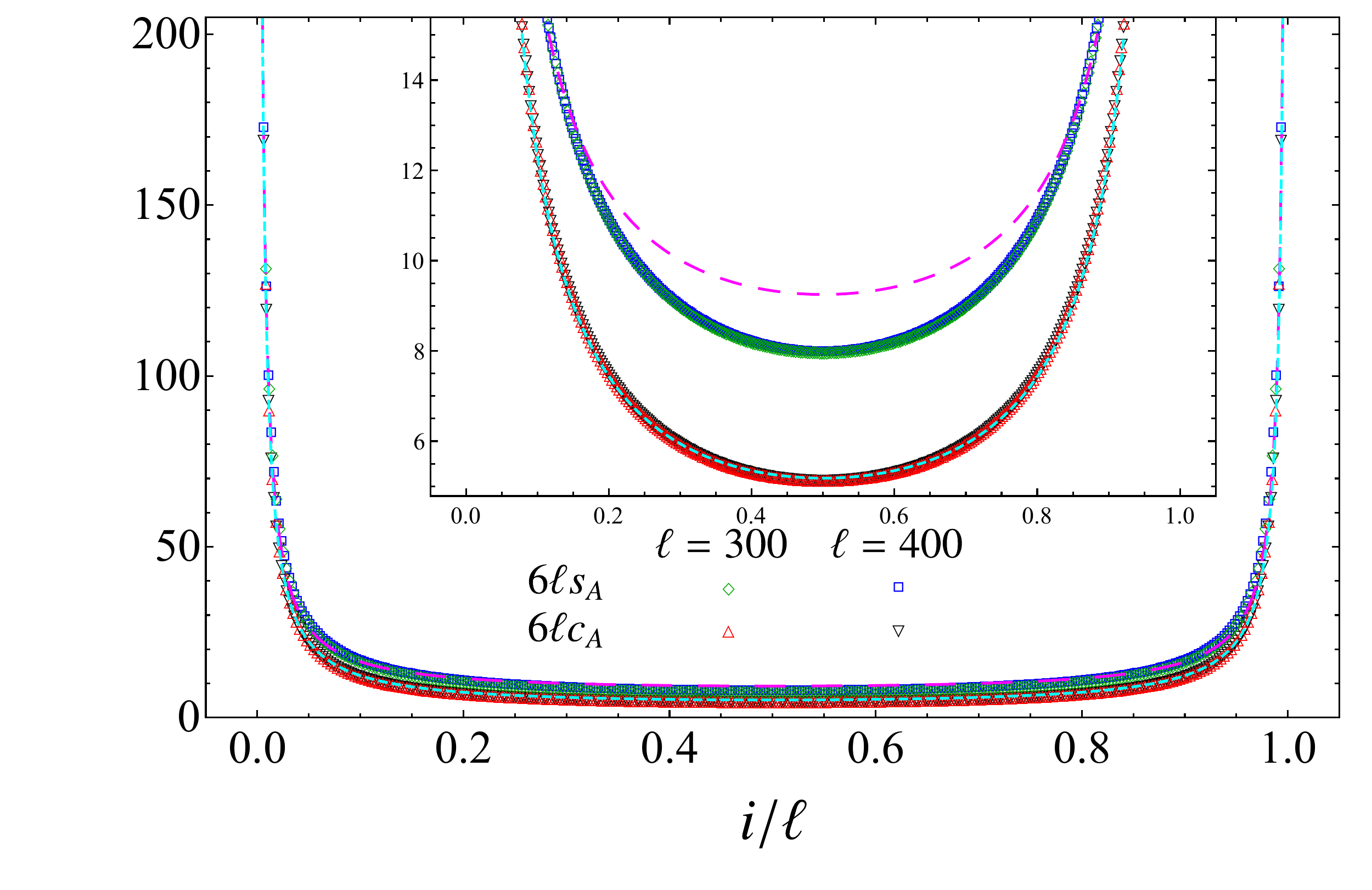}}
\subfigure
{\hspace{-.2cm}
\includegraphics[width=.57\textwidth]{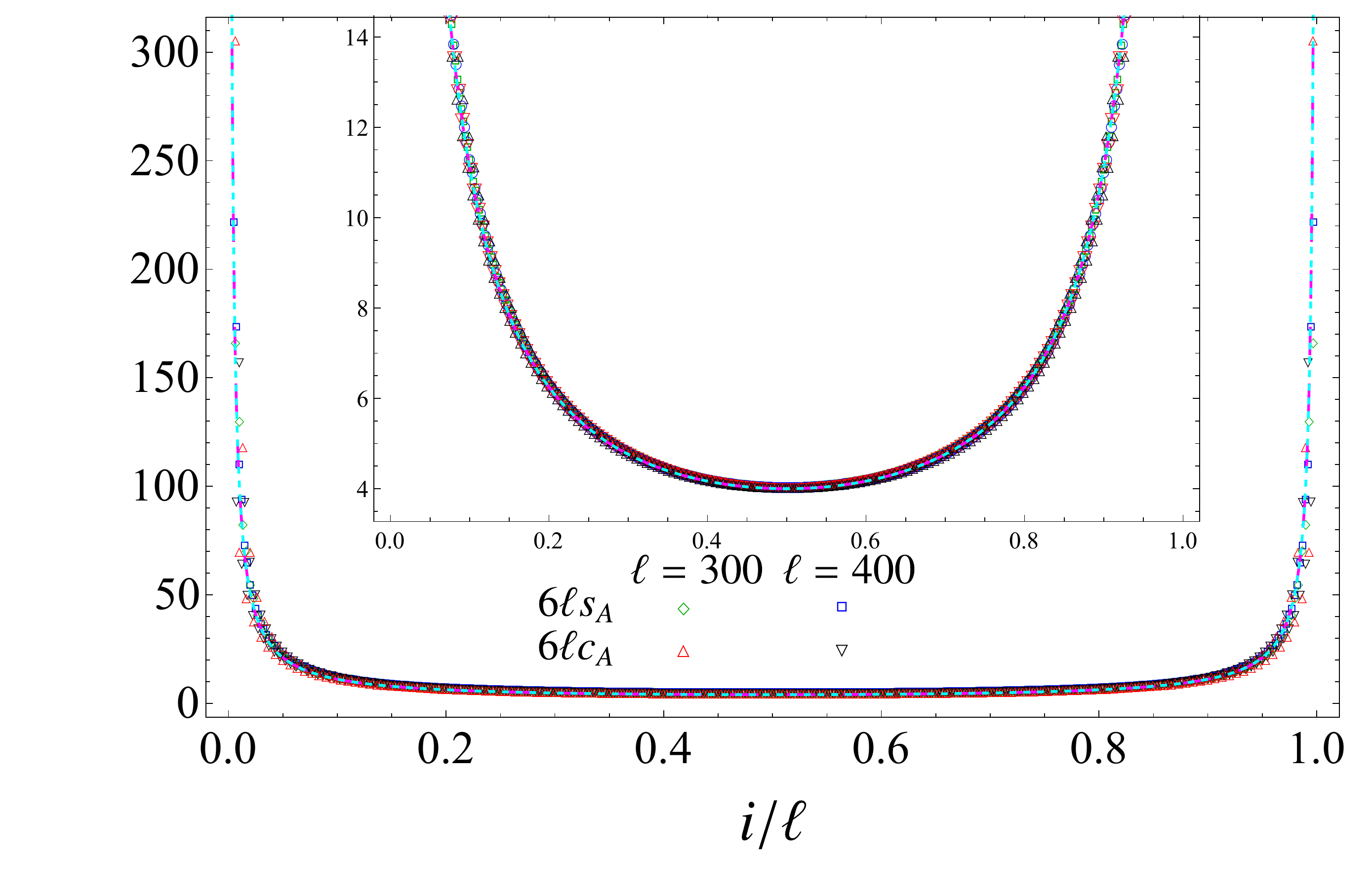}}
\subfigure
{
\hspace{-1.6cm}\includegraphics[width=.57\textwidth]{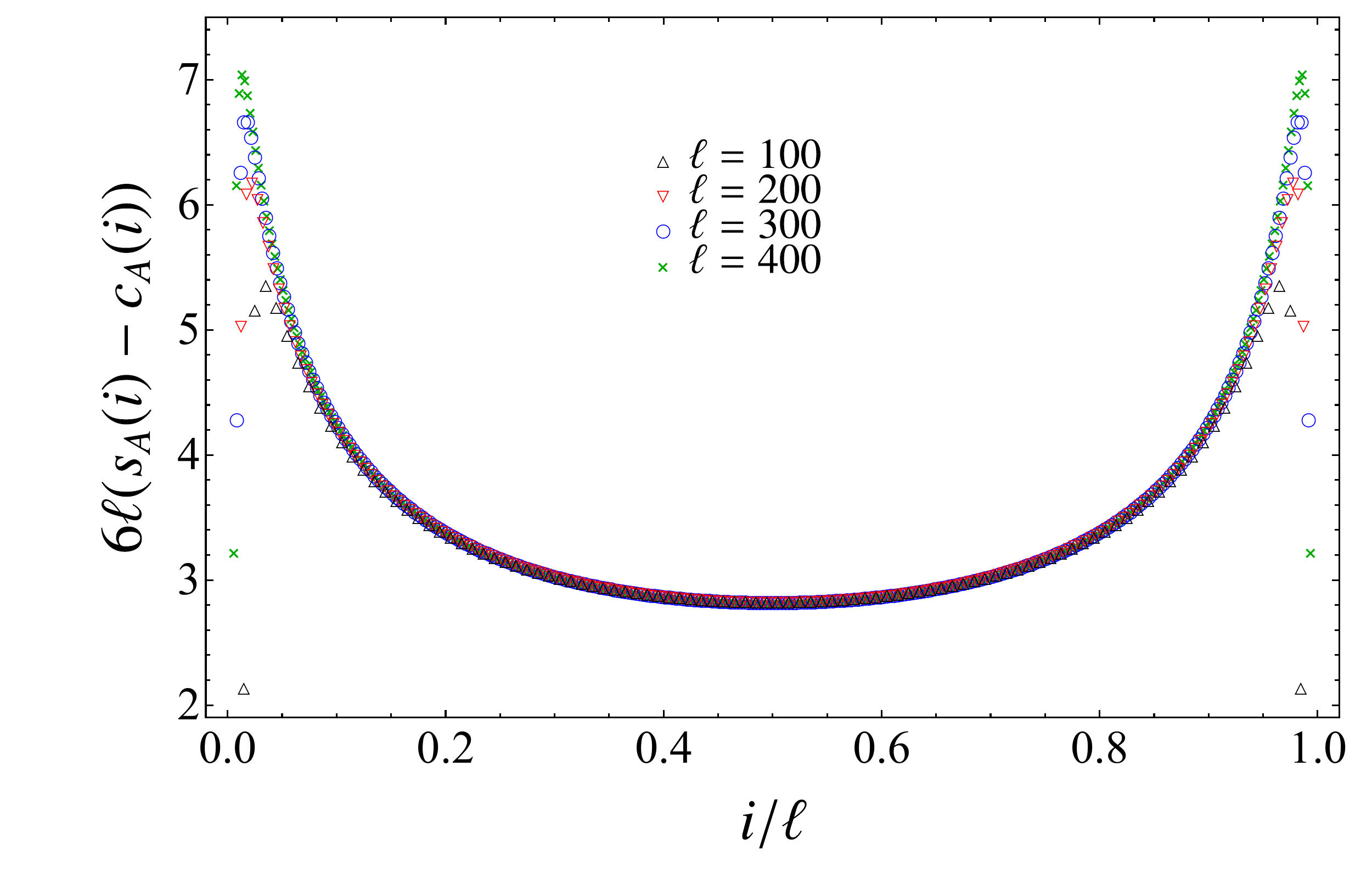}}
\subfigure
{\hspace{-.2cm}
\includegraphics[width=.57\textwidth]{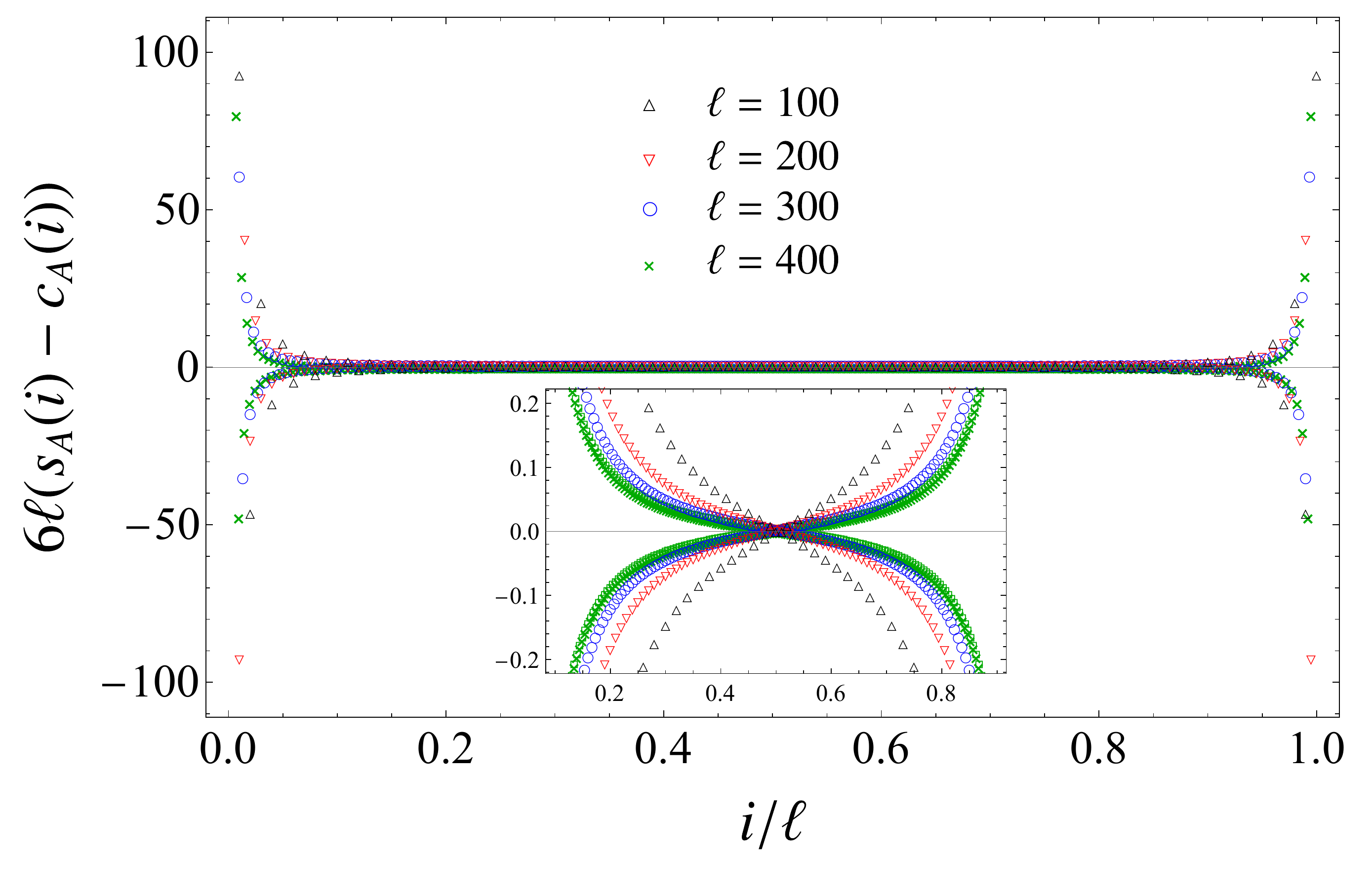}}
\vspace{-.2cm}
\caption{
Contour functions for the entanglement entropy and for the capacity of entanglement of a block made by $\ell$ sites in an infinite harmonic chain (left panels) and free fermionic chain (right panels) reported for various values of $\ell$. In the left panel we set $\omega\ell=10^{-10}$. The dashed lines in the top panels represent (\ref{cont-s-c inverval}) with $c=1$ and different additive constant. 
}
\vspace{.5cm}
\label{fig:contourCFT}
\end{figure}

As for the free lattice models that we are considering,  in the Appendix\;\ref{app-lattice} 
we construct the corresponding contour functions for the capacity of entanglement 
by adapting the constructions of the contour functions of the entanglement entropies
discussed in \cite{Chen_2014, Coser:2017dtb}.
The numerical results for these contour functions are displayed in Fig.\,\ref{fig:contourCFT},
both for the harmonic chain (left panels) and for the free fermionic chain (right panels),
where the dashed lines correspond to the curves obtained from CFT with $c=1$. For the dashed curves in the top left panel, the constants $c_1'$ and $[\partial^2_n(\log c_n)]\big|_{n=1}$ in (\ref{cont-s-c inverval}) have been fitted, while for the top right panel they have been set as predicted in \cite{Arias22Monotones}.
In the top panels we observe a nice agreement between the CFT curve and the lattice data points 
all over the interval for the fermionic case (right panel)
and only in the region close to the endpoints for the bosonic case (left panel).
The latter observation is made quantitative in the bottom left panel, where 
the difference between the contour functions 
of the entanglement entropy and of the capacity of entanglement is considered.
This quantity is non vanishing in the central part of the interval and the nice collapse
observed for its data points provides a curves that would be interesting to reproduce through a CFT analysis. 
In the fermionic case (bottom right panel), this quantity displays oscillations 
in the parity of the integer parameter labelling the sites,
whose amplitudes decrease with $\ell$.

We find it worth investigating also the temporal evolution of the contour function for the capacity of entanglement after a global quantum quench.

As for the contour function of the entanglement entropy after a global quantum quench,
by employing the quasi-particle picture, the following formula has been studied \cite{DiGiulio:2019lpb}
\be
\label{QPcontourguess}
s_A(x,t)
=
\frac{1}{2}\,\bigg[\,\int_{x<2|v_k|t<\ell} \!\! \tilde{s}(k)\, dk\,
+
\int_{\ell\,-\, x<2|v_k|t<\ell} \!\! \tilde{s}(k)\, dk\,
\bigg]
+\!
\int_{2|v_k|t>\ell} \!\! \tilde{s}(k)\, dk + f_0(x)\,,
\ee
where $x\in A$ and $A$ is a block of $\ell$ consecutive sites in an infinite chain. The function $\tilde{s}(k)$ has been introduced in (\ref{entropy_QPpict}), 
the velocity of the excitations with quasi-momentum $k$ is $v_k$ 
and $f_0(x)$ satisfying 
\be
\int _A  f_0(x) \,dx = S_A\big|_{t=0}\;,
\ee
must be added because
the quasi-particle picture does not take into account the value of the entanglement entropy at the initial time $t=0$.
When the post-quench evolution is determined by a CFT Hamiltonian, 
the following expression has been proposed \cite{DiGiulio:2019lpb}
\be
\label{func f0}
f_0(x)=\frac{\pi c}{3\tau_0} 
\left[\,
\frac{1}{\sinh(2\pi  x/\tau_0)}
+ 
\frac{1}{\sinh(2\pi  (\ell-x)/\tau_0)} 
\right] ,
\ee
where $\tau_0$ is the parameter introduced in Sec.\,\ref{subsec:CFTquench},
which is not known a priori from the lattice and it can be obtained by fitting 
the linear growth of the entanglement entropy 
(see Fig.\,\ref{fig:HC_CEvsEEvsQPformulaforEE} and Fig.\,\ref{fig:Cvstfermions} for the quench in the two models).

By adapting (\ref{QPcontourguess}), 
it is natural to write the following expression for 
the contour function of the capacity of entanglement 
\be
\label{QPcontourguess_capacity}
c_A(x,t)
=
\frac{1}{2}\,\bigg[\,\int_{x<2|v_k|t<\ell} \!\! \tilde{c}(k)\, dk\,
+
\int_{\ell\,-\, x<2|v_k|t<\ell} \!\! \tilde{c}(k)\, dk\,
\bigg]
+\!
\int_{2|v_k|t>\ell} \!\! \tilde{c}(k)\, dk + \tilde{f}_0(x)\,,
\ee
where $\tilde{c}(k)$ has been introduced in (\ref{capacity_QPpict})
and $\tilde{f}_0(x)$ satisfies
\be
\int_A  \tilde{f}_0(x) \,dx = C_A\big|_{t=0}\;.
\ee
By adapting the derivation of (\ref{func f0}) to the case of the capacity of entanglement
(as done e.g. for (\ref{contour_ent_cap_CFT})),
one expects $\tilde{f}_0(x)=f_0(x)+C$, 
where $C$ is non universal constant.

\begin{figure}[t!]
\vspace{-.2cm}
\hspace{-1.1cm}
\centering
\includegraphics[width=1.05\textwidth]{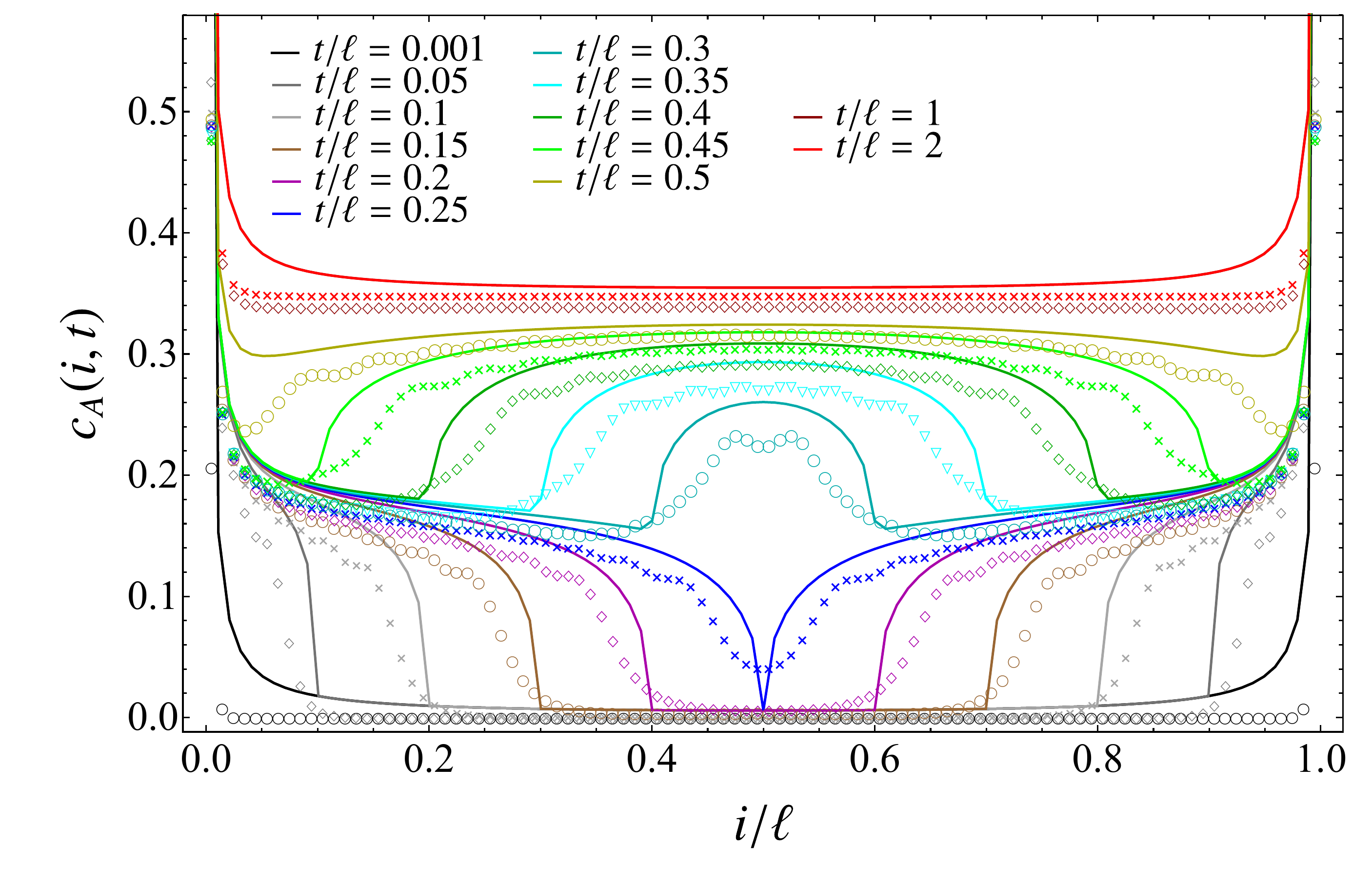}
\vspace{-.4cm}
\caption{
Temporal evolution of the contour of the capacity of entanglement of an interval made by $\ell=100$ sites 
after the mass quench in the infinite harmonic chain. 
In the quench considered $\omega_0=1$ and $\omega=0$. 
The quasi-particle formula provides the solid lines in all the panels. 
}
\vspace{.5cm}
\label{fig:contour_quenchHC}
\end{figure}

\begin{figure}[t!]
\vspace{.2cm}
\hspace{2.7cm}
 \subfigure
 {
\includegraphics[width=.57\textwidth]{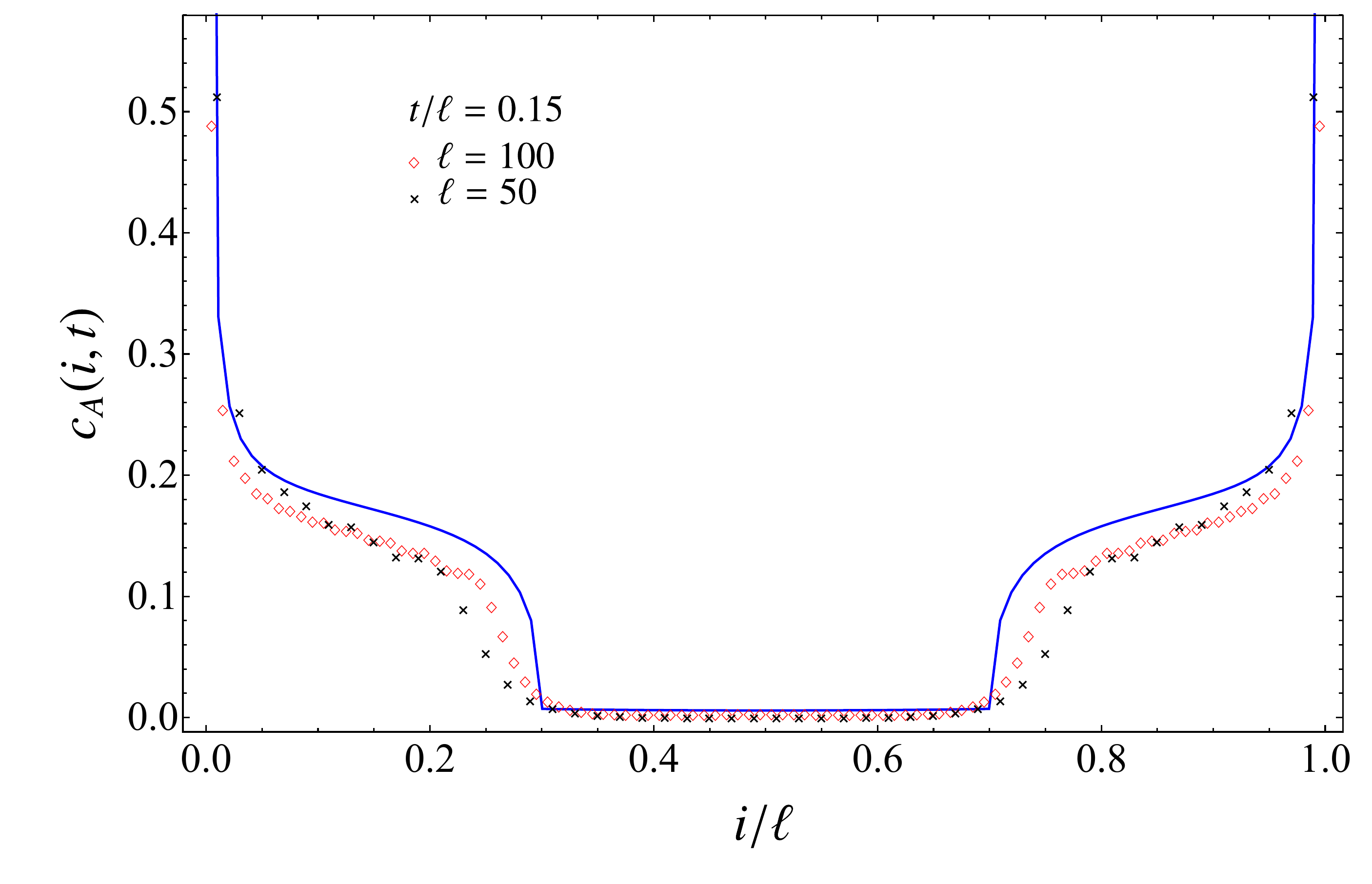}
 }
 \\
\subfigure
{
 \hspace{-1.4cm}
 \includegraphics[width=.57\textwidth]{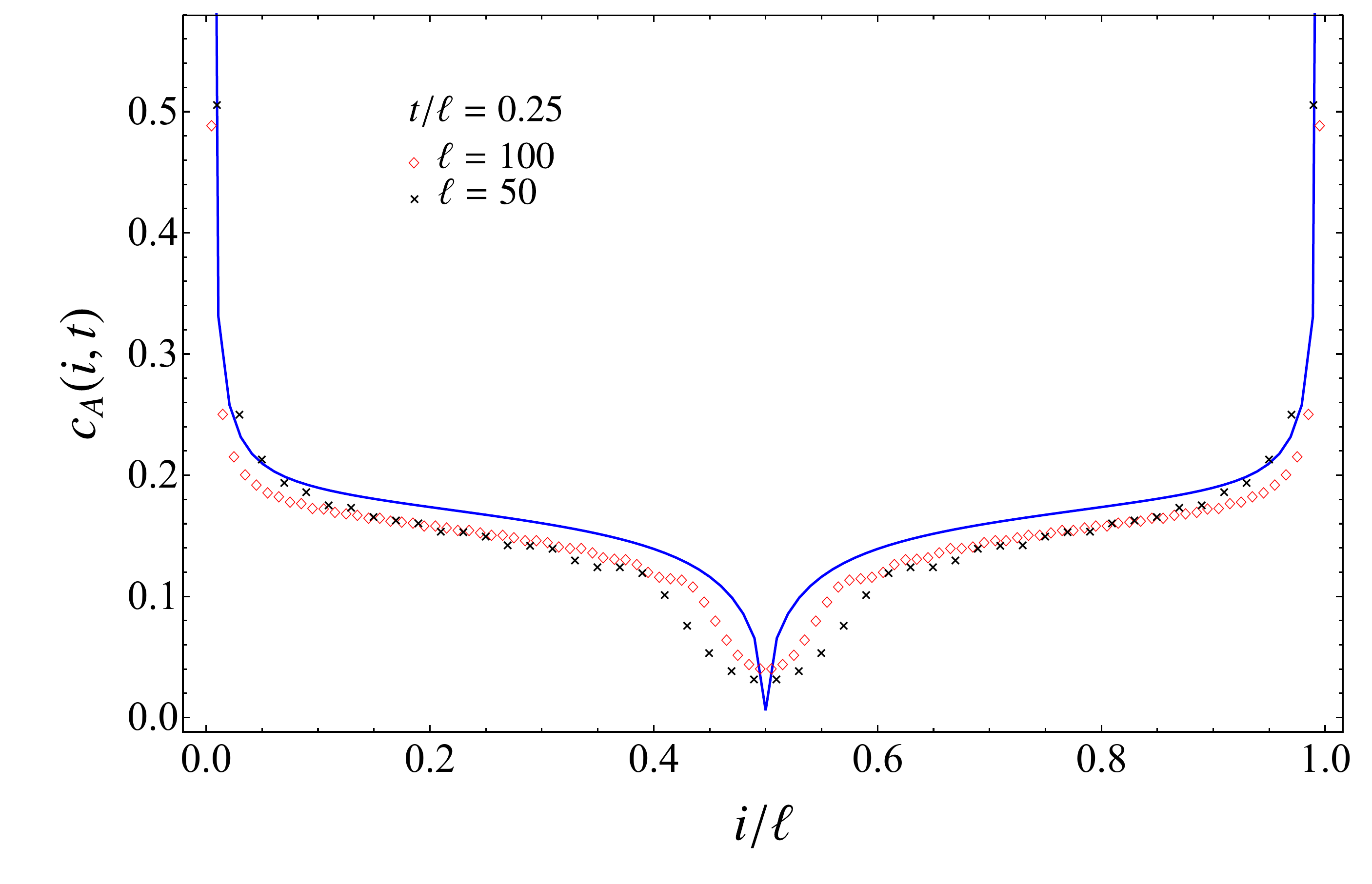}
}
\subfigure{
\hspace{-.7cm}
\includegraphics[width=.57\textwidth]{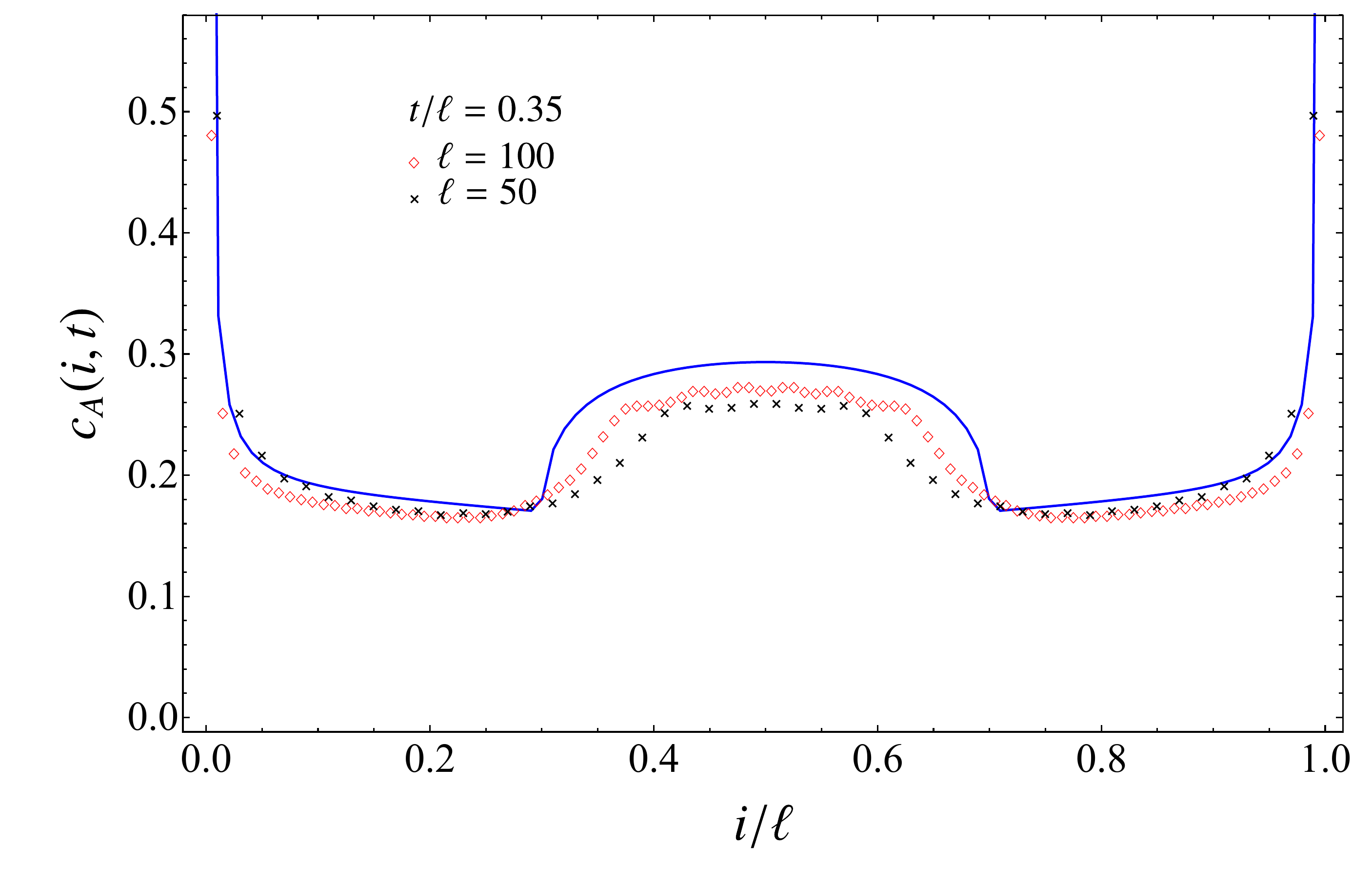}
}
\vspace{-.5cm}
\caption{
Temporal evolution of the contour of the capacity of entanglement of an interval made by $\ell$ sites 
after the mass quench in the infinite harmonic chain. 
In the quench considered $\omega_0=1$ and $\omega=0$. 
The quasi-particle formula provides the solid lines in all the panels. 
}
\vspace{-.2cm}
\label{fig:contour_quenchHC_differentell}
\end{figure}

For the global quench of the free fermionic chain described in Sec.\,\ref{subsec:quenchFF},
we can explore the contour function of the capacity of entanglement in the asymptotic regime $t\to\infty$
by adapting the results of \cite{ep-07-local-quench,DiGiulio:2019lpb}.
This leads to 
\be
\label{contours large time}
s_A(i)=\frac{2}{\ell +1}\sum_{k=1}^\ell s(\zeta_k) \big[\sin(i\theta_k)\big]^2\,,
\;\;\qquad\;\;
c_A(i)=\frac{2}{\ell +1}\sum_{k=1}^\ell c(\zeta_k) \big[\sin(i\theta_k)\big]^2\,,
\ee
where $\ell$ is the number of consecutive sites in $A$,
the functions $s(y)$ and $c(y)$ are defined in (\ref{sx def fermions}) and (\ref{cx def fermions}) respectively and
\be
\label{spes fermions large time}
\theta_k=\frac{\pi k}{\ell+1}\,,
\;\;\qquad\;\;
\zeta_k=\frac{1+\cos\theta_k}{2}\,.
\ee
It is worth taking the limit $\ell\to\infty$ of (\ref{contours large time})
because it allows to capture the behaviour of $s_A(i)$ and $c_A(i)$ close to one of the endpoints. 
This limit can be studied by substituting the sums over $k$ with an integral
(i.e. replacing $\frac{1}{\ell +1}\sum_{k=1}^\ell f(\theta_k)\to\int_{0}^\pi \frac{d\theta}{\pi}f(\theta)$ for any given function $f$).
In \cite{DiGiulio:2019lpb} 
it has been found that $s_A(i)$ in (\ref{contours large time}) becomes
\be
s_A(i)=\log 4 -1 +\frac{1}{2i(4i^2-1)}\,.
\ee
From (\ref{spes fermions large time}) and (\ref{cx def fermions}), for the limit $\ell \to \infty$ of $c_A(i)$ in (\ref{contours large time}) we obtain
\be
c_A(i)=\frac{\pi^2}{8} - 1 - 
\frac{1}{4\pi}
\int_{0}^\pi \!
\big[\log\!\big(\tan(\theta/2)\big)\big]^2 \, \big[2\cos(2 i \theta)-\cos(2 (i+1) \theta)-\cos(2 (i-1) \theta)\big]\,
d\theta \,.
\ee

\begin{figure}[htbp!]
\vspace{-.2cm}
\hspace{-1.1cm}
\centering
\includegraphics[width=1.05\textwidth]{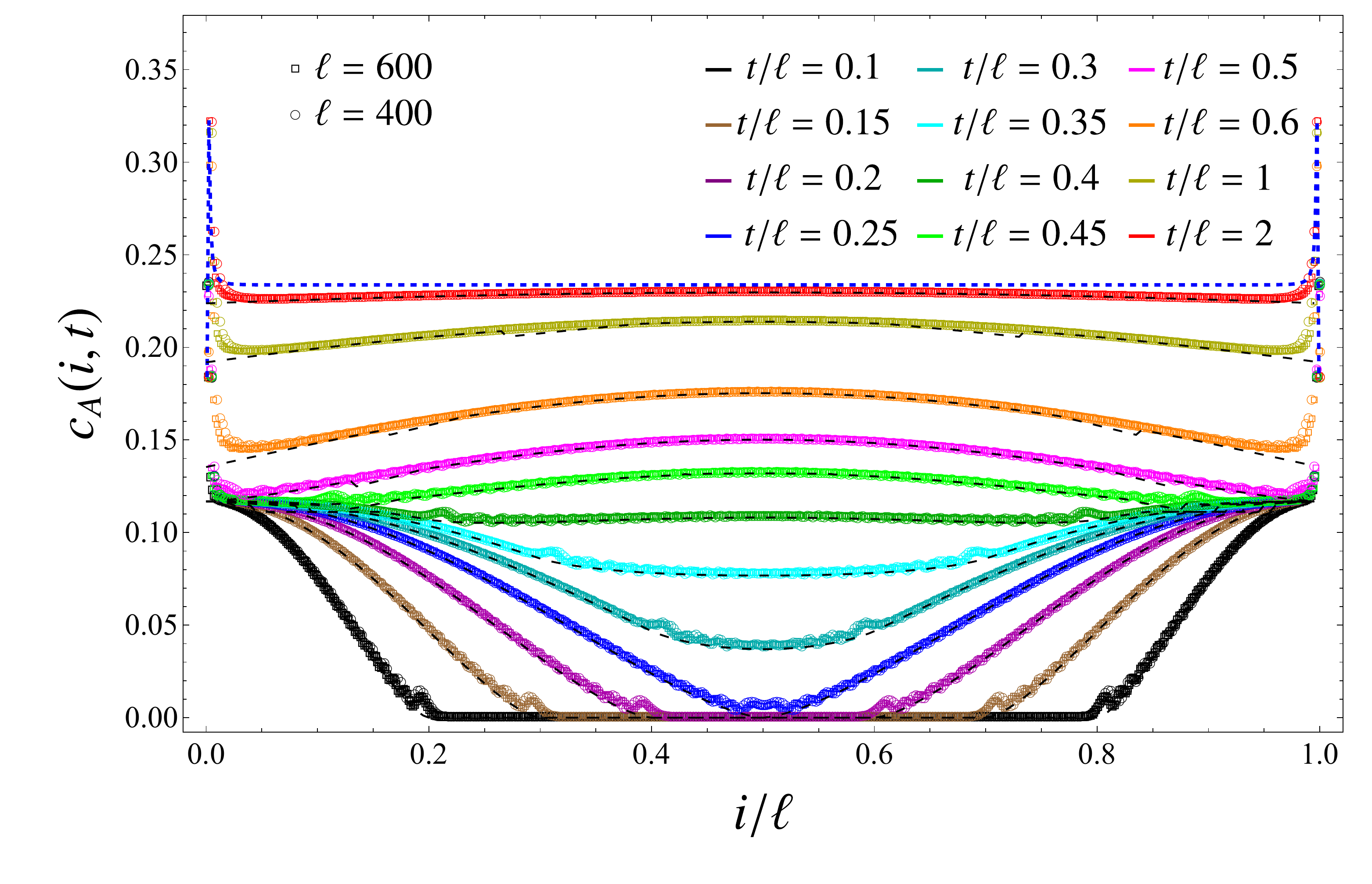}
\\
\hspace{-1.1cm}
\centering
\includegraphics[width=1.05\textwidth]{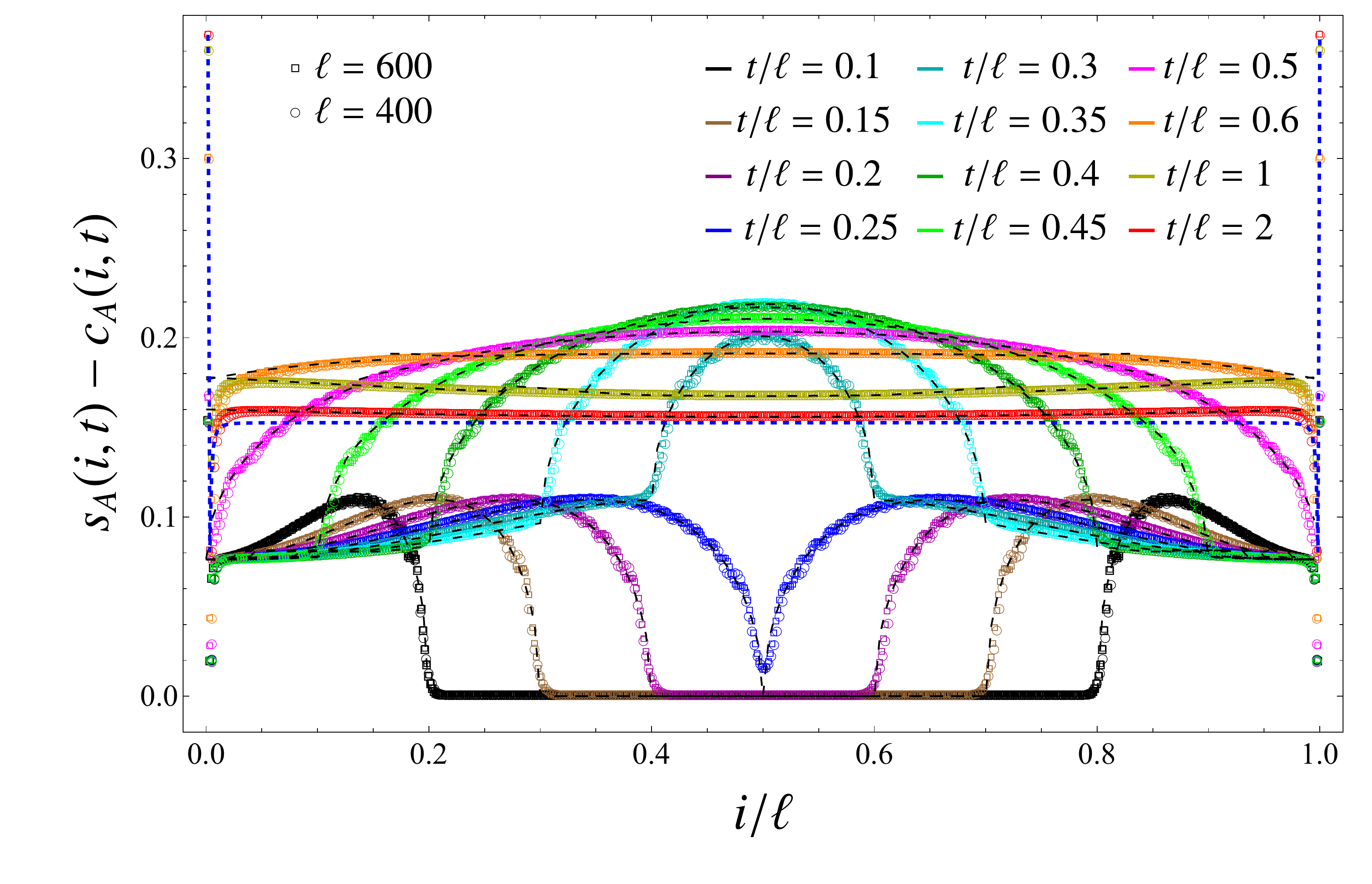}
\vspace{-.4cm}
\caption{
Temporal evolution of the contour function of the capacity of entanglement $c_A(i,t)$ (top panel) 
and of $s_A(i,t) - c_A(i,t)$
after the global quench in the free fermionic chain described in Sec.\,\ref{subsec:quenchFF}.
}
\vspace{.5cm}
\label{fig:contour_quenchFF}
\end{figure}

In Fig.\,\ref{fig:contour_quenchHC} and Fig.\,\ref{fig:contour_quenchHC_differentell} we show the numerical results for the temporal evolution of the contour function of the capacity of entanglement 
after the global quantum quench in the harmonic chain described in Sec.\,\ref{sec:quenchHC} and for a block made by $\ell=100$ sites,
obtained through the procedure discussed in the Appendix \ref{app-lattice}.
These numerical results are compared with the formula obtained from (\ref{QPcontourguess_capacity}), 
(\ref{h-c-function_HC}), (\ref{nk}), (\ref{vk}) and (\ref{func f0}),
which is represented by solid lines.
The function $\tilde{f}_0$ is (\ref{func f0}) in all the panels of Fig.\,\ref{fig:contour_quenchHC} and Fig.\,\ref{fig:contour_quenchHC_differentell}.
In Fig.\,\ref{fig:contour_quenchHC} we show the data corresponding to $\ell=100$.
We remark that $c_A(i,t)$ displays
the same qualitative behaviour observed for $s_A(i,t)$ in \cite{Chen_2014,DiGiulio:2019lpb}:
two fronts start from the endpoints of the interval moving at the same velocity but in opposite directions 
towards the center of the interval, where they meet and then superpose. 
In Fig.\,\ref{fig:contour_quenchHC_differentell} we report the numerical data for two values of $\ell$.
The corresponding curves do not collapse on the quasi-particle formula, 
like for the quench of the fermionic chain (see Fig.\,\ref{fig:contour_quenchFF}), where we have access to large enough values of $\ell$.

In Fig.\,\ref{fig:contour_quenchFF} we show the temporal evolution of the contour of the capacity of entanglement 
after the global quench in the free fermionic chain discussed in Sec.\;\ref{subsec:quenchFF}.
The numerical results are obtained as discussed in Appendix \ref{app-lattice} and correspond to two values of $\ell$. 
A remarkable agreement is observed with the formula 
computed from (\ref{QPcontourguess_capacity}), (\ref{h-c-function_HC}), (\ref{occupation Fermi GGE}) 
and (\ref{velocity Fermi GGE}) with $\tilde{f}_0=0$, which is represented by the black dashed lines.
The qualitative behaviour described above for the global quench in harmonic chains (see Fig.\,\ref{fig:contour_quenchHC})
is observed also for the temporal evolution of $c_A(i,t)$ in this fermionic quench. 
The curves corresponding to $s_A(i,t)$ for this quench have been reported in Fig.\,21 of \cite{DiGiulio:2019lpb}.
We find it instructive comparing these two quantities.  
For the contour of the entanglement entropy, the CFT analysis in \cite{DiGiulio:2019lpb} suggests that the fronts are almost vertical when the velocity of all the excitations is equal to 1 (see Fig.\,1 in \cite{DiGiulio:2019lpb}), while the analysis in \cite{Chen_2014,DiGiulio:2019lpb} shows that in the lattice models where the velocity distribution is non-trivial the fronts become less steep. Such a steepness decreases as we have less quasi-particles moving with velocity close to 1. 
This argument holds also for the contour for the capacity of entanglement.
 Thus, from the right panel of Fig.\,\ref{fig:densities_quench}, 
 we expect that the fronts in the temporal evolution of the contour for the entanglement entropy 
 are steeper than the ones of the contour for the capacity of entanglement. 
 This expectation is confirmed when the top panel of Fig.\,\ref{fig:contour_quenchFF} is compared with Fig.\,21 of \cite{DiGiulio:2019lpb}.

The top panel of Fig.\,\ref{fig:contour_quenchFF} shows also that,
for large values of $t/\ell$, 
the capacity saturates to a constant value and, correspondingly, the contour converges towards the limiting curve giving by (\ref{contours large time}) (blue dashed line).
When the post-quench evolution is determined by a CFT Hamiltonian, $v_k=1$ for any $k$ the saturation occurs exactly at $t/\ell=1/2$.
If the velocities $v_k$ are distributed in a non trivial way, the transition from the linear growth to the saturation regime occurs in a smoother way. 
From  top panel of Fig.\,\ref{fig:contour_quenchFF} we observe that for $t/\ell\simeq 1$ the saturation has not reached yet. 
Comparing this behaviour with the one of the contour for the entanglement entropy shown in Fig.\,21 of \cite{DiGiulio:2019lpb},
we notice that $s_A(i,t)$ reaches the asymptotic curve earlier than $c_A(i,t)$.
This can be explained through the considerations reported in the discussion of the right panel of Fig.\,\ref{fig:densities_quench}.

In the bottom panel of Fig.\,\ref{fig:contour_quenchFF} 
the temporal evolution of $s_A(i,t) - c_A(i,t)$ is reported.
These curves deserve further investigations.

\section{Symmetry-resolved capacity of entanglement}
\label{sec:SymResCapacity}

In this section we introduce the symmetry resolution for the 
the capacity of entanglement and for the $n$-th moments of shifted modular Hamiltonian
by adapting the analysis discussed in \cite{Goldstein:2017bua, Xavier:2018kqb} for the entanglement entropy.


Consider a system endowed with a $U(1)$ global symmetry generated by a charge $Q$ and a spatial bipartition $A \cup B$.
When the whole system is in an eigenstate $|\Omega\rangle$ of $Q$, 
its density matrix $\rho=|\Omega\rangle\langle\Omega|$ satisfies $[\rho,Q]=0$. 
Assume that $Q=Q_A\oplus Q_B$, where $Q_A$ and $Q_B$ 
denote the restriction of the charge operator to $A$ and $B$ respectively. 
Taking the trace of $[\rho,Q]=0$ over $B$, one obtains $[\rho_A,Q_A]=0$, 
which implies the following block diagonal structure for $\rho_A$
\be
\label{eq:decompositionrho}
\rho_A=\bigoplus_q p(q)\, \rho_A(q)\,,
\ee
where each block $\rho_A(q)$ corresponds to an eigenvalue $q$ of $Q_A$. 
The quantity $p(q)$ is the probability of finding $q$ in a measurement of $Q_A$ in the reduced density matrix $\rho_A$;
hence $p(q)=\mathrm{Tr}\Pi_q\rho_A$, 
where $\Pi_q$ is the projector onto the eigenspace of $Q_A$ with eigenvalue $q$.
The normalisation of each block implies that $\mathrm{Tr}\rho_A(q)=1$.
Since this normalisation condition holds, 
it is natural to define the so called symmetry-resolved entanglement entropies, 
i.e. the analogous of the entanglement entropies for each block 
\be
\label{SREEs}
S_A^{(n)}(q)=\frac{1}{1-n}\log\textrm{Tr}
\big[
\rho_A(q)^n
\big]\,,
\;\;\qquad\;\;
S_A(q)=
-\,\textrm{Tr}\big[\rho_A(q)\log\rho_A(q)\big]\,.
\ee
We find it natural to consider 
\be
\label{SRCapacity}
C_A(q)
=
\mathrm{Tr} \Big( \rho_A(q)\big[\log\rho_A(q)\big]^2\Big)
-
\Big[
\textrm{Tr}\big(\rho_A(q)\log\rho_A(q)\big)\Big]^2
=\,
\partial_n^2
\left[(1-n)S_A^{(n)}(q)
\right] \! \Big|_{n=1}\,,
\ee
and 
\bea
\label{SR-M_n}
M^{(n)}_A(q;b_n)
&=&
\mathrm{Tr}\Big[ \,
\rho_A(q) \big(
-\log \rho_A(q)  + b_n\big)^n
\Big]
-
b_n^n 
\\
\rule{0pt}{.7cm}
&=& 
e^{b_n}(-1)^n  \frac{d^n}{d\alpha^n} 
\Big[\exp\Big\{\!-\alpha \,b + (1-\alpha)S_A^{(\alpha)} (q)\Big\}\Big]\Big|_{\alpha =1, b=b_n}
\! - b_n^n\,,
\nonumber
\eea
that can be interpreted respectively as the 
symmetry-resolved capacity of entanglement 
and the symmetry-resolved moments of shifted modular Hamiltonian.
In (\ref{SR-M_n}) we find it convenient to highlight the dependence on the constant $b_n$.

The expression (\ref{SR-M_n}) reduces to the $S_A(q)$ in (\ref{SREEs}) when $n=1$.
Instead, 
when $n=2$ and $b_n=1$, we have that  (\ref{SR-M_n})
provides the symmetry-resolved version of (\ref{Mdefinition}) up to an additive constant.
%
From (\ref{SRCapacity}) and (\ref{SR-M_n}), it is straightforward to realise that $S_A^{(n)}(q)$ allows to compute also $C_A(q)$ and $M_A^{(n)}(q;b_n)$.


Remarkably, the entanglement entropy can be decomposed as a sum of the contributions from the different symmetry sectors.
Indeed,  from (\ref{eq:decompositionrho}) in (\ref{defentropy}) 
and the fact that the trace of a block diagonal matrix  is the sum of the traces of each block, one finds 
\bea
S_A
&=&
- \sum_q
\mathrm{Tr}\Big[ 
p(q)\,\rho_A(q)\,\log\! \big( p(q)\rho_A(q)\big)
\Big]
\\
&=&
\sum_q
\bigg\{\! -\mathrm{Tr}\Big[ p(q)\,\rho_A(q)\, \log \rho_A(q)\Big]
-
\mathrm{Tr}\Big[ p(q)\,\rho_A(q)\, \log p(q)\Big]
\bigg\}
\\
&=&
\label{decomp_SA}
\sum_q p(q) \, S_A(q)-\sum_q p(q)\log p(q)
\,\equiv\,
\sum_q p(q)\, S_A(q)+\sum_q h(q)\,,
\eea
where 
\be 
\label{classicalShannon}
h(q)\, \equiv\, 
-\, p(q)\, \log p(q)
\ee
is the Shannon entropy associated to the probability distribution $p(q)$ and we have also exploited that $\textrm{Tr}\rho_A(q)=1$.
The first and the second terms in (\ref{decomp_SA}),
which have been called respectively configurational and number entanglement entropies \cite{Lukin-19},
respectively quantify the entanglement within symmetry sectors and fluctuations thereof.
The configurational and the number entanglement entropies have been measured in experiments involving 
a system of interacting bosons with disorder \cite{Lukin-19}. 
This fact motivates to study these two quantities and to look for their possible extensions. 

The analogue of (\ref{decomp_SA}) for the R\'enyi entropies cannot be written 
because $S_A^{(n)}$ does not have the form $\mathrm{Tr}\left[f(\rho_A)\right]$, for some function $f$. 
%
Since $C_A$ defined in (\ref{defcapacity}) contains $S_A^2$, where $S_A$ is written as in (\ref{decomp_SA}),
we conclude that the capacity of entanglement cannot be written as a sum over the charge sectors. 
Instead, this can be done for the moments of the shifted modular Hamiltonian,
which are written as traces of specific functions of the reduced density matrix. 

Plugging the decomposition (\ref{eq:decompositionrho}) into the definition (\ref{Mn Tr Fn}) of the total $M_A^{(n)}$, we obtain
\be
\label{MAn_decomposition1}
M_A^{(n)}=\sum_q
 \textrm{Tr}\,
 \Big\{\,
 p(q) \,\rho_A(q)\, \big[\! -\log \left(p(q)\rho_A(q)\right)  + b_n\,\big]^n
\Big\}
 -b_n^n\,.
\ee
When $n=2$, from (\ref{MAn_decomposition1}) we have
\bea
M^{(2)}_A
&=&
\sum_q 
\mathrm{Tr}\Big[
p(q)\,\rho_A(q)
\Big(\!
\log [p(q)\rho_A(q)]
-
2 b_2
\Big)
\log [p(q)\rho_A(q)]
\Big]
\\
\rule{0pt}{.55cm}
&=&
\sum_q 
\Big\{
p(q)\,
\mathrm{Tr}\Big[
\rho_A(q) 
\Big(\!\log \rho_A(q)-2 b_2\Big)
\log \rho_A(q)
\Big]
\\
& & \hspace{.9cm}
+\,
2\, p(q)\log p(q) \,
\mathrm{Tr}\big[
\rho_A(q)\log \rho_A(q)
\big]
+
 p(q)\big(\log p(q)\big)^2-
 2 \,b_2 \,p(q)\,\log p(q)
 \Big\}
 \nonumber
\\
\rule{0pt}{.7cm}
&=&
\sum_q 
\Big(
p(q)\, M^{(2)}_A(q;b_2)
+2 \,h(q)S_A(q)
+m^{(2)}(q;b_2)
\Big)\,,
\label{sectordecompos MA}
\eea
where
\be
\label{mq_genericn}
m^{(n)}(q;a) \,=\,
 p(q)\,
\Big\{ \big[a-\log p(q)\big]^n-a^n\Big\}\,.
\ee
Notice that, differently from (\ref{decomp_SA}), in (\ref{sectordecompos MA})
also the product between the symmetry-resolved entanglement entropy and its classical counterpart (\ref{classicalShannon}) occurs,
which does not depend on free parameter $b_2$.

When $n=3$, we find 
\bea
M_A^{(3)}
&=&
\textrm{Tr}\,
\Big\{
\rho_A
\Big[\!
-\left(\log \rho_A\right)^3+3\,b_3\left(\log \rho_A\right)^2-3\,b_3^2\log \rho_A
\Big]
\Big\}
\\
\rule{0pt}{.7cm}
&=&
\sum_q
\textrm{Tr}\,
\Big\{
\rho_A(q)\, p(q) \big(\log \rho_A(q)+\log p(q)\big)
\\
& & \hspace{1.4cm}
\times
\Big[
-\big(\log \rho_A(q)+\log p(q)\big)^2
+\,3b_3\big(\log \rho_A(q)+\log p(q)\big)
- 3b_3^2
\,\Big]
\Big\}
\nonumber
\\
\rule{0pt}{.7cm}
&&
\label{MA3_decomposition}
\hspace{-1.7cm}
=
\sum_q
\Big\{
p(q) \,M_A^{(3)}(q;b_3)
+3 m^{(2)}(q;b_3/2)\; S_A(q)
+3 h(q)\, M_A^{(2)}(q;b_3/2)
+m^{(3)}(q;b_3)
\Big\}\,,
\phantom{xxxxx}
\eea
where $m^{(n)}(q;a)$ is given by (\ref{mq_genericn}). 
Also in (\ref{MA3_decomposition}) some terms involve the symmetry-resolved  moments of order lower than $n=3$, but now they do depend on $b_3$.
We guess that the generalisation of (\ref{MAn_decomposition1}) to a generic value of $n$ reads
\be
\label{MAn_decomposition2}
M_A^{(n)}
=
\sum_q \bigg\{
p(q)\,M_A^{(n)}(q;b_n)+m^{(n)}(q;b_n)
+
\sum_{k=1}^{n-1}
\bigg[
\binom{n}{k}\,
m^{(k)}(q;b_n/2)
\;M_A^{(n-k)}(q;b_n/2)
\bigg]
\bigg\}\,,
\ee
where $m^{(n)}(q;a)$ is defined in (\ref{mq_genericn}).
It would be useful to provide a proof for (\ref{MAn_decomposition2}),
which has been checked for the first 700 positive integer values of $n$.

The functions $m^{(n)}(q;a)$ in (\ref{mq_genericn}) can be found by replacing $\rho_A(q)$ with $p(q)$ into the definition of  $M_A^{(n)}(q;a)$ given in (\ref{SR-M_n}).
Since $m^{(1)}(q;a)=h(q)$ in (\ref{classicalShannon}), we have that (\ref{mq_genericn}) provides a generalisation of the Shannon entropy $h(q)$ to higher values of $n$.
For a given $n$, the relation (\ref{MAn_decomposition2}) tells us that
$M_A^{(n)}$ can be written in terms of the symmetry-resolved moments $M_A^{(k)}(q;a)$ with $k\leqslant n$ and their classical counterparts. 
Notice that (\ref{MAn_decomposition2}) becomes (\ref{decomp_SA}) when $n=1$.


As mentioned in Sec.\,\ref{sec:intro}, $b_n \geqslant n-1$
in order for $M_A^{(n)}$ to be concave and therefore provide an entanglement monotone.
The decomposition (\ref{MAn_decomposition2}) can be employed to find additional constraints on $b_n$
by imposing that all the quantities involved in the decomposition are entanglement monotones.
This holds when $b_n\geq n-1$ and  $b_n/2\geq n-2$ are satisfied.
For  $n=1$, the relation (\ref{MAn_decomposition2}) is independent of $b_1$
and therefore it is not useful in this analysis. 
When $n=2$ and $n=3$, we have that $n-1\geqslant 2n-4$; hence we do not obtain constraints stronger than $b_n\geqslant n-1$, 
already considered in Sec.\,\ref{sec:intro}. 
Instead, when $n>3$, we have $2n-4> n-1$ and therefore concavity condition for (\ref{MAn_decomposition2}) gives the stronger constraint $b_n\geqslant 2n-4$.


In the remaining part of this section 
we explore the quantities defined above for the 
Luttinger liquid CFT with parameter $K$, which is equivalent to the free compact boson CFT with radius $R\propto 1/\sqrt{K}$.
For this model, the $U(1)$ conserved charge is the electric charge; hence $q$ is an integer number.
When $A$ is an interval with length $\ell$ and the entire system is in its ground state, 
the symmetry-resolved entanglement entropies have been computed in \cite{Goldstein:2017bua,Xavier:2018kqb}, finding 
\bea
\label{SREE_Luttinger}
S_A^{(n)}(q)
&=&
\frac{1}{6}\bigg(\frac{n+1}{n}\bigg)\log(\ell /\epsilon)
-
\frac{1}{2}\log\!\bigg(
\frac{2 K}{\pi}\, \log(\ell /\epsilon)
\bigg)+y_n+\dots\,,
\\
\rule{0pt}{.7cm}
S_A(q)&=&\frac{1}{3}\, \log(\ell /\epsilon)
-
\frac{1}{2}\log\!\bigg(
\frac{2 K}{\pi}\, \log(\ell /\epsilon)
\bigg)+y_S+\dots\,,
\label{SRvN_Luttinger}
\eea
where $y_n$ is a non-universal additive constant and the dots denote subleading terms as $\epsilon\to 0$ and $y_S\equiv y_1$. 
The entanglement equipartition observed in \cite{Xavier:2018kqb} corresponds to the fact that
the leading terms of $S_A^{(n)}(q)$ and $S_A(q)$ are independent of $q$.

From (\ref{SREE_Luttinger}) and (\ref{SRCapacity}) we can compute $C_A(q)$, finding 
\be
\label{SRCapacity_Luttinger}
C_A(q)= \frac{1}{3}\, \log(\ell /\epsilon)
+y_C+\dots\,,
\ee
where $y_C=\partial^2_n [(1-n)y_n ]|_{n=1}$.
Comparing (\ref{SRvN_Luttinger}) and (\ref{SRCapacity_Luttinger}), 
we have that $S_A(q)=C_A(q)$ at leading order,
but a subleading term of order $\log\log(\ell/\epsilon)$ breaks this equality.  
Moreover, since  the parameter $q$ occurs only in the subleading terms of the expansion for $\epsilon\to 0$ of  the symmetry-resolved capacity of entanglement,
it also displays equipartition in the sense of \cite{Xavier:2018kqb}.

As for the symmetry-resolved moments of shifted modular Hamiltonian, by plugging (\ref{SREE_Luttinger}) into (\ref{SR-M_n}) it is straightforward to obtain
\be
\label{SR_Mn Luttinger}
M^{(n)}_A(q;b_n)
\,=\,
\left(
\frac{\log (\ell / \epsilon)}{3} 
\right)^n
-
\frac{n}{2}
\left(
\frac{\log (\ell / \epsilon)}{3} 
\right)^{n-1}
\log\!\bigg(
\frac{2 K}{\pi}\, \log (\ell / \epsilon)
\bigg)
+
O\!\left[\big( \log (\ell / \epsilon) \big)^{n-1}\right] \,,
\ee
where the subleading terms depend on non-universal constants (like e.g. $\partial_\alpha^k(y_\alpha)|_{\alpha=1}$ for $k\leqslant n$) and on $b_n$.
The first two leading terms in (\ref{SR_Mn Luttinger}) are independent of the charge $q$, 
but a non trivial dependence on $q$ occurs in the subleading terms that we have neglected. 
For instance, by considering the simplest case given by $n=2$,
from (\ref{SREE_Luttinger}) and (\ref{SR-M_n}) for $n=2$, we find
\bea
\label{SR_MA_Luttinger}
M_A^{(2)}(q;b_2)&=&
\frac{\big[\log (\ell / \epsilon)\big]^2}{9}
-
\frac{1}{3}\log\!\bigg(
\frac{2 K}{\pi}\, \log (\ell / \epsilon)
\bigg)
\log (\ell / \epsilon)
+
a_1 \log (\ell / \epsilon)
\\
\rule{0pt}{.7cm}
&&+\;
\frac{1}{4}
\left[\log\!\bigg(
\frac{2 K}{\pi}\, \log (\ell / \epsilon)
\bigg)
\right]^2
+
a_2\log\!\bigg(
\frac{2 K}{\pi}\, \log (\ell / \epsilon)
\bigg)
+
O(1)\,,
\nonumber
\eea
with the constants $a_1$ and $a_2$ defined as 
\be
a_1=\frac{1}{3}
\left(
1+2y_S+2 b_2
\right) ,
\;\;\qquad\;\;
a_2
=
-y_S-b_2\,,
\ee
which contain both the non universal constant $y_S$ and $b_2$.

Let us discuss the validity of (\ref{MAn_decomposition2}) at the leading orders.
By plugging (\ref{SR_Mn Luttinger}) into the r.h.s. of (\ref{MAn_decomposition2}) and using that $\sum_q p(q)=1$, we obtain
\bea
\label{check sumoverq}
&&
\sum_q
\sum_{k=1}^{n-1}
\binom{n}{k}
\frac{m^{(k)}(q;b_n/2)}{2}
\left[
\frac{2}{3^{n-k}}\left(\log\frac{\ell}{\epsilon}\right)^{n-k}
-
\frac{n-k}{3^{n-k-1}}\left(\log\frac{\ell}{\epsilon}\right)^{n-k-1}
\log\!\bigg(
\frac{2 K}{\pi}\log\frac{\ell}{\epsilon}
\bigg)
\right]
\nonumber
\\
&&
+\,
\left(
\frac{\log (\ell / \epsilon)}{3} 
\right)^n
-
\frac{n}{2}
\left(
\frac{\log (\ell / \epsilon)}{3} 
\right)^{n-1}
\!\!
\log\!\bigg(
\frac{2 K}{\pi}\, \log (\ell / \epsilon)
\bigg)
+
\sum_q
m^{(n)}(q;b_n)
+\dots\,,
\eea
where the dots represent the terms that have been neglected in (\ref{SR_Mn Luttinger}).
In order to check (\ref{MAn_decomposition2}) up to  $O\big(\log(\ell)^{n-1}\log(\log\ell)\big)$, 
we need to know $p(q)$ for the specific model we are considering. 
For the free compactified massless scalar,
in the limit $\ell/\epsilon \to\infty$, is has been found that  \cite{Xavier:2018kqb}
\be
\label{pq_leading}
p(q)=\sqrt{\frac{\pi}{2 K \log(\ell/\epsilon)}}\;
e^{-\frac{\pi^2 q^2}{2 K \log(\ell/\epsilon)}}\,,
\ee
and that the sum over $q$ can be approximated by an integral over the real axis. 
The dependence on $q$ in (\ref{check sumoverq}) occurs only through $m^{(k)}(q;a)$, whose integrals over $q$ reads
\be
\int_{-\infty}^\infty m^{(k)}(q;a) \,dq
\,=\,
\frac{1}{2^k} \bigg[\log\!\bigg(
\frac{2 K}{\pi}\,  \log(\ell/\epsilon)
\bigg)\bigg]^k
+\,
O\Big(\big[\log\!\big(\log(\ell/\epsilon)\big)\big]^{k-1}\Big)\,.
\ee
By employing this result into (\ref{check sumoverq}), we observe that the largest contribution comes from the term with $k=1$ in the sum, 
which cancels the second term in the second line.
Thus, for (\ref{check sumoverq}) we have
\be
\label{MA-check}
\left(
\frac{\log (\ell / \epsilon)}{3} 
\right)^n
+
O\Big(\big(\log (\ell / \epsilon)\big)^{n-1}\Big)\,,
\ee
consistently with the leading term of $M^{(n)}_A(b_n)$ when $A$ is an interval in a CFT
(see (\ref{Mn LeadingCFT})) in its ground state.

\newpage
\section{Conclusions}
\label{sec:conclusions}

In this section we report some conclusive remarks, organizing the main aspects investigated in this work into paragraphs.
In each of these we summarize the main related findings of the manuscript and we discuss various avenues for interesting future explorations.
\\

\noindent
{\bf Comparing $C_A$ with other quantities from quantum information theory}
\\
In \cite{deBoer:2018mzv} it has been pointed out that if one considers a $1+1$-dimensional CFT either in its ground state or in a thermal state, when the subsystem is a single interval, the entanglement entropy and the capacity of entanglement have the same leading logarithmic behaviour in the subsystem size.   
This observation naturally suggests considering the difference $C_A - S_A$, which is a UV finite quantity. Moreover, since the universal logarithmic divergence cancels, this difference should encode non-universal features at the leading order in the subsystem size; to support this statement, in Sec.\,\ref{CFT} we have computed $C_A-S_A$ in various cases of interest. In Sec.\,\ref{subsec:twointerval} we have considered free bosonic and free Dirac CFTs, when the subsystem $A$ is made by two disjoint intervals. Interestingly, $C_A-S_A$ is able to discriminate between the two theories, given that it is constant for the fermionic theory (see Fig.\,\ref{fig:FF_2int_CEvsEE}) and is a non-trivial function of the cross-ratio in the bosonic case (see (\ref{entropy 2int compact}) and (\ref{capacity 2int compact})).
Free massive quantum field theories in the regime where the mass $m$ is much smaller than the inverse of the subsystem  size $\ell$ have been considered in Sec.\,\ref{subsec:cap-massive}. While in the case of CFTs $C_A-S_A$ is of order one in the subsystem size, terms depending on $m\ell\ll 1$ arise in the massive case and they have a different functional form in the bosonic and the fermionic theory. In particular, for the massive scalar field, $C_A-S_A$ exhibits a double-logarithmic divergence as $m\ell\to 0$ (see (\ref{SAmasslessscalarFT})), which is not present in the massive Dirac theory, as shown in (\ref{CapacityEntropyMassiveFermion}). 
Understanding more about the differences between capacity of entanglement and entanglement entropy is an important task, which deserves future investigations. 

The difference $C_A-S_A$ proves to be interesting also at the level of the contour functions. The CFT analysis of Sec.\,\ref{sec:contour} suggests that, computing the difference between the leading terms of $s_A(x)$ and $c_A(x)$ in (\ref{cont-s-c inverval}), non-universal contributions can be detected. The same conclusion can be drawn by looking at
the bottom panels of Fig.\,\ref{fig:contourCFT}, where $s_A(x)-c_A(x)$ is shown and different behaviours are observed for a single interval in a harmonic chain (bottom left panel) and for the same bipartition in a free fermionic chain (bottom right panel). For the fermionic chain, the curves of data points approach zero as the subsystem size grows, while for the harmonic chain the data points collapse on a curve, which provides a non-trivial prediction in the continuum limit. To the best of our knowledge, the expression of this function is not known and therefore it would be very interesting to derive it through QFT techniques. Notice that, because of non-universal effects, from our numerical analysis we cannot ensure that the difference of the contour functions is finite close to the endpoints, as the CFT analysis would suggest.
Also out of equilibrium, the difference between the contour functions exhibits a rich behaviour (cf. bottom panel of Fig.\,\ref{fig:contour_quenchFF}), which is still to be completely understood.

As pointed out in \cite{Arias22Monotones} and discussed in Sec.\,\ref{sec:intro}, the capacity of entanglement and the other cumulants of the entanglement Hamiltonian can be combined to give the entanglement monotones $M_A^{(n)}$ defined in (\ref{Mn Tr Fn}). Another central question that this manuscript addresses is whether the capacity of entanglement and $M_A^{(n)}$ are capable of capturing features that the R\'enyi entropies are not sensitive to.
In this respect, the capacity of entanglement of a block of consecutive sites in a free fermionic chain with non-vanishing chemical potential is studied in Sec.\,\ref{subsec:oscillations}. We find that $C_A$ exhibits oscillations in the subsystem size, with frequency proportional to the Fermi momentum of the system. This features are not present for the entanglement entropy, while they show up for $S_A^{(n)}$ with $n\geqslant 2 $ \cite{Calabrese:2009us,CalabreseEssler_10_XXchain}.
Moreover, the temporal evolution of $M_A^{(2)}$ after a global quench in free bosonic and fermionic chains is studied in Sec.\,\ref{sec:quench}. As shown in Fig.\,\ref{fig:HC_Mfuncafterquench} and Fig.\,\ref{fig:FF_Mfuncafterquench}, it exhibits an initial quadratic growth in time before the saturation regime, differently from the linear growth of the entanglement entropies. This can be traced back to the presence of a term involving $S_A^2$, which is dominant for large subsystem sizes.
 It would be insightful to expand these analyses to other models that allow to find properties of $M_A^{(n)}$ and $C_A$, which are not shared by $S_A^{(n)}$.	
\\

\noindent
{\bf Looking for new $c$-functions}
\\
In \cite{Casini:2006es} a $c$-function for relativistic QFTs has been constructed as the logarithmic derivative of the entanglement entropy with respect to the subsystem size. Exploiting only the Lorentz invariance of the theory and the strong subadditivity of $S_A,$ this function is shown to be decreasing along the RG flow, consistently with the $c$-theorem \cite{Zamolodchikov:1986gt}. Along this line, in Sec.\,\ref{massivebosons} we have introduced the two functions in (\ref{CfunctC}) and (\ref{CfunctMtilde}) from the capacity of entanglement and the entanglement monotone $M_A$ defined in (\ref{Mdefinition}) respectively. When evaluated for free bosonic and fermionic theories, these functions exhibit a decreasing behaviour in the RG parameters, namely the masses. This behaviour is shown in Fig.\,\ref{fig:Cfunctions}. Since $C_A$ and $M_A$ do not satisfy the strong subadditivity, the argument of \cite{Casini:2006es} does not apply. Currently, we have no a priori reasons for establishing the monotonicity of (\ref{CfunctC}) and (\ref{CfunctMtilde}) along the RG flows and therefore we dub them accidental $c$-functions.
In order to frame our analysis in a more general context, we could ask whether monotonic $c$-functions can be constructed using properties different from strong subadditivity. A similar question has been addressed in \cite{latorre2003ground,Orus:2005jq,Riera:2006vj}, where it has been argued that reduced density matrices follow a majorization order along RG flows. The arguments in \cite{latorre2003ground,Orus:2005jq,Riera:2006vj} are worth being expanded and made more rigorous; 
this might allow to conclude that all the infinitely many Schur concave functions of the entanglement spectrum are monotonically decreasing along the RG flows.
\\

\noindent
{\bf Temporal evolution after a quantum quench} 
\\
In Sec.\,\ref{sec:quench} we compute the temporal evolution of the capacity of entanglement after a global quantum quench in free bosonic and fermionic chains. As shown in Fig.\,\ref{fig:HC_CEvsEEvsQPformulaforEE} for the bosons and in Fig.\,\ref{fig:Cvstfermions} for the fermions, we find an initial linear growth and a subsequent saturation to an asymptotic value. Both these features are very well captured by the formula (\ref{capacity_QPpict}) based on the quasi-particle picture of \cite{Calabrese_2005}, while the correct asymptotic value is predicted using the generalized Gibbs ensemble as stationary state at large times.
Comparing the evolution of the capacity of entanglement with the one of the entanglement entropy, one observes that the slopes characterizing the linear regime of the two quantities are different (see Fig.\,\ref{fig:HC_CEvsEEvsQPformulaforEE} and Fig.\,\ref{fig:Cvstfermions}).
This finding is a fingerprint of the dependence of the parameter $\tau_0$ introduced in \cite{Calabrese_2005} on $n$ (see also the right panel of Fig.\,3 of  \cite{Coser:2014gsa} for a numerical analysis of this dependence). Indeed, if we assume that $\tau_0$ is independent of $n$, the CFT computation would lead to the same slope for the entanglement entropy and the capacity of entanglement \cite{deBoer:2018mzv}.
To understand better the origin of the different initial slopes, it would be desirable to obtain these results exploiting other field theoretical techniques.

A related question is whether the reduced density matrices satisfy a majorization order along the temporal evolution after a quench. Notice that the ordering $\rho_A(t)\succ \rho_A(t')$ when $t > t'$ is ruled out given that both the Schur concave quantities $S_A$ and $M_A$ evaluated along the temporal evolution are increasing in time (see Figs.\,\ref{fig:HC_CEvsEEvsQPformulaforEE}, \ref{fig:HC_Mfuncafterquench}, \ref{fig:Cvstfermions} and \ref{fig:FF_Mfuncafterquench}). 
On the other hand, the possible ordering given by $\rho_A(t)\succ \rho_A(t')$ when $t < t'$ cannot be excluded exploiting the results reported in this manuscript, requiring a more refined investigation. A promising approach could be studying the temporal evolution of the entanglement spectrum of these chains, expanding, for instance, the analyses in \cite{DiGiulio:2019lpb}.
\\

\noindent
{\bf Symmetry-resolved related issues}
\\
In Sec.\,\ref{sec:SymResCapacity} we have found that the entanglement monotones $M_A^{(n)}$ defined in (\ref{Mn Tr Fn}) decompose non-trivially, as shown in (\ref{MAn_decomposition2}), into the charge sectors of a theory with a global $U(1)$ symmetry. 
Remarkably, this decompostion holds for $M_A^{(n)}$ for any value of $n$, but not for the  R\'enyi entropies $S_A^{(n)}$ (unless $n=1$, when the entanglement entropy is retrieved).
This analysis motivates the study of symmetry-resolved monotones $M_A^{(n)}(q)$ as a tool for probing the various charge sectors of the theory. For instance, given two density matrices such that $\rho\succ \sigma$, we may ask whether this majorization order survives  once we decompose $\rho$ and $\sigma$ according to (\ref{eq:decompositionrho}). One can try to diagnose this aspect using $M_A^{(n)}(q)$ in (\ref{SR-M_n}), given that these quantities are entanglement monotones also in a fixed charge sector. We leave this analysis for future investigations.

Finally, we have studied the resolution of the capacity of entanglement in the various $U(1)$ sectors of a Luttinger liquid CFT. As one can see in (\ref{SRCapacity_Luttinger}), we have found that the symmetry-resolved capacity of entanglement is independent of the charge at leading order for small cutoff. This feature is the so-called equipartition of entanglement and has been observed also for the entanglement entropy and the R\'enyi entropies in the same model \cite{Goldstein:2017bua,Xavier:2018kqb}. A remarkable difference with respect to the symmetry-resolved entanglement entropies (see (\ref{SREE_Luttinger}) and (\ref{SRvN_Luttinger})) is that $C_A(q)$ does not exhibit any double logarithmic correction and therefore shows dependence on the non-universal properties of the theory only at order one in the small cutoff expansion. An interesting consequence is that the difference $C_A(q)-S_A(q)$ enjoys equipartition and encodes non-universal features at leading order. On the other hand, differently from what we have discussed at the beginning of this section for the difference between the total quantities, $C_A(q)-S_A(q)$ is not UV finite because of the aforementioned log-log divergence of $S_A(q)$. It would be interesting to find a quantity that not only is equally distributed among the charge sectors and displays non-universal features at leading order, but is also UV finite.

\section*{Acknowledgments}

We are grateful to Jan de Boer for collaboration at an earlier stage of this work and important discussions throughout its development. 
We thank Mihail Mintchev, Sara Murciano and Diego Pontello for useful conversations. 
GDG, EKV and ET acknowledge Galileo Galilei Institute for warm hospitality and financial support during part of this work
through the program {\it Reconstructing the Gravitational Hologram with Quantum Information}.
GDG acknowledges support by the Deutsche Forschungsgemeinschaft (DFG, German Research Foundation) 
under Germany’s Excellence Strategy through the Würzburg-Dresden Cluster of Excellence on Complexity and Topology in Quantum Matter - ct.qmat (EXC 2147, project-id 390858490).
EKV's research has been conducted within the framework of InstituteQ - the Finnish Quantum Institute, and ET's
research within the framework of the Trieste Institute for Theoretical Quantum Technology (TQT).
%


\appendix

\section{Free fermionic and bosonic lattice models}
\label{app-lattice}

In this appendix we review the numerical procedure to evaluate
the entanglement entropy, the capacity of entanglement and their contour functions 
for the free bosonic and fermionic lattice models that we are considering in this manuscript. 
Moreover, we 
derive analytic expressions for the correlators of the fermionic chain with Hamiltonian (\ref{eq:hamiltonianF}).

\subsection{Capacity of entanglement and its contour function}

In a free lattice model in a generic number of spatial dimensions described by a quadratic Hamiltonian,
consider a sublattice $A$ made by $\ell$ sites.
When the entire system is in a Gaussian state (e.g. the ground state), 
the entanglement entropy and the capacity of entanglement can be computed as follows 
\cite{Peschel03,Botero04,Amico:2007ag,Eisert:2008ur,ep-rev,CHrev,Weedbrook12b}
\be
\label{SA_CA_lattice_boson}
S_A=\sum_{k=1}^\ell s(\xi_k)\,,
\;\;\;\qquad\;\;\;
C_A=\sum_{k=1}^\ell c(\xi_k)\,,
\ee
where the explicit expressions of $s(y)$ and $c(y)$ 
and the nature of the eigenvalues $\xi_k$ depend on whether the lattice is made by fermions or bosons.

The contour functions $s_A(i)$ and $c_A(i)$  in (\ref{entropy contour_lattice})
can be constructed by associating $\ell$ real numbers $p_k(i)$ to every $\xi_k$, being $1\leqslant i, k\leqslant \ell$.
The function $p_k(i)$ is called mode participation function \cite{Botero04} and fulfils  the following conditions
\be
\label{pki condition}
\sum_{i=1}^{\ell}p_k(i)=1\,,
\;\;\;\;\qquad\;\;\;\;
p_k(i)\geqslant  0\,,
\ee
which tell us that $p_k(i)$ is a probability distribution for every $k$.
A mode participation function $p_k(i)$ allows 
to write the contour functions (\ref{entropy contour_lattice}) as follows
\be
\label{contour ee-cap from mpf}
s_A(i)
=
\sum_{k=1}^{\ell}
p_k(i)\, s(\xi_k)\,,
\;\;\;\qquad\;\;\;
c_A(i)
=
\sum_{k=1}^{\ell}
p_k(i)\, c(\xi_k)\,,
\ee
through the functions $s(y)$ and $c(y)$  occurring  in (\ref{SA_CA_lattice_boson}). 
The explicit expression of the mode participation function $p_k(i)$ depends on the model and it is not unique, even for a given model. 
Some reasonable constraints that $p_k(i)$ must satisfy 
have been introduced in \cite{Chen_2014}. However, they do not fix $p_k(i)$ uniquely.

In the following we employ  the mode participation functions 
proposed in \cite{Chen_2014} for the free fermionic lattices and  in  \cite{Coser:2017dtb} for the free bosonic lattices,
whose construction is reviewed in the forthcoming discussion.

\subsection{Harmonic lattice}
\label{subapp:entanglementHC}

The Hamiltonian of the harmonic lattice with nearest neighbours spring-like interactions reads
\be
\label{HL ham}
\widehat{H}_{\textrm{\tiny HL}} = 
\sum_{i} 
\left(
\frac{1}{2\mu}\,\hat{p}_i^2+\frac{\mu\omega^2}{2}\,\hat{q}_i^2\right) +
\sum_{\langle i,j \rangle} 
\frac{\lambda}{2}(\hat{q}_{i} -\hat{q}_j)^2\,,
\ee
where the hermitean operators  $\hat{q}_i$ and $\hat{p}_i$
satisfy the canonical commutation relations
$[\hat{q}_i , \hat{q}_j]=[\hat{p}_i , \hat{p}_j] = 0$ 
and $[\hat{q}_i , \hat{p}_j]= \textrm{i} \delta_{i,j}$. The Hamiltonian (\ref{HL ham}) generalises (\ref{HC ham}) to higher dimensional lattices.

Assuming that the entire system is in a Gaussian state and considering a spatial bipartition of the lattice 
into a subsystem $A$ made by $\ell$ sites and its complement, 
the entanglement properties are encoded into the reduced covariance matrix $\gamma_A$, which is defined as 
the following $(2\ell)\times (2\ell)$ symmetric and positive definite matrix
 \be
\label{reduced CM}
\gamma_A
\equiv
\bigg( 
\begin{array}{cc}
Q_A  &  M_A \\
M_A^{\textrm{t}}  &  P_A  \\
\end{array}   \bigg)\,,
\ee
in terms of the two point functions restricted to $A$, namely
$(Q_A)_{i,j}=\langle \hat{q}_i \hat{q}_j \rangle$, $(P_A)_{i,j}=\langle \hat{p}_i \hat{p}_j \rangle$ 
and $(M_A)_{i,j}=\textrm{Re}\big[\langle \hat{q}_i \hat{p}_j \rangle\big]$, with $i,j=1,\dots,\ell$.
In the time independent case,  $\gamma_A=Q_A\oplus P_A$.
Since $\gamma_A$ is a $(2\ell)\times (2\ell)$ real symmetric and positive definite matrix, 
its symplectic eigenvalues $\{\sigma_1,\dots,\sigma_\ell \}$ can be considered \cite{Bhatia07book}.

The moments $\mathrm{Tr}\rho_A^n$ are obtained from the symplectic eigenvalues $\sigma_k$'s as \cite{holevo}
\be
\log \mathrm{Tr}\rho_A^n=-\sum_{k=1}^\ell \log\left[
\left(\sigma_k +\frac{1}{2}\right)^n\,
-\,
\left(\sigma_k-\frac{1}{2}\right)^n\,
\right]\,.
\ee
From (\ref{defentropy}) and (\ref{defcapacity}), one finds that $S_A$ and $C_A$ are given by (\ref{SA_CA_lattice_boson}) with
\be
\xi_k = \sigma_k\,,
\ee
and
\bea
\label{sx def}
s(y)
&=&
\left(y +\frac{1}{2}\right)\log\left(y+\frac{1}{2}\right)-\left(y -\frac{1}{2}\right)\log\left(y -\frac{1}{2}\right),
\\
\rule{0pt}{.9cm}
\label{cx def}
c(y)
&=&
\left[\left(y +\frac{1}{2}\right)\log\left(y +\frac{1}{2}\right)-\left(y -\frac{1}{2}\right)\log\left(y -\frac{1}{2}\right)
\right]^2
\nonumber
\\
& &
-\,
\left[
\left(y+\frac{1}{2}\right)\left(\log\left(y +\frac{1}{2}\right)\right)^2-\left(y -\frac{1}{2}\right)\left(\log\left(y -\frac{1}{2}\right)\right)^2
\right]\,.
\eea
The above expressions have been used to obtain the numerical data reported in the left panel of Fig.\,\ref{fig:HC_1intmassive_CEvsEE}, 
in the top panel of Fig.\,\ref{fig:Cfunctions},
in Fig.\,\ref{fig:HC_CEvsEEvsQPformulaforEE} 
and Fig.\,\ref{fig:HC_Mfuncafterquench}.

We adopt the proposal made in \cite{Coser:2017dtb} for the mode participation function in free bosonic lattice models,
which is constructed as follows. 
Consider the Williamson's decomposition of $\gamma_A$, namely
\be
\gamma_A=W^{\textrm{t}} \, \mathcal{D} \, W\,,
\ee
where $W$ is a symplectic matrix and $\mathcal{D}=\textrm{diag}\big(\sigma_1,\dots,\sigma_\ell\big)\oplus \big(\sigma_1,\dots,\sigma_\ell\big)$ contains the symplectic eigenvalues.
Let us introduce the auxiliary matrix $K$ given by 
\be
K=
W(W^{\textrm{t}} W)^{-1/2}
=
(W W^{\textrm{t}})^{-1/2}W\,,
\ee
which can be decomposed into $\ell \times \ell$ blocks 
\be
K=
\,
\bigg( \begin{array}{cc}
 U_K &  Y_K \\
 Z_K &  V_K \\
\end{array}  \bigg)\,.
\ee
The mode participation function $p_k(i)$ proposed in \cite{Coser:2017dtb} reads
\be
\label{mpf}
p_k(i)=\frac{1}{2}\bigg(\big[(U_K)_{ki}\big]^2+\big[(Y_K)_{ki}\big]^2+\big[(Z_K)_{ki}\big]^2+\big[(V_K)_{ki}\big]^2 \bigg)\,.
\ee
By using (\ref{mpf}) and the symplectic spectrum of the reduced covariance matrix in (\ref{reduced CM}),
we can compute the contour functions for the entanglement entropy and the capacity of entanglement for a generic harmonic chain.
In the left panels of Fig.\,\ref{fig:contourCFT}, in Fig.\,\ref{fig:contour_quenchHC} and in Fig.\,\ref{fig:contour_quenchHC_differentell}
the numerical data for the contour functions have been obtained by employing (\ref{sx def}), (\ref{cx def}) and (\ref{mpf}) into (\ref{contour ee-cap from mpf}).

 \subsection{Fermionic lattice}
\label{subapp:fermionlattice}

In free fermionic lattices (see e.g. the Hamiltonians (\ref{Ham FF}), (\ref{xxtext}) and (\ref{eq:hamiltonianF})) in their ground state,
the moments $\mathrm{Tr}\rho_A^n$ can be computed 
from the eigenvalues $\nu_k$ of the $\ell\times \ell$ reduced correlation matrix $\mathcal{C}_A$, 
i.e. the matrix whose entries are $\langle \hat{c}_i^\dagger \hat{c}_j \rangle$, with $i,j=1,\dots,\ell$. 
The logarithm of $\mathrm{Tr}\rho_A^n$ is given by \cite{Peschel03,ep-rev}
\be
\label{trrhon_FFlattice}
\log \mathrm{Tr}\rho_A^n=\sum_{k=1}^\ell
\log\!\big[\nu_k^n+(1-\nu_k)^n\big]\,.
\ee
From (\ref{defentropy}), (\ref{defcapacity}) and (\ref{trrhon_FFlattice}),
one finds that the entanglement entropy and the capacity of entanglement can be computed
through (\ref{SA_CA_lattice_boson}) with
\be
\label{xik_nuk}
\xi_k=\nu_k\,,
\ee
 and
\bea
\label{sx def fermions}
s(y)
&=&
-\,y\log(y)-(1-y)\log(1-y)\,,
\\
\rule{0pt}{.7cm}
\label{cx def fermions}
c(y)
&=&
\Big[
y\big(\log y\big)^2+(1-y)\big(\log(1-y)\big)^2
\Big]
-
\big[
y\log y+(1-y)\log(1-y)
\big]^2\,.
\eea
These relations have been used to obtain the numerical data for free fermionic chains reported in Fig.\,\ref{fig:FF_2int_CEvsEE}, 
in the bottom panels of Fig.\,\ref{fig:Cfunctions}, in Fig.\,\ref{fig:Cvstfermions} and Fig.\,\ref{fig:FF_Mfuncafterquench}.

The contour functions for the free fermionic chains considered in this manuscript have been evaluated by employing
the proposal made in \cite{Chen_2014}.
Since the reduced correlation matrix $\mathcal{C}_A$ is hermitian,
it is diagonalised by a unitary matrix $\widetilde{U}$, 
which can be exploited to construct the following mode participation function \cite{Chen_2014}
\be
\label{mpf fermion A-case}
p_k(i)
=
   \big| \widetilde{U}_{k,i} \big|^2\,.
\ee
Notice that the $i$-th element of the diagonal of the matrix relation  
$\widetilde{U}^\dagger \widetilde{U} = \boldsymbol{1}$ gives the condition $\sum_{k=1}^\ell p_k(i)=1$ 
for $1\leqslant i \leqslant \ell$.
The contour for the entanglement entropy and for the capacity of entanglement in these free fermionic models are obtained 
by using (\ref{mpf fermion A-case}), (\ref{sx def fermions}), (\ref{cx def fermions}) and the eigenvalues $\nu_k$'s in (\ref{contour ee-cap from mpf}).
Numerical data points found through these expressions are 
shown in the right panels of Fig.\,\ref{fig:contourCFT} and in Fig.\,\ref{fig:contour_quenchFF}.

\subsection{Lattice correlators for the massive Dirac field}
\label{app:massiveDirac}

In this appendix we derive analytic expressions for the two-point correlators
of the free fermionic chain described by the Hamiltonian (\ref{eq:hamiltonianF}),
which provides the lattice discretisation of the massive Dirac fermion in $1+1$ dimensions. 

Some of the numerical results reported in Sec.\,\ref{subsec:cfunction} have been obtained by employing the 
two-point correlators of the model (\ref{eq:hamiltonianF}) in the thermodynamic limit $N \to \infty$,
which reads \cite{Casini:2005rm}
\bea
\label{corr massiveDirac Int}
\braket{\hat{c}^{\dagger}_j \hat{c}_{k}}
&=&
\frac{1}{2}\delta_{j-k,0}+(-1)^j \displaystyle \int_0^{\frac{1}{2}}\frac{\widetilde{m} \cos (2\pi x (j-k))}{\sqrt{\widetilde{m}^2+\sin(2\pi x)^2}} \, dx\,,
 \qquad \textrm{even $ |j-k|$}\,, 
\\
\label{corr massiveDirac Int-2}
\rule{0pt}{.8cm}
\braket{\hat{c}^{\dagger}_j \hat{c}_{k}}
&=&
\textrm{i} \displaystyle \int_0^{\frac{1}{2}} 
\frac{\sin (2\pi x )}{\sqrt{\widetilde{m}^2+\sin(2\pi x)^2}} \, \sin (2\pi x (j-k)) \, dx\,,
\qquad  \textrm{odd $ |j-k|$}\,.
\eea
The following integrals
\be
\label{I and tilde I}
I_{j,k}= \int_0^{\frac{1}{2}} \frac{ \cos (2\pi x (j-k))}{\sqrt{\widetilde{m}^2+\sin(2\pi x)^2}}\, dx\,,
\,\,\qquad\,\,
\widetilde{I}_{j,k}= \int_0^{\frac{1}{2}} \frac{\sin (2\pi x )}{\sqrt{\widetilde{m}^2+\sin(2\pi x)^2}}\, \sin (2\pi x (j-k))\, dx \,,
\ee
can be evaluated in terms of hypergeometric functions
by employing the following integral representation of the hypergeometric function $_2F_1$
\bea
\label{HypergeomIntReps}
\int_0^\pi \!\!
\frac{ \cos(n \theta)}{2\pi(1-a \cos\theta)^b} \,  d\theta\, =
& &
\\
& &
\hspace{-2.2cm}
=\;
 \frac{2^{b-1}\, \Gamma(n+b)}{n!\, a^b\, \Gamma(b)} 
 \left( \frac{1-\sqrt{1-a^2}}{a}\,\right)^{n+b}
  \!\! \, _2F_1 \!\left( b\, ,n+b\, ; n+1 \,; \bigg( \frac{1-\sqrt{1-a^2}}{a}\,\bigg)^2 \,\right),
  \nonumber
\eea
where $n$ is an integer number and $\Gamma(x)$ is the gamma function.

As for $I_{j,k}$,
by denoting by $|j-k|\equiv r$ and performing the change of variables $2\pi x =\tilde{\theta}$, 
for the first integral in (\ref{I and tilde I}) we obtain
\be
I_{j,k}
= \sqrt{2 a(\widetilde{m})}\int_0^{\pi}\frac{ \cos (r\tilde{\theta})}{\sqrt{1-a(\widetilde{m})\cos(2\tilde{\theta})}}\; \frac{d\tilde{\theta}}{2\pi} \,,
\;\;\qquad\;\;
a(\widetilde{m})\equiv \frac{1}{1+2\widetilde{m}^2}\,.
\ee
By adopting the integration variable $\tilde{\theta}=\theta/2$ 
and exploiting the symmetry of the integrand function in the integration domain,
we get
\bea
\label{I final expr}
I_{j,k}
&=&
\sqrt{2 a(\widetilde{m})}\int_0^{\pi}\frac{ \cos (\theta r/2)}{\sqrt{1-a(\widetilde{m})\cos(\theta)}}\, \frac{d\theta}{2\pi} 
\nonumber
\\
\rule{0pt}{.7cm}
&=&
\frac{\Gamma\big(\frac{r}{2}+\frac{1}{2}\big)}{\sqrt{\pi} \; \Gamma\big(\frac{r}{2}+1\big)}\;Z(\widetilde{m})^{\frac{r}{2}+\frac{1}{2}} 
\,_2F_1 \!\left( \frac{1}{2}\, ,\frac{r}{2}+\frac{1}{2}\, ; \frac{r}{2}+1 \,; Z(\widetilde{m})^2 \,\right),
\eea
where $Z(\widetilde{m})\equiv 1+2\widetilde{m}^2-2\widetilde{m}\sqrt{1+\widetilde{m}^2}$
and (\ref{HypergeomIntReps}) has been used in the special case where $n=r/2$  and $b=1/2$. 
Notice that setting $n=r/2$ is not inconsistent with (\ref{HypergeomIntReps}), 
which holds only for integer values of $n$; 
indeed, the integral $I_{j,k}$ enters in (\ref{corr massiveDirac Int}), 
where $r=|j-k|$ is even; hence $n$ is integer.

The integral $\widetilde{I}_{j,k}$ in (\ref{I and tilde I}) can be studied by observing that
\be 
\label{Itilde Itildepm}
\widetilde{I}_{j,k}=\frac{1}{2}\big(\widetilde{I}_{j,k}^- - \widetilde{I}_{j,k}^+\big)\,,
\ee
where
\be
\widetilde{I}_{j,k}^{\pm}\equiv\int_0^{\frac{1}{2}}
\frac{ \cos (2\pi x |j-k\pm 1|)}{\sqrt{\widetilde{m}^2+\sin(2\pi x)^2}}\, dx\,.
\ee
By introducing  $r^\pm\equiv|j-k\pm 1|$ and repeating the same calculation discussed above 
for $I_{j,k}$ (with $r$ replaced by $r^{\pm}$) we obtain
\be
\label{Itildepm final expr}
\widetilde{I}_{j,k}^{\pm}
=
\frac{\Gamma\big(\frac{r^\pm}{2}+\frac{1}{2}\big)}{\sqrt{\pi} \;\Gamma\big(\frac{r^\pm}{2}+1\big)}\;
Z(\widetilde{m})^{\frac{r^\pm}{2}+\frac{1}{2}} \;_2F_1 \!\left( \frac{1}{2}\, ,\frac{r^\pm}{2}+\frac{1}{2}\, ; \frac{r^\pm}{2}+1 \,; Z(\widetilde{m})^2 \,\right)\,.
\ee
Thus, (\ref{I final expr}), (\ref{Itilde Itildepm}) and (\ref{Itildepm final expr})
provide analytic expressions for the correlators (\ref{corr massiveDirac Int})  and (\ref{corr massiveDirac Int-2}),
which have been used to obtain the numerical data displayed  in the bottom left panel of Fig.\,\ref{fig:Cfunctions}.

\section{Three qubits playground}
\label{app:3qubits}

In this appendix we explore the relation between strong subadditivity 
and concavity for the second moment of shifted modular Hamiltonian $M_A^{(2)}$ defined in (\ref{Mn Tr Fn}). 
In this appendix, we slightly modify the notation by removing the label referring to the subsystem $A$ 
and highlighting the dependence of $M^{(2)}$ on $b_2$.
For this purpose, we consider three examples involving simple qubit systems.  
From the first two of them we find that  the strong subadditivity property of $M^{(2)}(b_2)$ does not necessarily require it to be a concave function of the reduced density matrix. In the last example we obtain that the opposite also holds, namely that there is a range of the parameter $b_2$ where concavity is satisfied but  the strong subadditivity is not.

In all the three cases studied here the total Hilbert space ${\cal {H}}$ can be decomposed as ${\cal {H}}={\cal {H}}_1\otimes{\cal {H}}_2\otimes{\cal {H}}_3$, where ${\cal {H}}_i=\mathbb{C}^2$, with $i=1,2,3$.
\\

{\it Example 1.}
Consider the three qubits in the W state $|W\rangle\in \mathcal{H}$ \cite{Czerwinski:2021rbl}, given by 
\be
|W\rangle=\frac{1}{\sqrt{3}}\left(|001\rangle+|010\rangle+|100\rangle\right)\,.
\ee
The corresponding total density matrix is given by
\be
\rho_W=|W\rangle\langle W|\,,
\ee
from which we compute the following reduced density matrices $\rho_{W,12}=\mathrm{Tr}_{_3}\rho_W,$ $\rho_{W,23}=\mathrm{Tr}_{_1}\rho_W$ and $\rho_{W,2}=\mathrm{Tr}_{_{13}}\rho_W$, which will be needed to discuss the strong subadditivity. They read
\bea
\rho_{W,12}
&=&
\rho_{W,23}
\,=\,
\frac{1}{3}
\Big(
|00\rangle\langle00|+|01\rangle\langle01|+|10\rangle\langle10|+|01\rangle\langle10|+|10\rangle\langle01|
\Big)\,,
\\
\rule{0pt}{.7cm}
\rho_{W,2}
&=&
\frac{2}{3}|0\rangle\langle0|+\frac{1}{3}|1\rangle\langle1|\,.
\eea
The idea now is to see whether
\be
M^{(2)}(\rho_W;b_2)+M^{(2)}(\rho_{W,2};b_2)-\big[M^{(2)}(\rho_{W,12};b_2)+M^{(2)}(\rho_{W,23};b_2)\big]
\leqslant 0 \, ,
\label{SSAM}
\ee
is satisfied or not. Using the form of $M^{(2)}(b_2)$ written in \eqref{Mn Tr Fn} for a general coefficient $b_2$ we see that the inequality (\ref{SSAM}) is satisfied as long as $b_2 \gtrsim -1.109$. 
Then there is a range in the parameter  $b_2$ (that is $-1.109\lesssim b_2<1 $) 
where the  strong subadditivity condition (\ref{SSAM}) is satisfied but the concavity condition is not (remember that it holds for $b_2\geq1$, as discussed in Sec.\,\ref{sec:intro}). 
This means that  the strong subadditivity property does not require the concavity in terms of $\rho$. 
Of course, using the coefficient $b_2=1$ as used for instance in \cite{boes2020}, this issue does not occur because for that value 
 the strong subadditivity holds and also the concavity condition is satisfied. 
This example shows that at the level of density matrices the strong subadditivity implies
that concavity is not true in general.
\\

{\it Example 2.}
We can do the same analysis for the following three qubits example used in \cite{Petz:2015} to show that Tsallis entropy does not satisfy  the strong subadditivity property. 
The total density matrix $\rho_{123}$ of the state is 
\bea
\rho_{123}
&=&
\frac{1}{4}
\Big(
|010\rangle\langle010|+|010\rangle\langle011|+|100\rangle\langle100|+|100\rangle\langle101|+|011\rangle\langle010|
\nn
\\
&&
\hspace{.6cm}
+ \,|011\rangle\langle011|+|101\rangle\langle100|+|101\rangle\langle101|\Big)\,,
\eea
from which its reduced density matrices can be obtained. It is easy to compute $M^{(2)}(b_2)$ in this case and see that the SSA inequality \eqref{SSAM} is satisfied for any value of the coefficient $b_2$. And as before one can see that for $b_2<1$ 
 the strong subadditivity holds but concavity does not. Thus, also in this case 
 the strong subadditivity does not imply concavity.
\\

{\it Example 3.}
In the previous examples it is clear that for $b_2=1$ the second moment of shifted modular Hamiltonian satisfies 
 the strong subadditivity and one is tempted to think that this will always happen. In the following example we will show that this is not the case.
 
In this exercise we have three free parameters to play with. The density matrix of the three qubits state is
\bea
\rho_{123}
&=&
\eta
\Big(
|000\rangle\langle000|+|000\rangle\langle111|+a|001\rangle\langle001|+b|010\rangle\langle010|+c|100\rangle\langle100| 
\\
\rule{0pt}{.6cm}
&& \hspace{.5cm}
+ \, a^{-1}|011\rangle\langle011|+b^{-1}|101\rangle\langle101|+c^{-1}|110\rangle\langle110|+|111\rangle\langle000|+|111\rangle\langle111|
\Big)
\nn\,,
\eea
with $\eta=\frac{1}{2+a+b+c+1/a+1/b+1/c}$ to have a well normalized density matrix. After computing the corresponding reduced density matrices and setting, for example, $a=1,c=1,b=1$ we can see that  the strong subadditivity condition (\ref{SSAM}) is satisfied as long as $-\frac12\left(b_2-96\log2\right)\log2\leq0$, which leads to $b_2\gtrsim 66.5$. This means that for $1\leqslant b_2\lesssim 66.5$ we have the concavity property satisfied but  the strong subadditivity does not hold. Also it shows that, for the value of $b_2$ considered in \cite{boes2020} and used in some parts of this manuscript, $M^{(2)}(b_2)$ does not satisfy the strong subadditivity for any state, but just in particular cases as we mentioned above.

Thus, two conclusions can be taken from this appendix. The first one is that  the strong subadditivity and concavity are not related properties at the level of density matrices for $M^{(2)}(b_2)$. The second conclusion is that $M^{(2)}(b_2)$ does not satisfy  the strong subadditivity for any density matrix and therefore one can say that it is not a strong subadditive quantity.


\section{Some results based on the corner transfer matrix}
\label{app:CTMdetails}

In this appendix we report derivations of some results presented in Sec.\,\ref{subsec:cfunction}
and also provide supplementary computations based on the corner transfer matrix.

\subsection{$S_A$ and $C_A$ from the corner transfer matrix}
\label{app:derivationCTM}

In the following we describe the derivation of  (\ref{EE_CTM_HC_elliptic-ep}), (\ref{EE_CTM_HC_elliptic}) and (\ref{CE_CTM_HC_elliptic}).
First one observes that (\ref{ZnCTM_HC}), (\ref{EE_CTM_HC}) and (\ref{CoE_CTM_HC}) can be written in terms of elliptic functions.
This can be done by introducing $q\equiv e^{-\varepsilon}$ and taking the derivative with respect to $n$ of (\ref{ZnCTM_HC}), 
which gives
\be
\label{partialZnCTM_HC}
\partial_n  \log \textrm{Tr}\rho_A^n
\,=\,
\sum_{j=0}^\infty 
\bigg[\,
\frac{\varepsilon (2j+1) \,q^{n(2j+1)}}{1-q^{n(2j+1)}} -\log\!\big(1-q^{(2j+1)}\Big)
\,\bigg]\,.
\ee
Following the computation reported in Appendix A of \cite{Eisler:2020lyn} with $q$ replaced by $q^n$, we find
\be
\label{partialZnCTM_HC_v2}
\partial_n  \log \textrm{Tr}\rho_A^n
\,=\,
\frac{1}{24}\bigg[\, 
\log\!\big(16\, \kappa'^4 / \kappa^2\big) - \frac{4(1+\kappa_n^2)}{\pi}
\,K(\kappa_n)K(\kappa_n')
\,\bigg]\,,
\ee
where $\kappa_n'\equiv\sqrt{1-\kappa_n^2}$ and $\kappa_n$ is implicitly defined as follows
\be
\label{eps-omega-peschel-n}
n\,\varepsilon\equiv \frac{\pi\,K\!\big(\sqrt{1-\kappa_n^2}\,\big)}{K(\kappa_n)}\,,
\ee
which implies that $\kappa_1=\kappa$, where $\kappa$ is defined in (\ref{eps-omega-peschel}). 
By evaluating (\ref{partialZnCTM_HC_v2}) for $n=1$, one obtains (\ref{EE_CTM_HC_elliptic-ep}), first found in \cite{ep-rev}.

In order to explore the capacity of entanglement, 
let us first notice that the definition of $q$ in terms of $\varepsilon$ and the relation (\ref{eps-omega-peschel-n}) lead to 
\be
\label{Thetafunckappa}
\kappa_n=\frac{\theta^2_2(q^n)}{\theta^2_3(q^n)}\,,
\,\,\qquad\,\,
\kappa_n'\equiv\sqrt{1-\kappa_n^2}= \frac{\theta^2_4(q^n)}{\theta^2_3(q^n)}\,,
\ee
in terms of the Jacobi theta functions $\theta_r(q)$ with $r=2,3,4$ \cite{thetafunc_book}, which give
\be
\label{Thetafunckappav2}
\frac{\kappa'_n}{\sqrt{\kappa_n}}=\frac{ \theta_4^2(q^n)}{\theta_2(q^n) \, \theta_3(q^n)}\,.
\ee
By using (\ref{eps-omega-peschel-n}), (\ref{Thetafunckappa}), (\ref{Thetafunckappav2}) in (\ref{partialZnCTM_HC}) 
and exploiting the fact that the elliptic integral $K$ can be written as $K(\kappa_n)=\frac{\pi}{2}\,\theta_3^2(q^n)$, we obtain
\be
\label{ZnCTM_HC_elliptic_step2}
\partial_n \log \textrm{Tr}\rho_A^n
\,=\,
\frac{1}{6}\bigg[\,
\log\!\bigg( \frac{ \theta_4^2(q)}{\theta_2(q)\theta_3(q)}\bigg)-\frac{\varepsilon}{4}\big( \theta_2^4(q^n)+\theta_3^4(q^n)\big)
+ \log 2
\,\bigg]\,.
\ee
Taking the limit $n\to 1$, changing the sign of both sides of (\ref{ZnCTM_HC_elliptic_step2}) and exploiting that $q=e^{-\varepsilon}$, 
we find (\ref{EE_CTM_HC_elliptic}) for the entanglement entropy.
Then, by applying (\ref{defcapacity}) to (\ref{ZnCTM_HC_elliptic_step2}), one finds the capacity of entanglement in (\ref{CE_CTM_HC_elliptic}).

\subsection{Majorization in the harmonic chain}
\label{app:majorizationHC}

The corner transfer matrix results of \cite{Peschel} can be exploited to show that
the entanglement spectrum of a harmonic chain on the line with frequency $\omega_2$ 
corresponding to the bipartition of the line into two half-lines 
majorizes the entanglement spectrum obtained for $\omega_1<\omega_2$.

In this subsection and also in the following one, 
we consider the harmonic chain (\ref{HC ham}) with $\mu=1$ and $\lambda=1$, hence $\tilde{\omega}=\omega$.
When the subsystem of the harmonic chain in its ground state is half chain, 
the elements of the entanglement spectrum are given by \cite{Peschel} (see Sec.\ref{sec:CTM} for more details)
\be
\label{ES HC PeschelChung}
\lambda(n_1,n_2,n_3,\dots;\omega)=\prod_{k=0}^{\infty} e^{-n_k(2k+1)\varepsilon}\big(1-e^{-(2k+1)\varepsilon}\big)\equiv \prod_{k=0}^{\infty} \lambda_k(n_k;\omega)=\prod_{k=0}^{\infty} e^{-n_k(2k+1)\varepsilon}\lambda_k(0;\omega)\,,
\ee
where $\varepsilon=\varepsilon(\omega)>0$ has been introduced in (\ref{eps-omega-peschel}).
The eigenvalues (\ref{ES HC PeschelChung}) 
depend by infinitely many occupation numbers $n_k$ that can take non-negative integer values. 
Denoting by $\boldsymbol{\lambda}(\omega)$ the collection of all the eigenvalues (\ref{ES HC PeschelChung}),
our aim is to prove that $\boldsymbol{\lambda}(\omega_2)\succ \boldsymbol{\lambda}(\omega_1)$ when $\omega_2>\omega_1$.
The factorised structure of (\ref{ES HC PeschelChung}) allows us to exploit the lemma discussed in  \cite{Orus:2005jq} claiming
\be
\label{OrusLemma}
\textrm{if}\;\;\boldsymbol{\lambda}_k(\omega_2)\succ \boldsymbol{\lambda}_k(\omega_1)\;\; \textrm{when}\;\;\omega_2>\omega_1 \;\;\forall k
\qquad
\Longrightarrow
\qquad
\boldsymbol{\lambda}(\omega_2)\succ \boldsymbol{\lambda}(\omega_1)\;\; \textrm{when}\;\;\omega_2>\omega_1\,,
\ee
where $\boldsymbol{\lambda}_k\equiv \{\lambda_k(n_k;\omega)\, ; \,n_k\geqslant 0 \}$;
hence we can focus on the $k$-th mode.

In order to prove the l.h.s. of (\ref{OrusLemma}), 
we apply directly the definition in the footnote \ref{footnote:majorization} to $\boldsymbol{\lambda}_k $, 
which contains eigenvalues already ordered as $\lambda_k(0;\omega)>\lambda_k(1;\omega)>\lambda_k(2;\omega)>\dots$ 
(which is a consequence of (\ref{ES HC PeschelChung})) and satisfies the normalisation condition
\be
\sum_{n_k=0}^{\infty}\lambda_k(n_k;\omega)=1\,,
\;\;\qquad\;\;
\forall\, \omega, k\geqslant0\,.
\ee
Thus, we need to show that
\be
\label{derivative partial sum}
\frac{d}{d\omega}\sum_{n_k=0}^{N}\lambda_k(n_k;\omega)=\sum_{n_k=0}^{N}\frac{d}{d\omega}\lambda_k(n_k;\omega)\geqslant 0 \,,
\;\;\qquad\;\;
\forall\, \omega, k, N\geqslant0\,,
\ee
where, by using (\ref{ES HC PeschelChung}), we have that
\be
\label{derivative lambdak}
\frac{d}{d\omega}\lambda_k(n_k;\omega)
\,=\,
(2k+1) X\, \frac{d\varepsilon}{d\omega} \,\Big[ X^{n_k}-(1-X)\,n_k\, X^{n_k-1} \Big]\,,
\;\;\qquad\;\;
X\equiv e^{-(2k+1)\varepsilon}\,,
\ee
and $X\in[0,1]$, from $\omega,k\geqslant 0$.
Plugging (\ref{derivative lambdak}) into (\ref{derivative partial sum}) and exploiting the following relations 
\be
\sum_{n_k=0}^N X^{n_k}=\frac{1-X^{N+1}}{1-X}\,,
\,\,\qquad\,\,
\sum_{n_k=0}^N n_k X^{n_k-1}=\frac{1-(1+N)X^N+N X^{N+1}}{(1-X)^2}\,,
\ee
which is valid for $N\geq0$ and $X\in[0,1]$, we obtain 
\be
\frac{d}{d\omega}\sum_{n_k=0}^{N}\lambda_k(n_k;\omega)
=
(2k+1)(N+1)\,e^{-(2k+1)(N+1)\varepsilon}\,\frac{d\varepsilon}{d\omega}\,,
\ee
where also the relation between $X$ and $\varepsilon$ has been employed. 

Since $\varepsilon(\omega)$ in (\ref{eps-omega-peschel}) is a monotonically increasing function of $\omega$ when $\omega>0$, 
we have proved (\ref{derivative partial sum}) and, from (\ref{OrusLemma}), we have 
\be
\boldsymbol{\lambda}(\omega_2)\succ \boldsymbol{\lambda}(\omega_1)\,,
\;\;\;\qquad\;\;\;
\omega_2>\omega_1\,.
\ee
This tells us that a generic quantity defined as a Schur concave function of the entanglement spectrum 
is decreasing in the parameter $\omega$.
Since $S_A$ and $M_A^{(n)}$ defined in (\ref{Mn Tr Fn}) (with $b_n\geqslant n-1$) are Schur concave functions of the entanglement spectrum 
\cite{GeomQuantumStates_book,boes2020}, we conclude that
\be
\label{decreasing S and M}
\frac{d S_A}{d\omega}<0\,,
\;\;\;\qquad\;\;\;
\frac{d M_A^{(n)}}{d\omega}<0\,.
\ee
The relations in (\ref{decreasing S and M}) are confirmed by the behaviour of the curves in the right panel of Fig.\,\ref{fig:HC_1intmassive_CEvsEE}.

\subsection{Critical regime}
\label{sec:criticalCTM}

In the following we compute the critical limit of the results obtained in Sec.\,\ref{sec:CTM} 
through the corner transfer matrix  techniques  
and compare them with some results obtained from quantum field theory  \cite{ep-rev}.

Close to the critical point, the correlation length becomes large  $1\ll \xi <\infty$. 
In the harmonic chain the frequency $\omega = \xi^{-1}$; hence $\omega\ll 1$ close to the critical point. 
 By expanding (\ref{eps-omega-peschel}) in this regime, we 
 have that $\log \omega \simeq - \pi^2 /\varepsilon+ O(1)$, which implies 
 $\log \xi \simeq \pi^2 / \varepsilon + O(1)$,
 where $\varepsilon$ is defined in (\ref{eps-omega-peschel}).
Thus, the critical limit is achieved when $\varepsilon\to 0$.

In order to take $\varepsilon\to 0$ in (\ref{EE_CTM_HC}) and (\ref{CoE_CTM_HC}), 
we exploit the generalised Poisson resummation formula 
\be
\label{PoissonResummation}
\displaystyle \sum_{j=-\infty}^{\infty} f(|\varepsilon(b j+a) |)=\dfrac{2}{\varepsilon b}\displaystyle \sum_{k=-\infty}^{\infty}\hat{f} \left( \dfrac{2\pi k}{\varepsilon b}\right)e^{2\pi i k a/b}\,,
\ee
where
\begin{equation}
\label{eq:cosineFourier}
\hat{f}(y)=\displaystyle \int_0^{\infty}f(x) \cos (yx) \, dx\,,
\end{equation}
as done in \cite{cal2010} (see also \cite{mdc-20} for the application to the harmonic chain).

In the following we employ (\ref{PoissonResummation}) 
for $a=1/2$, $b=1$ and $f(x)=f_n(x)\equiv\log\!\big(1-e^{-2nx}\big)$. 
First one rewrites (\ref{ZnCTM_HC}) as
\bea
\label{eq:second}
 \log \textrm{Tr}\rho_A^n
&=&
\sum_{j=0}^{\infty} \big[ n f_{1}(\varepsilon(j+1/2))-f_{n}(\varepsilon(j+1/2)) \big]
\nonumber
\\
&=&
\frac{1}{2} \sum_{j=-\infty}^{\infty} \!\!
\big[ \,n f_{1}|(\varepsilon(j+1/2)|)-f_{n}(|\varepsilon(j+1/2))| \,\big]\,,
\eea
where the cosine Fourier transform (\ref{eq:cosineFourier}) of $f_n(x)$ is given by
\be
\label{CosFTf}
\hat{f}_n(y)= \frac{n}{y^2}-\frac{\pi \coth\big[\pi y/(2n)\big]}{2y}\,.
\ee
Then, by applying (\ref{PoissonResummation}) for (\ref{eq:second}) and using (\ref{CosFTf}), we obtain
\be
\label{eq:third}
 \log \textrm{Tr}\rho_A^n
=
\dfrac{1}{4 }\sum_{k=-\infty}^\infty \frac{(-1)^k}{ k } \, \bigg( \! \coth\big[\pi^2 k /(n\varepsilon)\big] -n \coth\big[\pi^2 k/\varepsilon\big]\bigg)\,.
\ee
Isolating the $k=0$ contribution, that gives the leading contribution in $1/\varepsilon$ when $\varepsilon\to 0$, 
and exploiting the fact that the argument of the sum is even in $k$, we obtain
\be
\label{eq:fourth}
 \log \textrm{Tr}\rho_A^n
\,=\,
\frac{\pi^2}{12 \,\varepsilon }\bigg(\frac{1}{n}-n\bigg)
+
\frac{1}{2 }\sum_{k=1}^\infty \frac{(-1)^k}{ k } \, \bigg(\! \coth\big[\pi^2 k /(n\varepsilon)\big] -n \coth\big[\pi^2 k/\varepsilon\big]\bigg)\,,
\ee
where $\coth(x)\simeq 1/x+x/3 +O(x^3)$ as $x\to 0$ has been used to get the first term. 
Taking $\varepsilon\to 0$ in the remaining sum and using that $\sum_{k=1}^\infty\frac{(-1)^k}{ k }=-\log 2$, we arrive to
\be
\log\textrm{Tr}\rho_A^n
=
\frac{\pi^2}{12 \,\varepsilon }\bigg(\frac{1}{n}-n\bigg)
-
(1-n)\frac{\log 2}{2}
+
O(\varepsilon)\,.
\ee
Then, from (\ref{defentropy}) and (\ref{defcapacity}) for the leading term of $S_A$ and $C_A$ we find
\be
\label{EntCapCriticalLimit}
S_A = C_A= \frac{\pi^2}{6 \varepsilon}
\simeq 
\frac{1}{6}\log \xi \,,
\ee
where (\ref{eps-omega-peschel}) has been exploited.
The critical limit of the entanglement entropy can be found also by taking $\omega\to 0$ in (\ref{EE_CTM_HC_elliptic-ep}), as done in \cite{ep-rev}.

These results can be compared with the corresponding ones found through quantum field theory methods.
Consider a massive field theory with mass $m$  that becomes a CFT with central charge $c$ as $m$ vanishes. 
When this model is in its ground state and the subsystem is the half-line, 
we have that \cite{Calabrese:2004eu} 
\be
\log\mathrm{Tr}\rho_A^n=-\frac{c}{12}
\bigg(\frac{1}{n}-n\bigg)
\log(m \epsilon)\,,
\ee
where $\epsilon$ is the UV cutoff.
From (\ref{defentropy}), (\ref{defcapacity}) and  $\xi \simeq m^{-1}$, at leading order one obtains
\be
\label{EntCapMassiveFT}
S_A=C_A =\frac{c}{6}\log\left(\frac{\xi}{\epsilon}\right),
\ee
where the result for $S_A$ has been found in \cite{Calabrese:2004eu}.
The massive harmonic chain in the continuum limit is the 
massive Klein-Gordon field theory in 1+1 dimensions, which becomes a specific CFT with $c=1$ in the massless limit. 
Setting $c=1$ in (\ref{EntCapMassiveFT}), we retrieve (\ref{EntCapCriticalLimit}) as expected,
once the UV cutoff $\epsilon$ is identified with the lattice spacing $a=1$.

\subsection{Capacity of entanglement in the XXZ chain}
\label{XXZCTM}

The corner transfer matrix allows to compute the capacity of entanglement also in the XXZ chain in the antiferromagnetic regime.
Similarly to the harmonic chain (see Sec.\,\ref{sec:CTM}),
we find that the lattice results for capacity of entanglement and entanglement entropy are different
and that  only in the critical limit  the leading terms of these two quantities coincide, 
consistently with the massive field theory predictions discussed at the end of Sec.\,\ref{sec:criticalCTM}.

The Hamiltonian of the anisotropic Heisenberg model (also called XXZ chain) is
\be
\label{eq:XXZ Hamiltonian}
H_{\textrm{\tiny XXZ}}
\,=\,
\sum_{j}
\left(
\sigma_j^x \sigma_{j+1}^x
+
\sigma_j^y \sigma_{j+1}^y
+
\Delta
\sigma_j^z \sigma_{j+1}^z
\right),
\ee
where $\sigma^\alpha$ with $\alpha=x,y,z$ are the Pauli matrices. 
The model has a quantum critical point for $\Delta=1$, it is gapless when $|\Delta|<1$ and gapped when $|\Delta|>1$. 
We consider this model in the antiferromagnetic gapped regime with $\Delta>1$.

Considering the bipartition of the infinite chain  into two half-chains, 
in \cite{PeschelKaulke99} it has been found that the entanglement Hamiltonian takes the form (\ref{EH_CTM}) with
\begin{equation}
\label{eq:singleparticlelevelXXZ}
\varepsilon_j=2 \varepsilon j\,,
\;\;\;\qquad\;\;\;
\varepsilon = \mathrm{arccosh}(\Delta)\,,
\end{equation}
and $n_j$ fermionic number operators.
This leads to \cite{cal2010}
\begin{equation}
\label{eq:firstXXZ}
\log  \textrm{Tr}\rho_A^n
=
\sum_{j=0}^{\infty}\log \big( 1+e^{-2 j n\varepsilon}\big)
-
\sum_{j=0}^{\infty} n\log \big( 1+e^{-2j \varepsilon} \big)\,,
\end{equation}
which implies
\be
\label{EE_CTM_XXZ}
S_A=- \partial_n \big[\log  \textrm{Tr}\rho_A^n\big] \! \big|_{n=1}
=
\sum_{j=0}^\infty
\left[ \, \frac{2\varepsilon j}{e^{2\varepsilon j}+1} +\log\! \big(1+e^{-2\varepsilon j}\big) \, \right]\,.
\ee
As for the capacity of entanglement, 
by applying the definition (\ref{defcapacity}) to (\ref{eq:firstXXZ}), we find
\be
C_A= \partial_n^2 \big[\log  \textrm{Tr}\rho_A^n\big]  \! \big|_{n=1}
=
\sum_{j=0}^{\infty} \bigg[ \frac{j\varepsilon}{\cosh(j\varepsilon)}
\bigg]^2\,.
\ee
Thus, the results for capacity of entanglement and entanglement entropy are different in the XXZ chain,
like in the harmonic chain considered in Sec.\,\ref{sec:CTM}.

The critical regime for this model corresponds to $\Delta\to 1$, where the system approaches its gapless phase. 
In terms of $\varepsilon$ defined in (\ref{eq:singleparticlelevelXXZ}), this limit is $\varepsilon\to 0$.
The relation between the correlation length of the model and $\varepsilon$ is \cite{baxterbook}
\begin{equation}
\label{eq:xifunctioneps}
\log \xi
\simeq
\frac{\pi^2}{2 \varepsilon} + O(1)\,,
\end{equation}
that gives $\xi \gg 1$ when $\varepsilon \to 0$, as expected.
In this regime,
by using the Poisson resummation formula, the critical limit of $\log \textrm{Tr}\rho_A^n$ gives \cite{cal2010}
\be
\log  \textrm{Tr}\rho_A^n
=\frac{\pi^2}{24 \varepsilon} \bigg(\frac{1}{n}-n\bigg) +(1-n) \frac{\log 2}{2}+O(\varepsilon)\,,
\ee
which gives the entanglement entropy \cite{Calabrese:2004eu}
\be
\label{EE_XXZ_crit}
S_A=
- \partial_n  \big[\log \textrm{Tr}\rho_A^n\big]\! \big|_{n=1}
=
\frac{\pi^2}{12 \varepsilon}  + \frac{\log 2}{2}+O(\varepsilon)
=
\frac{1}{6}\log\xi
+ \frac{\log 2}{2}+\dots\,,
\ee
and the capacity of entanglement 
\be
\label{CE_XXZ_crit}
C_A
=
 \partial^2_n \big[\log  \textrm{Tr}\rho_A^n\big]\! \big|_{n=1}
 =\frac{\pi^2}{12 \varepsilon} +O(\varepsilon)
=
\frac{1}{6}\log\xi+\dots\,,
\ee
where in the last step of (\ref{EE_XXZ_crit}) and (\ref{CE_XXZ_crit}) the relation (\ref{eq:xifunctioneps}) has been exploited. 
Thus, $S_A$ and $C_A$ have the same leading term in the critical regime,
similarly to harmonic chain (see the appendix \ref{sec:criticalCTM}).

In the continuum limit, the critical XXZ chain is described by a compact free bosonic field theory with central charge $c=1$,
where the compactification radius  is related to the parameter $\Delta$ of the lattice model \cite{Fradkin_book}. 
Comparing (\ref{CE_XXZ_crit}) with (\ref{EntCapMassiveFT}) with $c=1$, we have that $C_A$
in the critical limit of this lattice model matches the result expected from the underlying massive field theory.

\newpage

\bibliographystyle{nb}

\bibliography{bibTex_Capacity_Final-v2}

\end{document}